\documentclass[aps,pra,amssymb,onecolumn,showpacs,a4paper]{revtex4-1}

\usepackage{amsmath,amssymb,epsfig,graphicx,color,framed}
\usepackage{hyperref}
\usepackage{amsfonts} 
\usepackage{amsmath}
\usepackage{amssymb}
\usepackage{graphicx}
\usepackage{hyperref}
\usepackage{color}
\usepackage{natbib}  
\usepackage{bm} 
\usepackage{isomath} 
\usepackage{braket}
\usepackage{booktabs}

\newcommand{\vect}[1]{\vectorsym{#1}} 

\definecolor{dblue}{RGB}{0,0,0.8}

\newcounter{lecture}
\begin{document}

\title{Quantum kinetic theory of flux-carrying Brownian particles}

\author{Antonio A. Valido}
\email{alejandro.valido@urjc.es}

\affiliation{Nonlinear Dynamics, Chaos and Complex Systems Group,
Departamento de F\'isica, Universidad Rey Juan Carlos
Tulip\'an s/n, 28933 M\'ostoles, Madrid, Spain}

\date{\today}


\begin{abstract}
We develop the kinetic theory of the flux-carrying Brownian motion recently introduced in the context of open quantum systems. This model constitutes an effective description of two-dimensional dissipative particles violating both time-reversal and parity that is consistent with standard thermodynamics. By making use of an appropriate Breit-Wigner approximation, we derive the general form of its quantum kinetic equation for weak system-environment coupling. This encompasses the well-known Kramers equation of conventional Brownian motion as a particular instance. The influence of the underlying chiral symmetry is essentially twofold: the anomalous diffusive tensor picks up antisymmtretic components, and the drift term has an additional contribution which plays the role of an environmental torque acting upon the system particles. These yield an unconventional fluid dynamics that is absent in the standard (two-dimensional) Brownian motion subject to an external magnetic field or an active torque. For instance, the quantum single-particle system displays a dissipationless vortex flow in sharp contrast with ordinary diffusive fluids. We also provide preliminary results concerning the relevant hydrodynamics quantities, including the fluid vorticity and the vorticity flux, for the dilute scenario near thermal equilibrium. In particular, the flux-carrying effects manifest as vorticity sources in the Kelvin's circulation equation. Conversely, the energy kinetic density remains unchanged and the usual Boyle's law is recovered up to a reformulation of the kinetic temperature.
\end{abstract}

\maketitle
\tableofcontents

\section{Introduction}
The study of chiral fluids, understanding chirality in the sense of broken time reversal or broken
parity symmetry (or both), has drawn an incipient attention in several areas of physics as they host a rich variety of phenomena \cite{ganeshan20171,kaminski20141,lucas20141}, such as incompressibility \cite{bahcall19911}, topological waves \cite{souslov20191} or nondissipative viscosity \cite{avron19951} (coined odd viscosity). Notable examples include the analysis of triangular anomalies in relativistic quantum field theory \cite{son20131,son20132}, the chiral anomaly in condensed matter physics \cite{sekine20171}, or the odd viscosity of active Brownian particles in nonequilibirum statistical physics \cite{banerjee20171,han20201,hargus20201,hargus20201,Klymko20171,epstein20191}. In the realm of two dimensional (2D) spatial systems special attention has been devoted to the quantum Hall fluid (QHF) \cite{zee19951}, which is a quantum gas of electrons moving in a plane traversed by an external magnetic field. Importantly, the (Abelian) Chern-Simons (CS) theory has proved crucial in the understanding of the QHF \cite{dunne19991}, i.e. the CS action turns to be an essential ingredient in an effective description \cite{zee20121}. Indeed, the CS theory has been an active area of research over the last decades since it provides a field theoretic formulation for a vast range of physical phenomena beyond fluid mechanics \cite{jackiw20041}. Broadly speaking, the successes of the latter substantially relies on the fact that it encapsulates the notion of flux attachment responsible for the anyonic quasi-particle statistics \cite{wilczek19821}: that is, the particles are bound to a pseudomagnetic flux tube enable to induce an Aharonov-Bohm (AB) phase that carries out the statistics transmutation. Concretely, this concept permits to reinterpret the QHF as a 2D quantum gas of electrons tied to an even number of flux quanta \cite{jain19891}, termed composite particles. To controllably explore the basic physics behind the QHF, efforts have been made to envisage experimental platforms implementing the flux attachment idea. In particular, it was shown that identical impurities may behave as flux-tube-charged-particle composites owing to their interaction with the surrounding 2D bosonic bath \cite{yakaboylu20181,yakaboylu20201}. Simultaneously, it was found that a CS gauge field effectively emerges in a Bose-Einstein condensate coupled to a synthetic gauge field \cite{valenti20201} (see also \cite{correggi20191}).

Motivated by the relevance of the flux attachment notion in the treatment of 2D quantum many-particle systems, the CS theory was recently applied to open quantum systems in \cite{valido20191}. More precisely, by just demanding space-time locality as well as local U(1)-gauge invariance, the authors devised a rather fundamental microscopic description of the 2D Brownian motion (dubbed flux-carrying Brownian motion) in the context of the nonrelativistic Maxwell-Chern-Simons (MCS) electrodynamics \cite{deser19821}. This generalizes the conventional dissipative models \cite{weiss20121,grabert19881,caldeira20141} (e.g. the famous Caldeira-Leggett heat bath \cite{caldeira19831,caldeira19832}) to incorporate the flux attachment concept and to account for chirality from first principles (i.e. the microscopic constituents behave as "chiral" particles). Remarkably enough, the flux-carrying Brownian particle exhibits magneticlike properties beyond the celebrated Landau diamagnetism theory \cite{valido20201}, so that it may shed further light on the influence of chirality on the kinetics and hydrodynamics of 2D quantum fluids. Let us stress that this novel microscopic description is significantly distinct from most previous treatments within open quantum system theory to the best of our knowledge \cite{weiss20121,caldeira19831}, in particular, those related to the search for observable signatures of the Berry curvature \cite{misaki20181,yao20171} or spatial noncommutative effects  \cite{ghorashi20131,santos20171,halder20171,cobanera20161}.

\subsection{Goals and methods}

The main goal of our paper is to address the open quantum system dynamics of the $N$ flux-carrying Brownian particles endowed with harmonic interparticle interactions and confined by certain harmonic potential. To do so, we consider, as usual, an initial tensor-product state between the particles and the MCS environment: the particle system is assumed to be in an arbitrary state, whereas the environment is in a thermal equilibrium state at certain temperature. Under this prescription, we derive both the quantum kinetic equation, also known as quantum master equation, and the hydrodynamic conservation laws in the phase-space Wigner framework: the quantum system state will be represented by the Wigner quasiprobability distribution function, which replaces the familiar distribution function in the classical kinetic theory of gases \cite{schieve20091}. This framework has been extensively used in the kinetic study of the conventional Brownian motion in the realm of atomic physics, quantum optics \cite{agarwal19711,fleming20111,weiss20121}, as well as quantum field theory \cite{calzetta19881,calzetta20001,boyanovsky20051}. Our approach borrows from previous works \cite{calzetta19881,calzetta20001,calzetta20031,boyanovsky20051} in the context of path integral formalism, specifically we derive the quantum master equation by starting from the so-called nonequilibrium generating functional. The reason to follow this route are twofold: first, it will provide us valuable intuition to understand the consequences of the flux attachment upon the quantum kinetics and hydrodynamics, and second, the path integral formalism has played a major role in the description of quantum Brownian systems \cite{caldeira19831,caldeira19832,weiss20121,grabert19881}. Thanks to the separability property of the initial state, the nonequilibrium generating functional  characteristic of the flux-carrying Brownian particles can be readily obtained from the partition function provided in Ref.\cite{valido20201} after performing the Wick rotation \cite{weiss20121}. Once this is done, we then switch to the phase-space Wigner framework by following the prescription presented in Refs.\cite{calzetta20031,fleming20111,anisimov20091}. 

After the discussion of the general dissipative scenario, we focus the attention in the weak coupling regime, i.e. when the coupling between the environment and the particle system is small relative to the strength of the harmonic interaction (both the interparticle interaction and confining potential). Concretely, we carried out a Breit-Wigner approximation that retrieves a Fokker-Plank-type equation for the Wigner function of the $N$ flux-carrying particles. This encompasses the conventional Brownian motion (in the Markovian regimen) as a particular instance. We further obtain a mean-field kinetic equation by truncating the Bogolyubov-Born-Gree-Kirkwood-Yvon (BBGKY) hierarchy \cite{chavanis20061}, and finally derive the quantum balance equations in the low density limit, that is when the hydrodynamics collision terms or collective effects can be neglected. 

\subsection{Brief summary of results}

\begin{itemize}
    \item We derive for first time the quantum kinetic equation of the $N$ flux-carrying Brownian particles. By following a perturbative approach of the flux-attachement effects, we show that the quantum kinetic equation simultaneously contains an antysimmetric diffusive component and an environmental torque that closely resemblances the Lorentz force appearing in classical Brownian systems in presence of a constant magnetic field \cite{abdoli20201,abdoli20202,satpathi20191}, the mutual advection of 2D point vortices \cite{chavanis20081} or the active torque included in the paradigmatic description of chiral active fluids composed of dumbbells \cite{han20201,epstein20191,hargus20201,banerjee20171}. Yet, our results are substantially distinct as the flux-carrying Brownian particles exhibit vortex-like fluid dynamics under equilibrium conditions in contrast to the vast majority of previous chiral systems \cite{lucas20141,vuijk20191,banerjee20171}. Interestingly, we show that the single flux-carrying particle displays a persistent vortex flow at thermal equilibrium in the quantum regime.
    
    \item We extensively examine the hydrodynamic conservation laws characteristic of the flux-carrying Brownian motion in the dilute scenario (i.e. in the low density limit). For concreteness, we obtain the quantum balance equations for the number density, the stream velocity, the kinetic energy density, fluid vorticity and the circulation flux. We derive an extended Kelvin's circulation equation where the flux attachment effects play a major role as vorticity sources. Remarkably, we show that these are responsible for establishing a dissipationless chiral flow without the need of an external magnetic field \cite{abdoli20201,abdoli20202} or an intrinsic angular momentum \cite{markovich20211}. The kinetic energy density is, however, unaffected by the flux attachment effects and we get the usual balance equation from the standard Brownian motion.  
    
    \item The single-particle scenario is studied in detail. We show that the flux-carrying Brownian particle is subject to an environmental flux-like noise which is responsible for the aforementioned antisymmetric diffusive coefficient. Such flux noise is absent in previous treatments exploiting either an external/synthetic magnetic field \cite{misaki20181,cobanera20161,santos20171,yao20171,yakaboylu20181,yakaboylu20201} or an intrinsic angular momentum \cite{han20201,hargus20201,hargus20201,Klymko20171,epstein20191,banerjee20171}. Notably, we further show that the flow density contains a vortex-like transport coefficient that has none counterpart on the Fick's law of conventional diffusion \cite{mayorga20021}. We numerically compute the latter for an initial Gaussian state and encounter a similar pattern to the Lorentz flux found in the conventional Brownian motion in presence of an external magnetic field. Despite of this persistent vorticity, the flux-carrying Brownian motion retrieves the familiar Boyle's law up to a reformulation of the so-called kinetic temperature \cite{lagos20111}.
    
    \item We conclude that the novel flux-attachment effects rely on an environmental AB phase factor that turns into the aforementioned flux-like noise at the microscopic level. Concretely, we find out that the equilibrium features of the flux-carrying Brownian fluid are granted by an extended fluctuation-dissipation relation between the flux-noise and the environmental torque: intuitively, this implies that the strength of the injected energy in the particle system is balanced by the quantum thermal fluctuations responsible for the equilibrium dynamics. We show that such environmental AB phase at thermal equilibrium is completely wiped off in the high temperature limit in agreement with quantum mechanics arguments, and thereby, we eventually recover the equilibrium dynamics characteristic of the conventional Brownian motion in the classical regime. In particular, in the single particle scenario we prove that the novel vortex flow cancels in the high temperature limit, as well as the fluid vorticity and circulation flux. 
    
\end{itemize}

\subsection{Notation and organization}
All the open system dynamics occurs in the $x-y$ plane, with $\vect z$ denoting the unit vector perpendicular to the plane of motion. The indices $i, j, k, · · · = 1, 2, . . . ,N$ label the particles, whereas the Greek indices are reserved for the spatial components of either 2D vectors or $2\times2$ tensors (e.g. $\alpha,\lambda,... =1,2$). Vectors and matrices shall be denoted indifferently by bold letters, e.g. $\vect I_{n}$ shall represent the $n\times n$ identity matrix. The phase-space variables of the $N$-particle system is denoted by means of a 4$N$-dimensional vector $\vect x=(\vect q_{1},\cdots,\vect q_{N},\vect p_{1},\cdots,\vect p_{N})^{T}\in \mathbb{R}^{4N}$, where $\vect q_{i}\in \mathbb{R}^2$ stems for the spatial displacement of the $i$-particle with respect to a central position $\bar{\vect q}_{i}$, and $\vect p_{i}\in \mathbb{R}^2$ represents its canonical conjugate momentum. The quantum operators are distinguished from the phase-space variables by a hat symbol: for instance, the 2D position and momentum operator vectors of the $i$-particle are denoted by $\hat{\vect q_{i}}$ and $\hat{\vect p_{i}}$, respectively. Furthermore, $\check{f}$ and $\tilde{f}$ represent the real-time and imaginary-time Fourier transforms of certain function $f$, respectively. We shall also use the notation "flux" and "Brow" to label the flux-attachment and Brownian effects, respectively.

The text is intended to be accessible to readers from both condensed matter physics and quantum information communities, which explains a degree of redundancy and the presence of material which may be skipped by experts. Sec. \ref{Sec_FCBM} provides a quick introduction to the MCS microscopic description and an extensive comparison with previous treatments based on extended Caldeira-Leggett models, active Brownian motion or 2D point vortices. Sec. \ref{Sec_QBE} is completely devoted to illustrate the essential results found in the single-particle scenario: explicit calculations of the diffusive coefficients, equilibrium state and hydrodynamics quantities near thermal equilibrium are presented in \ref{Subsec_QKLT}, \ref{Sub_QHSP} and \ref{subsecQHALT}. The $N$-particle flux-carrying Brownian system is treated in full generality in Sec. \ref{Sec_QKT}, which contains the technical and detailed derivation of the quantum kinetic equation by starting from the nonequilibrium generating functional. Section \ref{Subsec_FPPC} focus the attention on the weak system-environment coupling regime and the Fokker-Planck-type equation is obtained. The hydrodynamics description of the flux-carrying Brownian motion is addressed in Sec. \ref{subsecHD}. Finally, we summarize and draw the main conclusions in Sec. \ref{OCR}. 

\section{Flux-carrying Brownian particles}\label{Sec_FCBM}
Before presenting our results, we briefly illustrate the microscopic model that characterizes the flux-carrying Brownian particles and highlight the differences with previous treatments where the microscopic components are chiral particles (e.g. active Brownian particles \cite{han20201,hargus20201,hargus20201,Klymko20171,epstein20191} or 2D point vortices \cite{chavanis20081}).

The open quantum system dynamics is captured by the recently introduced MCS description \cite{valido20191,valido20201}, which essentially distinguishes from the standard Brownian motion \cite{weiss20121,grabert19881,caldeira20141} in the fact that a dynamical pseudomagnetic flux tube is attached to each system particle. This must not be confused with the ordinary flux notion from the standard Maxwell electrodynamics (e.g. due to external magnetic fields). To see how arise such flux attachment, it is convenient to pay attention to the MCS electrodynamics. Let us consider a system composed of $N$ charged point particles minimally interacting with the MCS elctromagnetic field in the infinite $x-y$ plane. The Lagrangian density reads \cite{dunne19991,deser19821}
\begin{eqnarray}
L_{MCS}&=& \frac{1}{2}\big(\vect E^{2}-B^{2}\big)+\frac{\kappa}{2}\epsilon^{\mu\nu\lambda}A_{\mu}\partial_{\nu}A_{\lambda}+\vect{A} \cdot\vect{J}+A_{0}\rho, 
\label{LDAMCS}
\end{eqnarray}
with $\epsilon^{\mu\nu\lambda}$ being the completely antisymmetric tensor (i.e. $\epsilon_{012}=1$ and $\epsilon_{ij}=\epsilon_{0ij}$) and $\kappa \in \mathbb{R}$ being the so-called CS constant. The first term in the righ-hand side of Eq. (\ref{LDAMCS}) corresponds to the usual Maxwell kinetic term, whilst the second term represents the CS action. Here, $B$ and $\vect E$ represents the magnetic and electric fields (i.e. $B=\epsilon_{\alpha\beta}\partial_{\alpha}A_{\beta}$ and $E_{\alpha}=-\dot{A}_{\alpha}-\partial_{\alpha}A_{0}$), whereas $\rho$ and $\vect{J}$ are respectively the charge and current densities of the particle system. From Eq.(\ref{LDAMCS}) follows an extended Gauss law \cite{deser19821}, i.e. 
\begin{equation}
    \nabla\cdot\vect E-\kappa B+\rho=0,
    \label{GaussL}
\end{equation}
where explicitly appears the magnetic field. By appealing to the Stokes-Gauss theorem (and due to the electric field decays asymptotically as the photons get massive) once we have integrated (\ref{GaussL}) over the infinite plane, we obtain the pseudomagnetic flux associated to the particle system,
\begin{equation}
    \Phi_{CS}=\int d \vect x B(\vect x)=\frac{qN}{\kappa},
    \label{FluxL}
\end{equation}
where $q$ is the fundamental charge of the system particles. Equation (\ref{FluxL}) reveals that each system particle formally consists of a magneticlike flux along the z-axis and with strength proportional to $q/\kappa$ \cite{deser19821,dunne19991}. Notice that this is exclusively due to the CS action introduced in (\ref{LDAMCS}), otherwise it completely disappears in the pure Maxwell electrodynamics (i.e. when one takes $\kappa\rightarrow 0$ in Eq.(\ref{GaussL})). According to basic electrodynamics, one may expect that the presence of a magneticlike flux gives rise to a magnetiglike vector potential (for instance, static charges are able to generate simultaneously electric and magneticlike fields in MCS electrodynamics \cite{moura20011}). Indeed, the authors in Ref.\cite{valido20201} show that flux-carrying Brownian particles are subject to a dynamical pseudomagnetic field at the microscopic level, so they can be though of as "dressed" dissipative particles in much the same fashion as composite particles in condensed matter physics. Here, it is important to emphasize that such pseudomagnetic field completely arises from the coupling with the environment, rather than from an external source as occurs in the conventional Brownian motion in presence of an auxiliary magnetic field.  

The MCS description, which is the basis of the present work, is consistent with the (non-relativistic) MCS electrodynamics in the long-wavelength (i.e. dipole approximation \cite{flick20191}) and low-energy regime (i.e. small displacement approximation), where the MCS electromagnetic field plays the role of a heat bath \cite{valido20191}. This is further discussed in in the following section.

\subsection{Microscopic model}\label{Sub_MICm}
Our open quantum system consists of an array of $N$ interacting particles of identical mass $m$ which are constrained to move in the $x-y$ plane around equilibrium positions $\bar{\vect q}_{i} \in \mathbb{R}^2$ by certain harmonic potential. The confining harmonic potential as well as the harmonic interpaticle interaction is accounted by a $2N\times2N$ symmetric matrix $\vect U$ in the phase space. The MCS electromagnetic field can be viewed as an ensemble of harmonic oscillators with masses $m_{\vect k}$ and excitation frequencies $\omega_{\vect k}$ with $\vect k\in 2\pi/L \ \mathbb{Z}^2$ given by the dispersion relation,
\begin{equation}
 \omega_{\vect k}^2=c^2|\vect k|^2+\kappa^2,
 \label{EDRW}
\end{equation}
where $c$ is the speed of light (or the sound velocity of the MCS environment in a more general dissipative scenario), and  $L$ is a characteristic length of the MCS environment. In its simplest version, it was shown in Ref.\cite{valido20191} that the dissipative dynamics is captured by the following microscopic model expressed in terms of the Lagrangian (see App.\ref{app1} for further details),
\begin{equation}
\hat{\mathcal{L}}=\hat{\mathcal{L}}_{\text{Sys}}+\hat{\mathcal{L}}_{\text{MCS,I}},
\label{EAE}
\end{equation}
with
\begin{align}
\hat{\mathcal{L}}_{\text{Sys}}&=\sum_{i=1}^{N}\frac{m}{2}\bigg(\dot{\hat{\vect q}}_i^2-\sum_{j=1}^{N}\sum_{\alpha,\lambda=1,2}(U_{\text{ren}})_{\alpha\lambda}^{ij}\hat q^{\alpha}_{i}\hat q^{\lambda}_{j}\bigg),
\label{ALSys} \\
\hat{\mathcal{L}}_{\text{MCS,I}}&=\sum_{\vect k\in \mathbb{R}^2}\frac{m_{\vect k}}{2}\Bigg(\dot{\hat{x}}_{\vect k}^2-2\sum_{j=1}^{N}\sum_{\alpha=1,2}\frac{g_{\alpha}(\vect k, \bar{\vect q}_{j}) }{m_{\vect k}}\dot{\hat{x}}_{\vect k}\hat q^{\alpha}_{j} -\omega_{\vect k}^2\left(\hat{x}_{\vect k}-\sum_{j=1}^{N}\sum_{\alpha=1,2}\frac{l_{\alpha}(\vect k, \bar{\vect q}_{j})}{\omega_{\vect k}m_{\vect k}}\hat q^{\alpha}_{j}\right)^2\Bigg),
\label{ALMCSI}
\end{align}
where $\hat x_{\vect k}$ represents the quantum position operator of the environmental $\vect k$-mode, we have introduced the auxiliary coupling coefficients (cf. Eqs. (\ref{gCC}) and (\ref{lCC})) and $(\vect U_{\text{ren}})_{\alpha\lambda}^{ij}$ stands for the elements of the $2N\times 2N$ matrix encoding the renormalized harmonic potential, i.e.
\begin{equation}
\hat U_{\text{ren}}(\hat{\vect q}_{1},\cdots,\hat{\vect q}_{N}) =\frac{m}{2}\sum_{i,j=1}^{N}\sum_{\alpha,\lambda=1,2}\big(U_{\alpha\lambda}^{ij} -\phi_{\alpha\lambda}(\Delta\bar{\vect q}_{ij})\big)\hat q^{\alpha}_{i}\hat q^{\lambda}_{j},\label{HCPR}
\end{equation}
where $\Delta\bar{\vect q}_{ij}=\bar{\vect q}_{i}-\bar{\vect q}_{j}$ quantifies the average distance between the oscillators $i$ and $j$, and $U_{\alpha\lambda}^{ij}$ is a shorthand notation for the matrix elements of $\vect U$. Aside the gapped spectrum (\ref{EDRW}), it follows from (\ref{HCPR}) that the CS action introduces a potential renormalization as well, namely $\phi_{\alpha\lambda}(\Delta\bar{\vect q}_{ij})$ (cf. Eq.(\ref{QRP1})). In the phase-space Wigner framework, the latter is encoded by the $2N\times2N$ renomalized potential matrix, say $\vect U_{\text{ren}}$, with elements given by
\begin{equation}
    ( \vect U_{\text{ren}})_{\alpha\lambda}^{ij}=U_{\alpha\lambda}^{ij}-\phi_{\alpha\lambda}(\Delta\bar{\vect q}_{ij}).
    \label{RPME}
\end{equation}
Importantly, the CS action is responsible for the second term in the right hand side of Eq.(\ref{ALMCSI}), which manifests the underlying time-reversal symmetry breaking at the Lagrangian level. This represents a bilinear interaction term of type velocity-position between the environmental and system particle degrees of freedom. To the best of our knowledge, this term has none counterpart in previous microscopic dissipative descriptions, for instance, it is absent in the famous Caldeira-Leggett model and recent extended treatments \cite{yao20171}. We will return to this point in the following section.

As stated in the introduction, our starting point is the partition function derived from (\ref{EAE}) when the coupled system-environment complex, composed of the 2D harmonic particles and the environmental MCS field, is in a canonical equilibrium state at inverse $\beta=1/k_{B}T$ (that is $e^{-\beta \hat H_{\text{Sys}-MCS}}$ up to normalization, with $\hat H_{\text{Sys}-MCS}$ being the Hamiltonian obtained from (\ref{EAE}) by the Legendre transform). This was computed in Ref.\cite{valido20201} and it permitted to extensively analyse the free energy, internal energy, the entropy and the heat capacity of the flux-carrying Brownian motion.  Let us stress that the motivation here is substantially distinct from Ref.\cite{valido20201}, since the present work is in the line to study the open quantum dynamics when the coupled system-environment complex is initially in a tensor-product state $\hat \rho_{0}\otimes \hat \rho_{\beta}$: the particle system is in an arbitrary state $\hat \rho_{0}$, while the environment is supposed to be in a canonical equilibrium state $\hat \rho_{\beta}$ characterized by the inverse temperature $\beta$ (that is $e^{-\beta \hat H_{MCS}}$ up to normalization, with $\hat H_{MCS}$ being the Hamiltonian obtained from (\ref{ALMCSI}) by the Legendre transform after ignoring the system-environment coupling).

The partition function, which characterizes the quantum thermodynamic properties of the flux-carrying Brownian particles, is better illustrated in terms of the imaginary-time path integral \cite{valido20201} (the interesting reader can find further details in App \ref{app1}),
\begin{equation}
    Z_{\text{Brow-flux}}=\oint \mathcal{D}\underline{\vect q}(\cdot)\ \text{exp}\left\lbrace-\big(S_{\text{Brow}}^{(E)}[\{\vect q(\cdot)\}]+S_{\text{flux}}^{(E)}[\{\vect q(\cdot)\}]\big)/\hbar\right\rbrace,
    \label{ZIFBP}
\end{equation}
with $\underline{\vect q}(\tau)=\prod_{i=1}^{N} \vect q_{i}(\tau)$, and $\tau$ being the imaginary time \cite{weiss20121}. The first term in the right-hand side of (\ref{ZIFBP}), denoted by $S_{\text{Brow}}^{(E)}$, is the familiar effective action describing the conventional Brownian motion (cf. Eq.(\ref{EASBR})), while the second term, denoted by $S_{\text{flux}}^{(E)}$, completely emerges from the flux attachment and, therefore, encapsulates the environmental CS effects at the thermodynamic level. This can be compactly expressed as follows
\begin{align}
S_{\text{flux}}^{(E)}[\{\vect q\}]=\sum_{i,j=1}^{N}&\sum_{\alpha,\lambda=1,2}\int_0^{\hbar\beta}d\tau \ \int_{0}^\tau d\tau' \ \Big( -\phi_{\alpha\lambda}(\Delta\bar{\vect q}_{ij})\ \delta(\tau-\tau') \nonumber \\
&+\text{Re}\ \check{\Lambda}^\parallel_{\alpha\lambda}(\tau-\tau',\Delta\bar{\vect q}_{ij})+i\ \text{Im}\ \check{\Lambda}^\perp_{\alpha\lambda}(\tau-\tau',\Delta\bar{\vect q}_{ij})\Big)q_i^{\alpha}(\tau) q_j^{\lambda}(\tau'),\label{EASF}
\end{align}
where $ \check{\vect \Lambda}^{||}$ and $ \check{\vect\Lambda}^{\perp}$ are, respectively, the imaginary-time Fourier transforms of the longitudinal and transverse dynamical susceptibilities (cf. Eqs. (\ref{SDCII}) and (\ref{SDCIII})). It turns out that $ \check{\vect\Lambda}^{||}$ is a real functional like the conventional dissipation kernel (which is given by Eq.(\ref{SDCI})). Since the real part of the effective action is related to relaxation \cite{weiss20121}, the longitudinal dynamical susceptibility is not expected to enrich the open quantum system dynamics beyond the standard Brownian motion. Indeed, in Sec.\ref{Subsec_FPPC} we shall see for the weak system-environment coupling that the flux-carrying effects stemming from $ \check{\vect\Lambda}^{||}$ can be recast into the dissipation kernel, and thus, they just modify the friction tensor. By contrast, the pure imaginary term $ \check{\vect\Lambda}^{\perp}$ in (\ref{EASF}) plays the role of a AB phase-like factor in the Boltzmann weight, thereby it must represent a dissipationless environmental mechanism. For instance, $ \check{\vect\Lambda}^{\perp}$ recalls the ordinary Hall action of two-dimensional particles found in the dissipative Hofstadter model \cite{callan19921,novais20051} or the dissipationless Hall term in the phonon effective action of a crystal hosting a gapped time-reversal symmetry-breaking electronic state \cite{barkeshli20121}. However, it is important to realize that the phase source here arises out from the dynamical pseudomagnetic field mentioned in the previous section: that is, it is consequence of the environmental interaction with the dynamical gauge field representing the MCS environment \cite{valido20201}, rather than being generated by a synthetic magnetic or auxiliary gauge field \cite{doniach20011}. We also notice that (\ref{EASF}) has none clear topological significance since the attached flux is not necessarily quantized \cite{deser19821} (e.g. a proportionality to a topological winding number), though it has to do with special topology of 2D systems \cite{tanaka20151}. By beginning from Eq.(\ref{ZIFBP}), we shall derive the quantum master and balance equations in Sec. \ref{Sec_QKT}. We anticipate that the bipartite structure of the Euclidean action in terms of the standard Brownian and the flux-carrying (\ref{EASF}) actions is preserved along the kinetic and hydrodynamic equations for weak system-environment coupling, so hereafter we will be able to clearly distinguishes the contribution due to the flux attachment. 

As a final remark, we would like to emphasise that the model (\ref{EAE}) eventually returns the relaxation of the particle system into a thermal equilibrium state (with temperature set by the MCS environment) provided the CS coupling constant in (\ref{LDAMCS}) remains small in comparison with the confining potential \cite{valido20191,valido20201} (e.g. for a harmonic particle of width $\sigma$ and frequency $\omega_{\text{ren}}$ it is found the subsidiary condition $
e^2\kappa^2/m \sigma \omega_{\text{ren}}\ll 1$). This will allow us in Sec. \ref{Subsec_FPPC} to perform a second-order perturbative analysis of the flux-attachment effects relying on an appropriate Breit-Wigner approximation. In the opposite limit, when the pure CS action dominates over the Maxwell action in (\ref{LDAMCS}), there would be no environmental dynamics supporting an irreversible transference of energy coming from the reduced system because the CS action has a vanishing Hamiltonian \cite{dunne19991}. In other words, the pure topological electrodynamics alone is not consistent with a dissipative scenario \cite{valido20201}. From here one may conclude that the Brownian action $S_{\text{Brow}}^{(E)}$ (responsible for the dissipative dynamics) results from the Maxwell action in (\ref{LDAMCS}), whilst the flux-attachment action $S_{\text{flux}}^{(E)}$ completely relies on the CS action. From quantum mechanical arguments follows that $\check{\vect \Lambda}^{\perp}$ solely introduces pure quantum effects in the equilibrium dynamics, and thus, its influence over the kinetics is expected to be completely wiped out at thermal equilibrium in the classical regime \cite{valido20201}. In particular, we shall show for the single-particle scenario in Sec. \ref{Subsec_QKLT} that the usual Brownian kinetics \cite{weiss20121,grabert19881,caldeira20141} is eventually recovered in the high temperature limit.

\subsection{Comparison with previous works based on extended Caldeira-Leggett models}\label{CLcomp}

The Lagragian (\ref{EAE}) is a fundamental as well as simple dissipative microscopic description that contains the essential ingredients demanded for a sensible 2D electrodynamic theory. We understand simplicity in the sense that the environmental degrees of freedom can be analytically integrated out to obtain the open quantum system dynamics, whereas its universality relies on the fact that the MCS electrodynamics is considered to be the most general (Abelian) gauge theory \cite{dunne19991,deser19821,pachos20121}. The latter means that our microscopic description preserves space-time locality and local $U(1)$ gauge invariance. Equivalently, the Lagrangian (\ref{EAE}) represents a minimal-coupling theory of the particle system and the MCS electromagnetic field acting as a heat bath \cite{valido20191}. This feature is common to the famous Caldeira-Leggett model \cite{ford19881,kohler20131}, also known as the independent harmonic oscillator model \cite{ford19881}, which is the foundation of the microscopic treatments of the conventional Brownian motion \cite{valido20131,caldeira19831,grabert19881,caldeira20141,devega20171}. The latter is fully contained in our microscopic description (\ref{EAE}) as a particular instance \cite{kohler20131}: the second term in Eq.(\ref{ALMCSI}) completely disappears by disregarding the CS action, and thus, Eq.(\ref{EAE}) returns the Lagrangian characteristic of the Caldeira-Leggett model \cite{weiss20121,caldeira19831,grabert19881,caldeira20141}. Indeed, we shall see that we recover the well-known results of the quantum kinetic theory of the standard Brownian motion when the flux attachment is turned off (i.e. $\kappa \rightarrow 0$) \cite{chavanis20101,mayorga20021,chandrasekhar19431,agarwal19711,vacchini20091,chavanis20061,chavanis20062,chavanis20111,fleming20111,hu19921}.

The Lagrangian (\ref{EAE}) also represents an alternative description to those treatments that go beyond the pure Caldeira-Leggett model by taking account the Berry curvature of internal degrees of freedom \cite{misaki20181,yao20171}, or by endowing the system particles with either spatial or momentum non-commutative relations \cite{cobanera20161,santos20171}. On one side, the authors from Ref.\cite{yao20171} provide an extended version of the independent-harmonic oscillator model by introducing time-reversal-symmetry-breaking interacting terms (e.g. $\hat{x}_{\vect k}\dot{\hat{q}}^{\alpha}_{i}$). Contrary to the model (\ref{EAE}), this does not guarantee the gauge invariance for a given choice of the system-environment coupling coefficient, and thus, this is not prevented from spurious predictions coming from the specific choice of the coordinate-reference system. Similarly, the authors from Ref.\cite{misaki20181} emulate the flux attachment by engineering an static AB phase-like via exploiting the Berry curvature due to the spin degrees of freedom. By contrast, the flux attachment encoded by $ \check{\vect\Lambda}^{\perp}$ is dynamic, which implies that the geometric phase effects studied here stem in the own open quantum dynamics. Recall that the present flux attachment arises from a pseudo-magnetic field that explicitly depends of the environmental degrees of freedom, which yields that $\check{\vect\Lambda}^{\perp}$ would be a function of the environmental spectral density (see Eq.(\ref{SDCIII})). On the other side, the authors in Ref.\cite{cobanera20161,ghorashi20131,santos20171,halder20171} introduce non-commutative relations between system particles degrees of freedom regardless of the dissipative dynamics. Instead, the CS action give rises to an (equal-time) non-commutative relation between the environmental degrees of freedom (see Eq.(\ref{MCSEF}) in the following section) \cite{valido20201,valido20191}. Overall, the novel effects studied here ultimately relies on the open quantum system dynamics itself, in contrast to the vast majority of preceding works based on the Caldeira-Leggett model (included the Brownian motion subject to an external magnetic field).  We also note that the treatment in \cite{yakaboylu20181,yakaboylu20201} differs from the one provided by Eq.(\ref{EAE}): the flux tube in the latter is self-induced by an emergent (external) gauge field within the classical Fr\"ohlich-Bogoliubov theory (which is a particular instance of the standard Maxwell electrodynamics), rather than a pseduomagnetic field as occurs in the present description. 

Finally, our work also bears important similarities with  those treatments addressing particles with an intrinsic angular momentum \cite{markovich20211,han20201,hargus20201,hargus20201,Klymko20171,epstein20191,chavanis20081}. Since these systems play an important role in chiral fluids, we devote the following section to elaborate on the similarities and differences with the flux-carrying Brownian particles.

\subsection{Comparison with active Brownian particles and 2D point vortices}\label{Sub_COMP2DV}

Let us carried out our discussion in terms of the generalized Langevin equation, as it is usually the starting point in the vast majority of previous treatments of active Brownian particles \cite{han20201,hargus20201,hargus20201,Klymko20171,epstein20191,chavanis20081} or 2D vortices \cite{chavanis20081}. In the single particle scenario (i.e. $N=1$), we find out in Sec. \ref{Subsec_FPPC} that this can be cast in the following form
\begin{equation}
m\ddot{\hat{\vect q}}+m\omega_{\text{ren}}^{2}\hat{\vect q}+m\gamma_{\text{Brow}}\dot{\hat{\vect q}}-m\Omega_{\text{flux}}^{2}\vect z\times \hat{\vect q} =\hat{\vect \xi}_{\text{Brow}}+\hat{\vect \xi}_{\text{flux}},
\label{QLEMASP}
\end{equation}
where $\omega_{\text{ren}}>0$, $\gamma_{\text{Brow}}>0$ and $\hat{\vect \xi}_{\text{Brow}}$ (with null average value) denote, respectively, the usual renormalized potential frequency, the (Stokes) friction coefficient, and the environmental fluctuating force that are common to the conventional Brownian motion (e.g. $\hat{\vect \xi}_{\text{Brow}}$ corresponds to the Gaussian white noise in the classical regime). Additionally, the open system dynamics of the flux-carrying Brownian particle involves a fourth term in the right hand side of (\ref{QLEMASP}), which has none counterpart in the microscopic descriptions discussed in the previous section \cite{kohler20131,yao20171}. This must not be confused with the Lorentz force produced by an external magnetic field $\vect B_{ext}$ (recall that the Lorentz force is proportional to $\dot{\hat{\vect{q}}}\times \vect B_{ext}$). Instead this represents an environmental torque of strength $\Omega_{\text{flux}}$ encoding the CS coupling  (that is, the CS effects cancel by taking $\Omega_{\text{flux}}\rightarrow 0$). Interestingly, this flux-attachment effect resemblances the active torque of distinct chiral particle systems. The best known examples include the active Brownian particles \cite{han20201,hargus20201,hargus20201,Klymko20171,epstein20191} or 2D point vortices \cite{chavanis20081}. Indeed, the Brownian terms of Eq.(\ref{QLEMASP}) together with such rotational force identically coincides with the generalized Langevin equation associated to the toy model recently studied in \cite{han20201}, which consists of frictional granular particles driven by large active torques with equal strength. Similarly, the stochastic equation of the Brownian motion of 2D point vortices in the overdamped limit, when the inertial term is negligible (i.e. $\ddot{\hat q}\approx 0$), takes a closed form to Eq.(\ref{QLEMA}) with advective term given by $m\omega_{\text{ren}}^2$ \cite{chavanis20081}. For the $N$-particle dissipative scenario, we shall see in Sec. \ref{Subsec_FPPC} that this flux contribution to the generalized Langevin equation also gives rise to an environmental-mediated interaction among the flux-carrying particles that recalls the long range interaction between the flux lines in a vortex liquid \cite{feigelman19931} or the transverse interaction due to self-spinning in chiral active fluids \cite{han20201,hargus20201}. Hence, one may expect that the flux-carrying Brownian particle will exhibit an intricate vortex-like dynamics that is completely beyond the conventional Brownian motion subject to an external magnetic field.

Beside the usual Brownian fluctuating force, the flux-carrying particle is subject to an additional environmental force $\hat{\vect \xi}_{\text{flux}}$ in contrast to the aforementioned dissipative chiral systems. We shall see in Sec. \ref{Sec_QKT} that the $\hat{\vect \xi}_{\text{flux}}$ completely stems from the transverse dynamical susceptibility $\check{\vect\Lambda}^{\perp}$, so it completely disappears when the CS action is disregarded. It was shown in detail in \cite{valido20191} that $\hat{\vect \xi}_{\text{flux}}$ identifies with a non-conservative electric field which satisfies a non-commutative relation, i.e.
\begin{equation}
\left[  \hat \xi^{\alpha}_{\text{flux}}, \hat \xi^{\beta}_{\text{flux}}\right]\propto -i \kappa \epsilon_{\alpha\beta},\label{MCSEF}
\end{equation}
recall that $\epsilon_{\alpha\beta}$ denotes the totally asymmetric tensor in 2D. In a nutshell, the fluctuations of the pseudomagnetic field owing to the flux attachment induced such electric field according to Faraday's law. It turns out that the mean average value of $\hat{\vect \xi}_{\text{flux}}$ cancels (i.e. $\left\langle\hat{\vect \xi}_{\text{flux}} \right\rangle_{\hat \rho_{\beta}}=0$), so that the mean average motion of the flux-carrying particle may evoke the vortex dynamics of the active Brownian systems. However, while the interesting applications of the vast majority of these systems imply highly nonequilibrium conditions \cite{vuijk20191,lucas20141,banerjee20171}, we shall show in Sec.\ref{Sec_QBE} that the flux-carrying Brownian motion is able to support vortex flow in thermal equilibrium conditions. As anticipated in the preceding sections, provided the system satisfies appropriate conditions (these are fully discussed in Sec. \ref{Sec_QKT}), Eq.(\ref{QLEMASP}) becomes asymptotically stable, and thus, the flux-carrying Brownian particle behaves ergodic \cite{weiss20121}. In our model this occurs because there is a complex interplay between $\hat{\vect \xi}_{\text{flux}}$ and the environmental torque. Concretely, upon considering that the MCS environment is in a canonical equilibrium state at inverse temperature $\beta$, we find out that the flux fluctuating force satisfies an extended fluctuation-dissipation relation (see Eq.(\ref{FDTHBW}) for further details), i.e. 
\begin{equation}
\left\langle \left\lbrace \hat \xi_{\text{flux}}^{\alpha}(t),\hat \xi_{\text{flux}}^{\beta}(t')\right\rbrace \right\rangle =- \epsilon_{\alpha\beta}2m\beta^{-1}\Omega_{\text{flux}}^{2}\coth\bigg(\frac{\pi (t-t')}{\hbar \beta}\bigg).\label{FDRF}
\end{equation}
The above relation exhibits the parity breaking and time-reversal asymmetry of the underlying microscopic model (\ref{EAE}): the time-reversal transformation $(t,t')\rightarrow (-t,-t')$ produces a change of sign since the hyperbolic cotangent is an odd function, as well as a 2D parity transformation $q_{x}\rightarrow -q_{x}$. Nonetheless, the fluctuation-dissipation relation remains invariant under a simultaneous parity and time-reversal transformation. The Markovian limit is recovered in time scales $1\ll  (t-t')/\hbar \beta$ (such that $\coth(\pi (t-t')/\hbar \beta)\rightarrow \text{sgn}(t-t')$), Eq.(\ref{FDRF}) then simplifies to 
\begin{equation}
\left\langle \left\lbrace \hat \xi_{\text{flux}}^{\alpha}(t),\hat \xi_{\text{flux}}^{\beta}(t')\right\rbrace \right\rangle =- \epsilon_{\alpha\beta}2m\beta^{-1}\Omega_{\text{flux}}^{2}\text{sgn}(t-t'),\label{FDRFML}
\end{equation}
with $\text{sgn}(t)$ denoting the sign function. According to the linear response theory, from Eqs. (\ref{FDRF}) and (\ref{FDRFML}) follows that $\hat{\vect \xi}_{\text{flux}}$ is responsible for a transverse linear-response coefficient. This point will be cleared in the next section when we deal with the single flux-carrying Brownian particle. Notably, the flux-carrying term in (\ref{FDRFML}) closely resemblances the ordinary Hall response of two-dimensional particles found in the dissipative Hofstadter model \cite{callan19921,novais20051}. Actually, the Fourier transform of the statistical correlator (\ref{FDRFML}) in the frequency domain identically coincides with the fluctuations of an antisymmetric 1/f noise of power spectrum $S(\omega) = (i\omega)^{-1}$ (to see this one must realize in (\ref{FDRM2}) that $\coth\left(\hbar \omega\beta/2\right)$ renders an inverse scaling for the power spectrum of the transversal correlations, while the spectral density contributes with a constant value in the Breit-Wigner approximation). Recalling that the $1/f$ power law is characteristic of the low-frequency magnetic flux noise in superconducting circuits \cite{anton20131,lee20081}, Eq.(\ref{FDRFML}) suggests that $\hat{\vect \xi}_{\text{flux}}$ can be thought of as a flux-like noise as well. 

In summary, the novel essential feature of the flux-carrying Brownian motion in the weak coupling regime is the additional fluctuating force that satisfies both the extended fluctuation-dissipation relation (\ref{FDRF}) and the (equal-time) non-commutative relation (\ref{MCSEF}). This can be regarded as a flux-like noise primarily generated by the dynamical flux attachment: that is, the erratic motion of the system particles gives rise to a fluctuating pseudomagnetic flux which in turn induces an intricate electromotive force responsible for $\hat{\vect \xi}_{\text{flux}}$. We shall see in the next section that the latter gives rises to a nondifussive transport process perpendicular to density gradients. As a consequence, the flux-carrying Brownian particles may display a vortex-like dynamics under thermal equilibrium conditions. This is in stark contrast with the vast majority of previous parity-violating or time-reversal-violating descriptions involving an intrinsic spinning force, such as active Brownian particles, where the detailed balance is broken due to the continuous injection of energy coming from the active torque \cite{banerjee20171}. 

\section{Quantum kinetic theory of the single flux-carrying Brownian particle}\label{Sec_QBE}

In order to illustrate the quantum kinetic theory that emerges from the microscopic description (\ref{EAE}), it is instructive to focus the attention on the single particle scenario (i.e. $N=1$) when the system-environment interaction is weak in comparison with the strength of both the confining potential or the harmonic interparticle interaction. This permit to make a clear expositions of our main results and to compare them with the case of the 2D conventional Brownian motion in presence of an external magnetic field \cite{vacchini20091} or an active torque \cite{hargus20201,han20201}. The detailed derivation of these results (quantum master equations, quantum balance equations, etc) and the quantum kinetic theory of a general dissipative scenario can be found in Sec. \ref{Sec_QKT}.

Let us consider the single flux-carrying Brownian particle with mass $m$, isotropic renormalized frequency $\omega_{\text{ren}}$, and isotropic friction coefficient $\gamma_{\text{Brow}}$. Its time evolution is governed by the extended Langevin equation (\ref{QLEMASP}). As illustrated in the preceding section, the strength of the CS effects shall be characterized by $\Omega_{\text{flux}}$.

\subsection{Quantum master equation in the weak coupling regime}
In our treatment, the quantum state at certain time $t$ is represented by a Wigner quasidistribution function, denoted by $W(\vect x,t)$, where $\vect x=(\vect q,\vect p)\in \mathbb{R}^4$ is a point of the particle phase space (notice that $\vect q=(q_{x},q_{y})$ and  $\vect p=(p_{x},p_{y})$). As detailed in Sec. \ref{Subsec_FPPC}, the quantum evolution of the single flux-carrying Brownian particle is dictated by an extended Kramers equation, 
\begin{align}
\bigg(\frac{\partial }{\partial t}+\frac{\vect p}{m}\cdot\frac{\partial}{\partial \vect q}-m\omega_{\text{ren}}^2\vect q\cdot\frac{\partial}{\partial \vect p} \bigg)W(\vect x,t)&=\frac{\partial}{\partial \vect p}\cdot\bigg(\gamma_{\text{Brow}} \vect p+2D^{(qp)}_{\text{Brow}}(t)\frac{\partial}{\partial \vect q}+ D^{(pp)}_{\text{Brow}}(t)\frac{\partial}{\partial \vect p} \bigg)W(\vect x,t)\nonumber \\
& +\frac{\partial}{\partial \vect p}\cdot\bigg(m\Omega_{\text{flux}}^2 \ \hat z\times\vect q-2D_{\text{flux}}(t) \ \vect z\times \frac{\partial}{\partial \vect q}\bigg)W(\vect x,t),\label{QBE1SP}
\end{align}
where the symbols $\cdot$ and $\times$ represent the matrix product and the cross product, respectively. Equivalently, one may verify that the quantum master equation (\ref{QBE1SP}) can be obtained from the extended Langevin equation (\ref{QLEMASP}) by following the standard derivation of the Fokker-Planck equation for the conventional Brownian motion \cite{fleming20111,agarwal19711}.

While the first line in the right hand side of Eq.(\ref{QBE1SP}) is the familiar Fokker-Planck collision operator from the conventional quantum Brownian motion \cite{agarwal19711,caldeira19832}, the second line corresponds to a new collision operator that models all the novel effects arising from the CS action. Aside the renormalized potential and friction terms, one may recognize the time-dependent decoherence coefficient $D^{(pp)}_{\text{Brow}}(t)$, and the system-to-bath diffusion coefficient (also known as the anomalous diffusion coefficient) $D^{(qp)}_{\text{Brow}}(t)$ characteristic of the standard Brownian motion \cite{fleming20111,agarwal19711,ford20011,hu19921}. Additionally, we may interpret the flux-carrying contribution as follows: the first term of the second line acts as a dissipationless rotational drift of strength $\Omega_{\text{flux}}$, whereas $D_{\text{flux}}(t)$ represents a transverse diffusive coefficient between perpendicular spatial degrees of freedom. The latter has none counterpart in previous treatments in the context of quantum kinetic theory \cite{schieve20091, vacchini20091,chavanis20101}, and its origin traces back to the flux-like noise anticipated in Sec. \ref{Sub_COMP2DV} (see Eq. (\ref{FDRF})): the linear response theory tells us that $D_{\text{flux}}(t)$ characterizes the linear reaction of the flux-carrying Brownian particles against $\hat{\vect \xi}_{\text{flux}}$. According to Eq. (\ref{MCSEF}), this represents a transverse reaction that closely recalls the ordinary Hall response in quantum Hall fluids \cite{zee19951}.

Alternatively, we could rewrite Eq.(\ref{QBE1SP}) in terms of the $4\times 4$ diffusive matrix, namely $\vect D_{BW}(t)$, encoding all the diffusion coefficients. The latter can split into separated Brownian and flux parts, i.e. 
\begin{equation}
\vect D_{BW}(t)=  \vect D_{\text{Brow}}(t)+\vect D_{\text{flux}}(t), 
\label{DFMT}
\end{equation}
where $\vect D_{\text{Brow}}(t)$ represents the customary diffusive matrix from the standard Brownian motion \cite{fleming20111,agarwal19711}, i.e.
 \begin{equation}
    \vect D_{\text{Brow}}(t)=\left(\begin{array}{cccc}
       0& 0  & D_{\text{Brow}}^{(qp)}(t) & 0 \\
       0&  0  &0 &D_{\text{Brow}}^{(qp)}(t)  \\
       D_{\text{Brow}}^{(qp)}(t)&0  & D_{\text{Brow}}^{(pp)}(t)  & 0\\
       0& D_{\text{Brow}}^{(qp)}(t)  & 0 & D_{\text{Brow}}^{(pp)}(t)\\
    \end{array}\right) \label{DBrowM}
 \end{equation}
whilst the flux contribution $\vect D_{\text{flux}}(t)$ is a symmetric anti-diagonal matrix, i.e.
 \begin{equation}
    \vect D_{\text{flux}}(t)=\left(\begin{array}{cccc}
       0& 0  & 0 & D_{\text{flux}}(t) \\
       0&  0  &-D_{\text{flux}}(t) &0 \\
       0&-D_{\text{flux}}(t)  & 0  & 0\\
        D_{\text{flux}}(t)& 0  & 0 & 0\\
    \end{array}\right). \label{DfluxM}
 \end{equation}

\begin{figure*}
\centering
\includegraphics[scale=0.28]{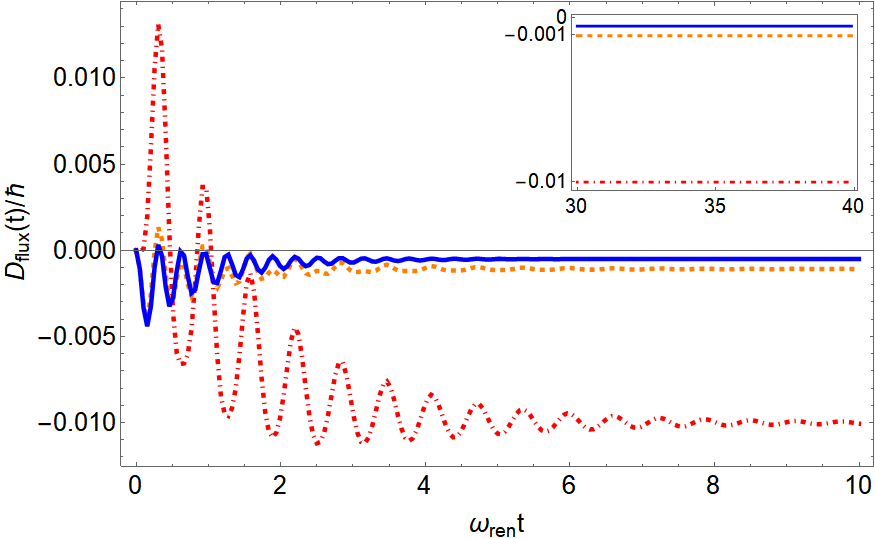}
\includegraphics[scale=0.37]{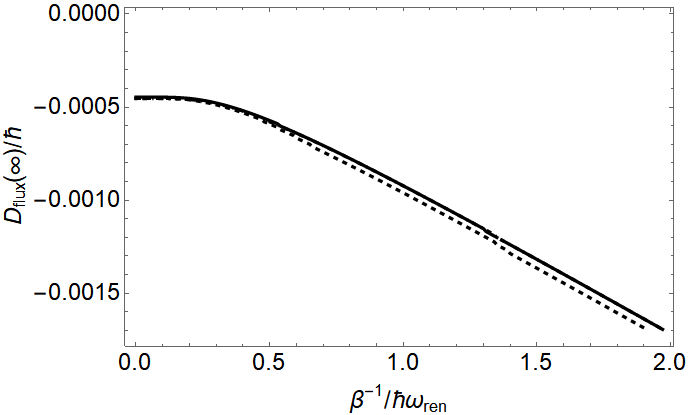}
\caption{(color online). (Left) The flux diffusive coefficient as a function of time for fixed damping coefficient $\gamma_{\text{Brow}}=0.1\omega_{ren}$ and three different values of the inverse temperatures: the solid blue, dashed orange and dot-dashed red lines correspond to $\beta=1000(\hbar\omega_{\text{ren}})^{-1}$, $\beta=(\hbar\omega_{\text{ren}})^{-1}$, and $\beta=0.1(\hbar\omega_{\text{ren}})^{-1}$, respectively. The inset shows the long time behavior. (Right) The asymptotic flux diffusive coefficient as a function of the inverse temperature for distinct values of the damping coefficient: the solid and dashed black lines correspond to $\gamma_{\text{Brow}}=0.1\omega_{\text{ren}}$, and $\gamma_{\text{Brow}}=0.2\omega_{\text{ren}}$, respectively. In both pictures, we have fixed $\omega_{\text{ren}}=10$, and $\Omega_{\text{flux}}=0.01\omega_{\text{ren}}$. \label{Fig1BW}} 
\end{figure*}

As expected, the conventional Brownian part returns the standard result, where $D_{\text{Brow}}^{(i)}(t)$ being the usual diffusion coefficients from the conventional 1D Brownian motion \cite{fleming20111,kumar20091,lombardo20051}. Instead, the flux diffusive coefficient is given by (at leading order in the CS effects)
\begin{equation}
    D_{\text{flux}}(t)=\frac{m\hbar\Omega_{\text{flux}} ^2}{\pi}\int_{0}^{\infty}  \frac{(\gamma_{\text{Brow}}\omega)^3F(t,\omega)}{\left(\gamma_{\text{Brow}} ^2 \omega ^2+\left(\omega ^2-\omega_{\text{ren}}^2\right)^2\right)^2}\coth\left(\frac{\hbar \omega\beta}{2}\right)\ d\omega,
    \label{EMDF1}
\end{equation}
where we have included the auxiliary function $F(t,\omega)$, this involve a lengthy expression that is not crucial for the following discussion. Clearly, from Eq. (\ref{DfluxM}) follows that the anomalous diffusion matrix acquires antisymmetric components as consequence of the flux-carrying effects, in contrast to the conventional Brownian situation (\ref{DBrowM}). This feature implies that the open quantum system dynamics can not be reduced to the problem of two independent 1D Brownian particles. Instead of providing the full expression of the flux diffusive coefficient, the left panel of the Fig. \ref{Fig1BW} depicts this as a function of time for several values of the inverse temperature. One may observe that the flux diffusive coefficient has an initial oscillatory transient behavior, which is a signature of the intrinsic non-Markovinity of the quantum system dynamics at low temperatures \cite{devega20171}. After this transient behavior, the diffusive coefficient converges to certain steady value in a time scale longer than the particle renormalized frequency, i.e. $\omega_{\text{ren}} t\gg 1$ (see the inset in fig. \ref{Fig1BW}). This is a trademark of the relaxation process, and the steady value characterizes the thermal equilibrium state. By paying further attention, one may appreciate that this time scale is roughly determined by $\gamma_{\text{Brow}}^{-1}$, which is usually refereed to as the late-time regime \cite{fleming20111}. We shall study the hydrodynamic properties in this regime, as it permits to substantially simplify the quantum kinetic analysis: the coefficient of the extended Kramer equation (\ref{QBE1SP}) becomes time independent.

Additionally, the time asymptotic behavior of the flux diffusive coefficient in terms of the inverse temperature is depicted by the right panel of the figure (\ref{Fig2BW}) for distinct choices of the friction parameter. One may notice that it slightly changes with the friction coefficient $\gamma_{\text{Brow}}$. Interestingly, the plot also reveals that $D_{\text{flux}}(\infty)$ grows linearly at high temperatures, which is consistent with the fluctuation-dissipation relation (\ref{FDRFML}). This behavior is also encountered for the ordinary diffusion coefficients in the conventional Brownian motion \cite{lombardo20051,fleming20111}. In the opposite temperature limit, the flux-carrying diffusion coefficient saturates to a non-vanishing value barely determined by $\gamma_{\text{Brow}}$. This value can be directly computed from the integral expressions (\ref{EMDF1}) by using standard contour integration techniques after taking the asymptotic time limit $t\rightarrow \infty$ (see Eq. (\ref{EMDF1AR}) in App. \ref{App_QBE}). Specifically, we obtain in the underdamped regime (i.e.$\gamma_{\text{Brow}}<2\omega_{\text{ren}}$) and high temperature limit (i.e. $\hbar \eta_{\text{Brow}} \beta \ll 1$)

\begin{align}
    D_{\text{flux}}(\infty)&=\frac{\Omega_{\text{flux}}^2}{\beta\zeta_{\text{Brow}}^3}\bigg(2 \Big(\omega_{\text{ren}}^2-\frac{\gamma_{\text{Brow}}^2}{2} \Big)\text{Re}\Big(\frac{1}{\eta_{\text{Brow}}}\Big)+\gamma_{\text{Brow}}\zeta_{\text{Brow}}\text{Im}\Big(\frac{1}{\eta_{\text{Brow}}}\Big)\bigg)+\mathcal{O}(\beta),    \label{EMDF1ARHT} 
\end{align}
and in the low temperature limit (i.e. $\hbar \eta_{\text{Brow}} \beta \gg 1$)
\begin{align}
    D_{\text{flux}}(\infty)&=-\frac{\hbar \Omega_{\text{flux}}^2}{2\zeta_{\text{Brow}}^3}\bigg[\Big(\omega_{\text{ren}}^2-\frac{\gamma_{\text{Brow}}^2}{2} \Big)\Big(1-\frac{i}{\pi}\log\Big(\frac{\eta_{\text{Brow}}}{\eta_{\text{Brow}}^{\dagger}}\Big) \Big) +  \frac{\gamma_{\text{Brow}}\zeta_{\text{Brow}}^2}{2\pi }\Big(1+\frac{\eta_{\text{Brow}}}{\eta_{\text{Brow}}^{\dagger}}\Big)\bigg] +\mathcal{O}(\beta^{-1}), \label{EMDF1ARLT} 
\end{align}
where 
\begin{equation}
    \eta_{\text{Brow}}=\sqrt{\omega_{\text{ren}}^2-\frac{1}{2}\Big(\gamma_{\text{Brow}}^2+ i\sqrt{\gamma_{\text{Brow}}^2(4\omega_{\text{ren}}^2-\gamma_{\text{Brow}}^2)}\Big)}.
    \label{AppSPR}
\end{equation}
From Eqs. (\ref{EMDF1ARHT}) and (\ref{EMDF1ARLT}) follows that the flux diffusion coefficient is mainly characterized by the strength $\Omega_{\text{flux}}^{2}$ of the flux-carrying effects, so that the flux diffusion coefficient does not depend of the sign of the CS constant ($\kappa$ can take either positive or negative continuous values in principle). Clearly, Eq. (\ref{QBE1SP}) retrieves the familiar Fokker-Planck equation characteristic of the standard Brownian motion \cite{agarwal19711,fleming20111} by disregarding the flux-attachment effects (i.e. $\Omega_{\text{flux}}\rightarrow 0$). 

\subsubsection{Quantum kinetics at late times}\label{Subsec_QKLT}

Let us draw some attention to the late-time dynamics. In Sec. \ref{Subsec_FPPC}, we show that the flux-carrying Brownian particle will relax to certain thermal equilibrium state provided the following condition is satisfied,
\begin{equation}
    \frac{\Omega_{\text{flux}}}{\omega_{\text{ren}}}< \frac{\gamma_{\text{Brow}}}{\omega_{\text{ren}}}\ll 1.\label{SDC1SPS}
\end{equation}
The physical intuition behind the above expression is that the environmental spectrum will well accommodate the renormalized frequency of the flux-carrying particle, making possible an irreversible energy transfer from it to the MCS environment at least in a finite time sufficiently larger than the natural time scale of the system $\omega_{\text{ren}}^{-1}$. Alternatively, Eq.(\ref{SDC1SPS}) establishes formally the weak coupling regimen between the system and environment. Hereafter we work within the parameter domain where (\ref{SDC1SPS}) holds.

\begin{figure*}
\centering
\includegraphics[scale=0.32]{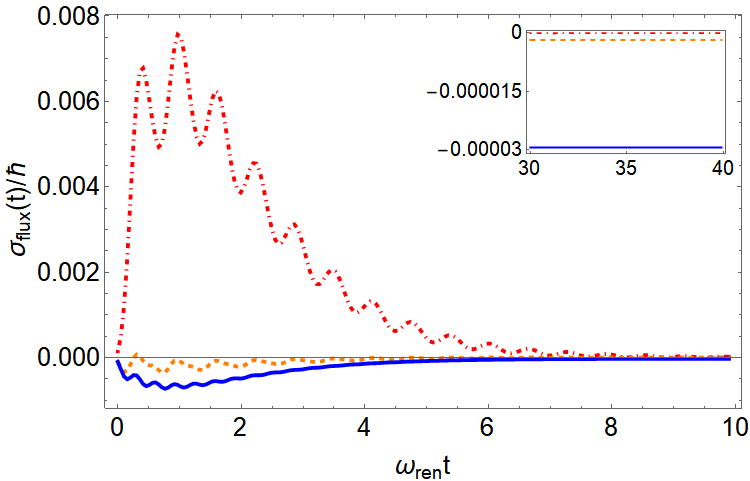}
\includegraphics[scale=0.37]{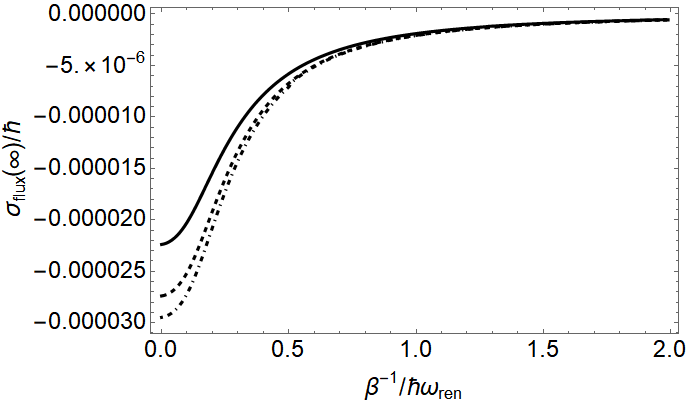}
\caption{(color online). (Left) The flux thermal-covariance coefficient as a function of time for three different values of the inverse temperature: the solid blue, dashed orange and dot-dashed red lines correspond to $\beta=1000(\hbar\omega_{\text{ren}})^{-1}$, $\beta=(\hbar\omega_{\text{ren}})^{-1}$, and $\beta=0.1(\hbar\omega_{\text{ren}})^{-1}$, respectively. We have fixed $\omega_{\text{ren}}=10$, $\gamma=0.1\omega_{\text{ren}}$, and $\Omega_{\text{flux}}=0.01\omega_{\text{ren}}$.  (Right) The flux thermal-covariance coefficient in the asymptotic time as a function of the inverse temperature for three different values of the damping coefficient: the solid, dashed and dot-dashed black lines correspond to $\gamma_{\text{Brow}}=0.1\omega_{\text{ren}}$, $\gamma_{\text{Brow}}=0.2\omega_{\text{ren}}$, and $\gamma_{\text{Brow}}=0.5\omega_{\text{ren}}$, respectively. We have fixed $\omega_{\text{ren}}=11$, and $\Omega_{\text{flux}}=0.01\omega_{\text{ren}}$. \label{Fig2BW}}
\end{figure*}

On the other side, since the quantum master equation (\ref{QBE1SP}) is quadratic in the position and momentum coordinates, the asymptotic state $W(\vect x,\infty)$ will be Gaussian, and its covariance matrix, denoted by $\vect V(\infty)$, will eventually converge to the so-called thermal covariance matrix, denoted by $\vect \sigma(t) $, once the stationary state has been reached \cite{fox19781,fleming20111,agarwal19711}, i.e. $\vect V(t)\rightarrow \vect \sigma(\infty)$ when $t\rightarrow \infty$. In the weak coupling regime, a similar decomposition to the diffusive matrix (\ref{DFMT}) can be made for the thermal covariance matrix, where the Brownian and flux parts can be clearly distinguished, i.e. $\vect \sigma(t)=\vect \sigma_{\text{Brow}}(t)+\vect \sigma_{\text{flux}}(t)$ with
\begin{equation}
    \vect \sigma_{\text{Brow}}(t)=\left(\begin{array}{cccc}
       \sigma_{\text{Brow}}^{(qq)}(t)  & 0 & \sigma_{\text{Brow}}^{(qp)}(t) & 0 \\
       0&  \sigma_{\text{Brow}}^{(qq)}(t)  & 0 &\sigma_{\text{Brow}}^{(qp)}(t) \\
        \sigma_{\text{Brow}}^{(qp)}(t)  & 0 &\sigma_{\text{Brow}}^{(pp)}(t) & 0\\
        0& \sigma_{\text{Brow}}^{(qp)}(t)  & 0 & \sigma_{\text{Brow}}^{(pp)}(t)\\
    \end{array}\right),
    \label{CTMB}
\end{equation}
where the matrix elements coincide with the well-known expressions from the damped harmonic oscillator \cite{fleming20111,lombardo20051}. It is important to realize in (\ref{CTMB}) that the matrix elements related to transversal degrees of freedom are identical, so (\ref{CTMB}) formally coincides with the thermal covariance matrix of two 1D Brownian particles. In contrast, we find that the flux-carrying contribution takes the form of a symmetric anti-diagonal matrix, i.e.
\begin{equation}
    \vect \sigma_{\text{flux}}(t)=\left(\begin{array}{cccc}
       0& 0 & 0 & \sigma_{\text{flux}}(t) \\
     0 &  0  &-\sigma_{\text{flux}}(t) &0 \\
       0&-\sigma_{\text{flux}}(t)  & 0  & \\
        \sigma_{\text{flux}}(t)& 0  & 0 & 0\\
    \end{array}\right).
    \label{CTMF}
\end{equation}
with matrix coefficients given by
\begin{equation}
\sigma_{\text{flux}}(t)=\frac{2 m\hbar\Omega_{\text{flux}} ^2}{\pi}\int_{0}^{\infty}  \frac{\omega   \left(\left(\omega ^2-\omega_{\text{ren}}^2\right)^2-\gamma_{\text{Brow}} ^2 \omega ^2\right)}{\left(\gamma_{\text{Brow}} ^2 \omega ^2+\left(\omega ^2-\omega_{\text{ren}}^2\right)^2\right)^2}f(t,\omega)\coth\left(\frac{\hbar \omega\beta}{2}\right)\ d\omega,
\label{EMSF1}
\end{equation}
where the explicit expression of $f(t,\omega)$ is given by Eq. (\ref{AEqf1}). The left panel of figure \ref{Fig2BW} depicts $\sigma_{\text{flux}}(t)$ as a function of time for several values of the inverse temperature, whereas the right panel plots its time asymptotic value $\sigma_{\text{flux}}(\infty)$ as a function of the inverse temperature for several values of the dissipative coefficent. Remarkably, one may appreciate that the flux-carrying contribution to the thermal covariance matrix (\ref{EMSF1ARHT}) eventually cancels in the high temperature limit. This immediately implies that the influence of the flux-carrying effects is irrelevant in the classical equilibrium state. This coincides with the result obtained from the standard Brownian motion under the influence of an static Berry curvature in the momentum space \cite{misaki20181}. To understand the latter we must recall that the transverse dynamical susceptibility responsible for these flux-carrying effects constitutes a AB phase-like factor to the Boltzmann weight of the partition function (\ref{ZIFBP}), so that we should recover the equilibrium state consistent with the classical statistical mechanics. A quick glance also indicates that, after an initial oscillatory transient behavior, $\sigma_{\text{flux}}(t)$ converges to certain steady values in a time scale longer than the particle renormalized frequency, as similarly occurs for the diffusion coefficient (\ref{EMDF1}). Here, it is important to realize that $\sigma_{\text{flux}}(\infty)$ represents the cross correlation between the quantum position and momentum operators in the equilibrium thermal state reached (i.e. $\sigma_{\text{flux}}= \left\langle\hat{q}_{x}\hat{p}_{y}+\hat{p}_{y}\hat{q}_{x} \right\rangle_{\hat \rho_{\beta}}=-\left\langle\hat{q}_{y}\hat{p}_{x}+\hat{p}_{x}\hat{q}_{y} \right\rangle_{\hat \rho_{\beta}}$), so that it equivalently determines the asymptotic average value of the angular momentum of the flux-carrying Brownian particle. 

Additionally, the time asymptotic value of the flux thermal-covariance coefficient can be directly computed from the integral expressions (\ref{EMSF1}) by using standard contour integration techniques once we have taken the asymptotic time limit $t\rightarrow \infty$ (see Eq. (\ref{EMSF1AR}) in App. \ref{App_QBE}). Concretely, we obtain in the underdamped regime (i.e.$\gamma_{\text{Brow}}<2\omega_0$) and high temperature limit (i.e. $\hbar \eta_{\text{Brow}} \beta \ll 1$)
\begin{align}
    \sigma_{\text{flux}}(\infty)&=\frac{\beta\hbar^2\Omega_{\text{flux}}^2}{24\zeta_{\text{Brow}}^3}\big(\gamma_{\text{Brow}}\text{Re}\ \eta_{\text{Brow}}+2\zeta_{\text{Brow}}\text{Im}\ \eta_{\text{Brow}}\big)+\mathcal{O}(\beta^2),     \label{EMSF1ARHT}
\end{align}
where $\eta_{\text{Brow}}$ is given by Eq. (\ref{AppSPR}). Similarly, in the low temperature limit (i.e. $\hbar \eta_{\text{Brow}} \beta \gg 1$) and underdamped regime, we arrive at
\begin{align}
    \sigma_{\text{flux}}(\infty)&=-\frac{\hbar\gamma_{\text{Brow}} \Omega_{\text{flux}}^2}{4\zeta_{\text{Brow}}^3}\bigg[1+\frac{i}{2\pi\gamma_{\text{Brow}}\eta_{\text{Brow}}^2}\Big(-\gamma_{\text{Brow}}^3+2\gamma_{\text{Brow}}(2\omega_{\text{ren}}^2-i\gamma_{\text{Brow}}\zeta_{\text{Brow}}), \nonumber \\
    &+4i\zeta_{\text{Brow}}(\omega^{2}_{\text{ren}}+|\eta_{\text{Brow}}|^2)-2\gamma_{\text{Brow}}(\eta_{\text{Brow}}^{\dagger})^{2}\log\Big(\frac{\eta_{\text{Brow}}}{\eta_{\text{Brow}}^{\dagger}}\Big)\Big)\bigg]  +\mathcal{O}(\beta^{-1}), \label{EMSF1ARLT}
\end{align}
where $\eta_{\text{Brow}}$ is given by Eq.(\ref{AppSPR}). Eq. (\ref{EMSF1ARHT}) proves that the flux-carrying effects completely disappear from  the  thermal  equilibrium  state  in  the  classical  regime (i.e. $\beta\rightarrow 0$), as expected from the above discussion. In the opposite temperature limit (i.e. $\beta\rightarrow \infty$), Eq. (\ref{EMSF1ARLT}) revels that the flux-carrying coefficient saturates to a non-vanishing value independent of the inverse temperature. As a result, the flux carrying Brownian particle is effectively endowed with a finite angular momentum in the quantum regime. One may also notice that $\sigma_{\text{flux}}(\infty)$ smoothly changes with the friction coefficient $\gamma_{\text{Brow}}$, which indicates that the flux-carrying effects are significantly robust to dissipative mechanisms.  This means that we must get a compromise between these dissipative and flux carrying effects in order to observe a vortex flow of the flux-carrying Brownian particle. This shall be studied in further detail in the next section.

\subsection{Quantum balance equations}\label{Sub_QHSP}

In this section we present the hydrodynamic conservation laws for the number density, the stream velocity, the kinetic energy density, the fluid vorticity and the vorticity flux. In Sec \ref{subsecHD} we treat the more general scenario of $N$ flux-carrying Brownian particles in the low density and weak coupling regime.

First, since the particle number is conserved, the familiar continuity equation is fully satisfied \cite{mayorga20021},
\begin{equation}
\frac{\partial n}{\partial t}+\nabla\cdot  \vect J=0.
\label{BEN1SP}
\end{equation}
where $\nabla$ stands for the gradient operator in the variable $\vect q$, and we have introduced the single particle number density,
\begin{equation}
 n (\vect q,t)= \int_{\mathbb{R}^2} W(\vect q,\vect p,t)\ d^2\vect p,
\label{NDDFSP}
\end{equation}
as well as the single particle flow density \cite{lagos20111, mayorga20021,chavanis20101},
\begin{equation}
    \vect J(\vect q,t)=n(\vect q,t) \vect u(\vect q,t)=\frac{1}{m}\int_{\mathbb{R}^2}  \vect p W(\vect q,\vect p,t)\ d^2\vect p, \label{MDFDSP}
\end{equation}
with $\vect u$ being the stream velocity. One may verify that Eq.(\ref{BEN1SP}) is obtained as a consequence of the momentum integral over the collision operators in (\ref{QBE1SP}) vanishes. By introducing the material or hydrodynamic derivative \cite{kreuzer19811}, i.e.
\begin{equation}
    \frac{d}{dt}=\frac{\partial}{\partial t}+\vect u \cdot \nabla,
    \nonumber
    \end{equation}
the continuity equation (\ref{BEN1SP}) can be alternatively expressed as 
\begin{equation}
    \frac{dn}{dt}+n\nabla\cdot\vect u=0.
    \label{BEN2SP}
\end{equation}
Now we may take the partial derivative of the definition (\ref{MDFDSP}) and insert the extended Kramers equation (\ref{QBE1SP}). By carrying out integration by parts over momentum space once we have replaced the hydrodynamic derivative, we arrive to the stream velocity balance equation (the interesting reader is refereed to Sec. \ref{subsecHD} for further details) 
\begin{equation}
n\frac{d \vect u}{dt}+\frac{1}{m}\nabla  \cdot\vect T=- \gamma_{\text{Brow}} (n\vect u)+\Omega_{\text{flux}}^2 \ \vect z\times(n \vect q),
\label{ENVSTSP}
\end{equation}
where the $2\times 2$ stress tensor $\vect T$ takes the form,
\begin{equation}
    \vect T=\vect P_{K}+\vect P_{\Phi_{\text{ren}}}+2\left(\begin{array}{cc}
     D^{(qp)}_{\text{Brow}}(t)   & D_{\text{flux}}(t) \\
      -D_{\text{flux}}(t)   & D^{(qp)}_{\text{Brow}}(t)
    \end{array}\right)n 
    \label{ENVST1}
\end{equation}
Here, we have identified the kinetic contribution to the (local) stress tensor \cite{mayorga20021,sonnenburg19911,lagos20111,Klymko20171}
\begin{align}
\vect P_{K}(\vect q,t)=\frac{1}{m}\int_{\mathbb{R}^{2}} (\vect p-m\vect u)\circ(\vect p-m\vect u)W(\vect q,\vect p,t)  \ d^2\vect p,
\label{STKCSP}
\end{align}
and due to the harmonic confining potential
\begin{align}
\vect P_{\Phi_{\text{ren}}}(\vect q,t)=\omega_{\text{ren}} n (\vect q,t)\vect I_{2},   \label{HydroPSP}
\end{align}
where $\vect I_{2}$ denotes the $2\times 2$ identity matrix. Notice that the diagonal elements of (\ref{STKCSP}) corresponds to the so-called hydrostatic pressure, which shall be denoted by $p_{K}(\vect q,t)$.

Equation (\ref{ENVSTSP}) represents an extension of the damped Euler equation characteristic of the conventional Brownian motion \cite{chavanis20101,mayorga20021}. As anticipated in Sec. \ref{Sub_COMP2DV}, the flux-carrying Brownian particles are subject to a rotational drift that closely resemblances the active stress tensor of a 2D fluid composed of chiral active dumbbells \cite{hargus20201}. We remark that the latter is absent in the stress tensor characteristic of the conventional Brownian motion subject to an external magnetic field \cite{lagos20111,czopnik200111,jimenez20061}. Additionally, the flux-like noise $\hat{\vect \xi}_{\text{flux}}$ is responsible for the antisymmetric components in $\vect T$, which constitutes a hallmark of the nondiffusive nature of the hydrodynamics. In time scales much larger than the characteristic damping rate $\gamma_{\text{Brow}}^{-1}$, namely the diffusive regime \cite{mayorga20021}, we may ignore the time derivative of the stream velocity field in Eq.(\ref{ENVSTSP}) and obtain an approximated expression for the flow density.  Besides the conventional contribution satisfying the Fick's law, we find that the flow density $\vect J$ in the diffusive regime contains an additional vortex-like component (i.e. $\vect J=\vect J_{\text{Brow}}+\vect J_{\text{flux}}$) which reads
\begin{align}
\vect  J_{\text{flux}}(\vect q,t)&=\frac{\Omega_{\text{flux}}^2}{\gamma_{\text{Brow}}}(\vect z\times \vect q)\ n(\vect q,t)+\frac{2D_{\text{flux}}(t)}{m\gamma_{\text{Brow}}}\big(\vect z\times \nabla n(\vect q,t)\big).
\label{FDFDR}
\end{align}
While the first term in the right hand side of (\ref{FDFDR}) plays an identical role to the active torque in chiral active fluids composed of dumbbells \cite{han20201,hargus20201,hargus20201,Klymko20171,epstein20191}, the second term indicates that $D_{\text{flux}}$ characterizes a nondiffusion transport process perpendicular to density gradients as anticipated above. Significantly, unlike Brownian particles subject to either an external magnetic field \cite{vuijk20191} or an active torque \cite{banerjee20171,lucas20141}, Eq. (\ref{FDFDR}) also reveals that the flux-carrying Brownian particle will display a persistent vortex flow regardless nonequilibrium conditions (such as temperature gradients or an intrinsic source of energy). One can draw a parallel with the persistent charge current due to a AB phase in mesoscopic systems at thermodynamic equilibrium \cite{loss19921}: they both prevail indefinitely in the low-temperature regime and cannot be dissipated. This is consistent with our preliminary observation that the novel effects due to the flux attachment are encoded in the transverse dynamical susceptibility which plays the role of an AB phase-like factor in the partition function (\ref{ZIFBP}).

We shall now derive the balance equation for the kinetic energy density, denoted by $e_{K}(\vect q_{1},t)$. This can be red off from the expression for the kinetic energy \cite{mayorga20021,chavanis20061,lagos20111}, i.e.
\begin{equation}
   E_{K}(\vect q,t)=n(\vect q,t)e_{K}(\vect q,t)=\frac{1}{2m}\int_{\mathbb{R}^2}  (\vect p-m\vect u)^2 W(\vect q,\vect p,t) \ d^2\vect p.
   \label{EDDSP}
\end{equation}
To set up the balance equation for (\ref{EDDSP}) we repeat a similar procedure as followed for (\ref{ENVSTSP}): that is, we replace the time derivative of (\ref{EDDSP}) by expression (\ref{QBE1SP}) and integrate over the momentum variables. After some transformation and integration by parts, we get
\begin{align}
\frac{\partial (ne_{K})}{\partial t}&=-\vect P_{K}\circ\big(\nabla \cdot \vect u\big) -\nabla\cdot\big(\vect J_{K}-(ne_{K})\vect u \big) -2\gamma_{\text{Brow}}(ne_{K}) + \frac{2n}{m} D_{\text{Brow}}^{(pp)}(t),\label{BECESP}
\end{align}
where we have introduced the energy flow vector density or heat current \cite{mayorga20021,lagos20111,chavanis20101},
\begin{equation}
  \vect J_{K}(\vect q,t)=\frac{1}{2m^{2}}\int_{\mathbb{R}^2}   (\vect p-m\vect u)^{2} (\vect p-m\vect u)W(\vect q,\vect p,t) \ d^2\vect p.
  \label{HCD1}
\end{equation}
Unlike previous parity-violating fluids \cite{lucas20141,kaminski20141}, one may appreciate that the kinetic energy density is apparently unaffected by the flux-carrying effects, such that we recover the usual balance equation from the conventional Brownian motion. Once again, this observation supports our argument that the flux-carrying contribution consists of a dissipationless environmental mechanism. 

We conclude this section by paying attention on the fluid vorticity and the circulation flux. In our 2D system, we understand the fluid vorticity as a pseudo-vector whose value is formally given by \cite{thorne20171,jackiw20041},
\begin{eqnarray}
\varpi =\big[ \nabla \times \vect u\big]_{z}=\frac{\partial u_y}{\partial  q_{x}}-\frac{\partial u_x}{\partial q_{y}},
\label{DFVORTSP}
\end{eqnarray} 
where $q_{x}$ and $q_{y}$ stand for the components of the spatial coordinate. To derive the desired evolution equation for the fluid vorticity, we take the curl of Eq.(\ref{ENVSTSP}).  By using the following vector identities holding in two dimensions \cite{thorne20171}
\begin{align}
    (\vect u\cdot\nabla)\vect u&=\nabla(u^2)/2-\vect u\times(\varpi \hat z), \nonumber \\
    \nabla\times(\vect u\times(\varpi \hat z) )&= -\varpi (\nabla\cdot \vect u)-(\vect u\cdot \nabla)\varpi,
    \label{MVDI}
\end{align}
after some manipulation we obtain an extended Helmholtz equation for the fluid vorticity of the flux-carrying Brownian particle,
\begin{align}
     \frac{\partial\varpi}{\partial t}&=-(\vect u\cdot\nabla)\varpi-\varpi (\nabla\cdot \vect u) -\frac{1}{m n^2}\big[\nabla\cdot (\vect P_{K}+\vect P_{\Phi_{\text{ren}}})\times\nabla n\big]_{z}-\gamma_{\text{Brow}}\varpi +2\Omega_{\text{flux}}^2  \nonumber \\
      &+\frac{2D_{\text{flux}}(t)}{m n}\bigg(\nabla^2 n-\frac{1}{n}(\nabla n)^2\bigg)
    \label{BEV1SP}
\end{align}
Upon further inspection of (\ref{BEV1SP}), one may identify the first term in the right hand side as the hydrodynamic derivative, whereas the second term can be replaced by making use of the continuity equation (\ref{BEN2SP}). This finally yields the desired expression for the vorticity balance equation,
\begin{align}
    \frac{d\varpi}{d t}&= 2\Omega_{\text{flux}}^2-\gamma_{\text{Brow}}\varpi+\frac{\varpi}{n}\frac{dn}{d t}-\frac{1}{m n^2}\big[\nabla\cdot (\vect P_{K}+\vect P_{\Phi_{\text{ren}}})\times\nabla n\big]_{z}  +\frac{2D_{\text{flux}}(t)}{m n}\bigg(\nabla^2 n-\frac{1}{n}(\nabla n)^2\bigg).
    \label{BEV2SP}
\end{align}
A quick glance to the first term in the right hand side of Eq. (\ref{BEV2SP}) reveals that the environmental torque plays the role of a time-independent, uniform vorticity source. This is peculiar to the cyclotron frequency in Brownian particles subject to an uniform external magnetic field \cite{abanov20131}. The second term corresponds to the dissipation of the vorticity and can be found in active Brownian descriptions \cite{banerjee20171}: it would be responsible for the equilibration of the system if the active torque cancels. Now, it is readily to obtain the counterpart of Kelvin's circulation equation from (\ref{ENVSTSP}). To do so, it is convenient to consider the common definition of the vorticity flux $\psi$ through a (simple connected) area in terms of the associated line integral around a counter-clock-wise contour $\mathcal{C}$ (with unit radii) \cite{jackiw20041}, i.e.
\begin{equation}
    \psi=\oint_{C} (\nabla\times\varpi) \cdot d \vect l.
    \label{EDC1SP}
\end{equation}
To derive the balance equation for $\psi$, we directly replace Eq. (\ref{BEV2SP}) in (\ref{EDC1SP}) and borrow the standard procedure from hydrodynamic theory \cite{thorne20171}. After some manipulation, we arrive at the following balance equation
\begin{align}
    \frac{d\psi}{d t}&= 2\pi\Omega_{\text{flux}}^2-\gamma_{\text{Brow}}\psi -\oint_{\mathcal{C}}\frac{dP}{n} -\frac{2D_{\text{flux}}(t)}{m}\oint_{\mathcal{C}}\frac{\nabla  n}{n} \times d\vect q .
    \label{EDC2SP}
\end{align}
By regarding the vorticity flux as the fluid counterpart of the magnetic flux definition, Eq. (\ref{EDC2SP}) indicates that $\Omega_{\text{flux}}$ can be thought of as the magnitude of a coarse-grained pseudomagnetic flux traversing the plane. Conversely, Eq.  (\ref{EDC2SP}) reveals that the flux-carrying effects challenge with the dissipative effects to establish a non-vanishing vorticity. Similarly to Eq. (\ref{FDFDR}), in the diffusive regime Eq. (\ref{EDC2SP}) retrieves (when $\dot{\psi}\approx 0$)
\begin{align}
    \psi&= \frac{2\pi\Omega_{\text{flux}}^2}{\gamma_{\text{Brow}}}-\frac{1}{\gamma_{\text{Brow}}}\oint_{\mathcal{C}}\frac{dP}{n} -\frac{2D_{\text{flux}}(t)}{m \gamma_{\text{Brow}}}\oint_{\mathcal{C}}\frac{\nabla  n}{n} \times d\vect q, 
    \label{EDC2DR}
\end{align}
which pinpoints that the environmental torque will enforce the flux-carrying Brownian particles to spin with an single angular speed of strength $\Omega_{\text{flux}}$. As somehow expected, this resemblances the situation of a weakly interacting chiral active gas \cite{banerjee20171}, where the active torque is dominant over the the inter-rotor interaction. We shall show in the following section that the second term in the right hand side of Eq. (\ref{EDC2DR}) cancels, and thereby, both the environmental torque and the transverse diffusive coefficient are responsible for the novel dissipationless vortex flow.

\subsubsection{Quantum hydrodynamics at late times}\label{subsecQHALT}

In this section we focus the attention on the hydrodynamics conservation laws at late times, that is, when the flux-carrying particle is near thermal equilibrium (recall that this occurs when the diffusion coefficients have already reached their steady values). Although this description does not correspond to the whole time evolution dictated by the balance equations, it provides the same asymptotic hydrodynamics. More precisely, we concentrate on the flux-carrying effects upon the flow density, stream velocity, fluid vorticity, energy density and pressure tensor in a time scale $\gamma_{\text{Brow}}^{-1}\ll t$. For sake of simplicity, we consider a radially symmetric initial Gaussian state with a simple covariance matrix $\vect V(0)=\vect I_{4}$ and zero mean values $\left\langle \vect x(0) \right\rangle_{\hat \rho_{\beta}}=0$ (this corresponds to the extensively studied coherent state in quantum optics). The reason to focus the attention on this kind of states is because it retrieves a purely diffusive flow (which is parallel to the particle density gradient) in the case of the conventional Brownian motion subject to an uniform magnetic field \cite{abdoli20201,abdoli20202}. Since the initial state is Gaussian, the time-evolved state $W(\vect x,t)$ will remain Gaussian as well \cite{agarwal19711,fleming20111,fox19781}, i.e.
\begin{equation}
  W(\vect x,t)=\frac{1}{(2\pi)^2\sqrt{\det{(\vect V(t))}}}  e^{-\frac{1}{2}\vect x^T\cdot (\vect V(t))^{-1}\cdot \vect x},
  \label{WFSG}
\end{equation}
and one can make use of standard Green's function methods to determine its covariance matrix $\vect V(t)$ (further details can be found in App. \ref{App_QBE}). Concretely, we find out
\begin{equation}
    \vect V(t)=\left(\begin{array}{cccc}
        V_{\text{Brow}}^{(qq)}(t)  & 0 & V_{\text{Brow}}^{(qp)}(t) & V_{\text{flux}}(t) \\
       0&   V_{\text{Brow}}^{(qq)}(t)  & - V_{\text{flux}}(t)  &V_{\text{Brow}}^{(qp)}(t) \\
        V_{\text{Brow}}^{(qp)}(t)  & - V_{\text{flux}}(t)  & V_{\text{Brow}}^{(pp)}(t) & 0\\
         V_{\text{flux}}(t) & V_{\text{Brow}}^{(qp)}(t)  & 0 &  V_{\text{Brow}}^{(pp)}(t)\\
    \end{array}\right), \label{CVMHSLT}
\end{equation}
where the matrix entries are given by Eq. (\ref{CVEQ}) in App.\ref{App_QBE}.

\begin{figure*}
\centering
\includegraphics[scale=0.18]{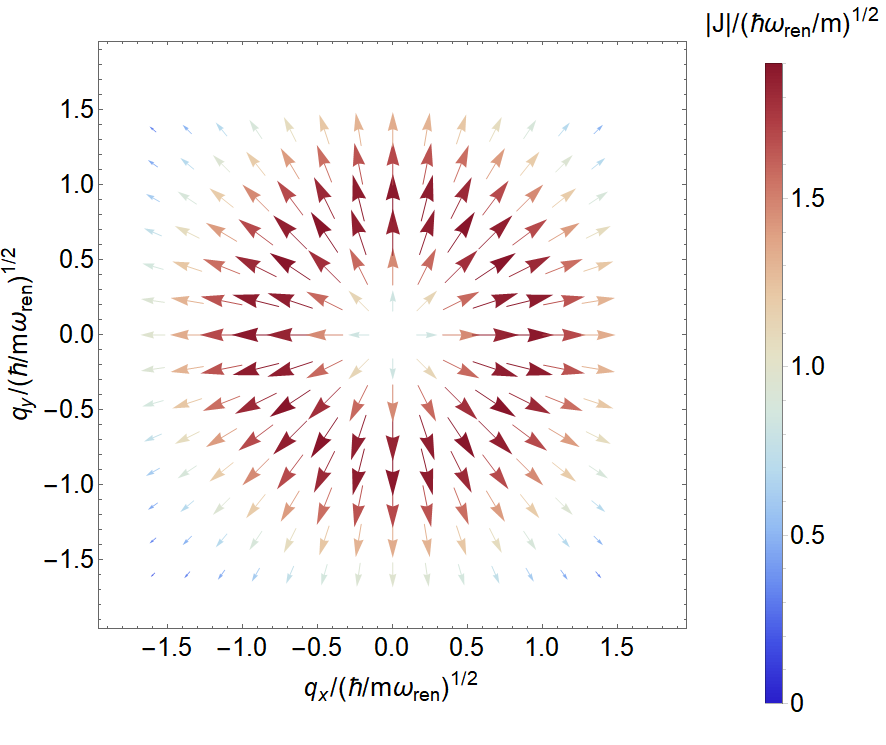}
\includegraphics[scale=0.18]{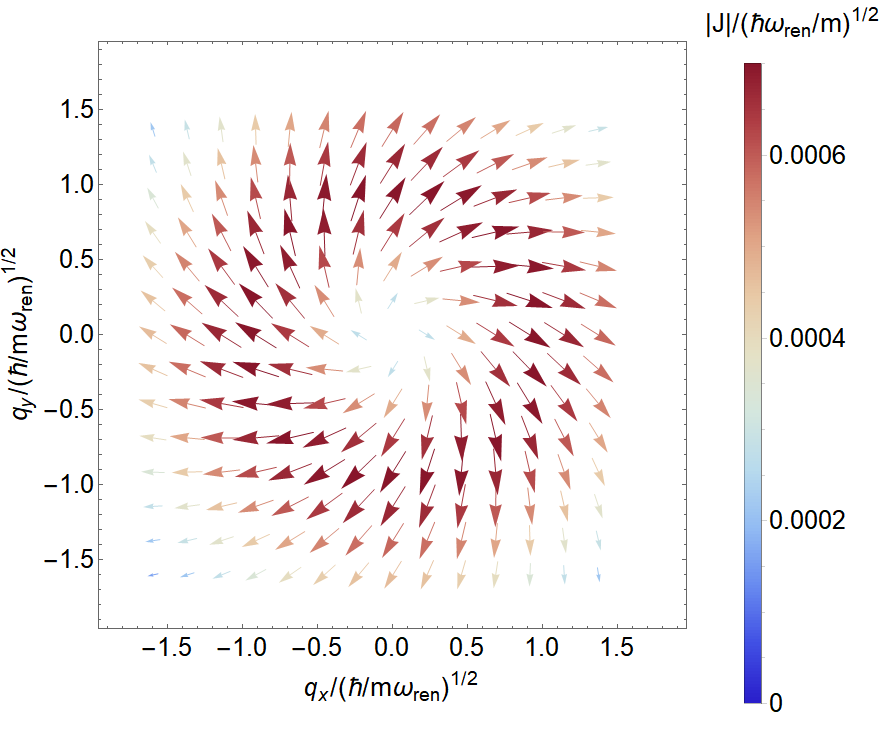}
\includegraphics[scale=0.18]{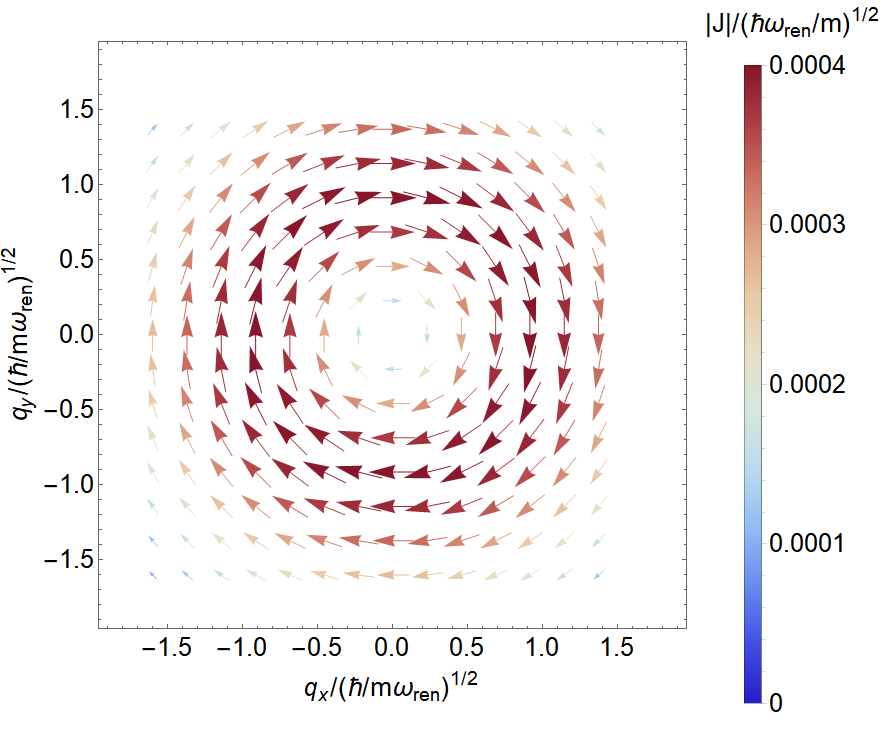}
\caption{(color online). Depiction of the flow density $\vect J(\vect q,t)$ on the low-temperature limit as a function of the position evaluated in three different times: (left) $t=1.65\omega_{\text{ren}}^{-1}$, (central) $t=115.5\omega_{\text{ren}}^{-1}$, and (right) $t=335.5\omega_{\text{ren}}^{-1}$. In all pictures, we have fixed $\omega_{\text{ren}}=11$, $\gamma=0.1\omega_{\text{ren}}$,  $\Omega_{\text{flux}}=0.01\omega_{\text{ren}}$ and $\beta=100(\hbar\omega_{\text{ren}})^{-1}$. Additionally, we have employed the asymotitc values of the thermal covariance matrix given by equations from (\ref{TCVCLTS1}) to (\ref{TCVCLTS3})  with $\omega_c=10\omega_{\text{ren}}$.\label{Fig1HS}} 
\end{figure*}

Let us start analysing the single-particle number density, this is obtained from definition (\ref{NDDFSP}) by substituting (\ref{WFSG}). As the integrand over momentum space is Gaussian, the integral can be computed exactly, i.e.
\begin{equation}
    n(\vect q,t)=\frac{e^{-\frac{|\vect q|^2 }{2V_{\text{Brow}}^{(qq)}(t)}}}{2\pi V_{\text{Brow}}^{(qq)}(t)},
    \label{HSNDF}
\end{equation}
which identically coincides with the well-known result for the standard Brownian motion for weak system-environment coupling. That is, the chirality features introduced by the flux-carrying action (\ref{EASF}) do not modify the initial symmetry of the spatial distribution of the system. This situation contrast with the conventional Brownian motion in presence of an external inhomogeneous magnetic field \cite{abdoli20202}, where the particle density evolution may be substantially influenced by nondifussive fluxes.

\begin{figure*}
\centering
\includegraphics[scale=0.24]{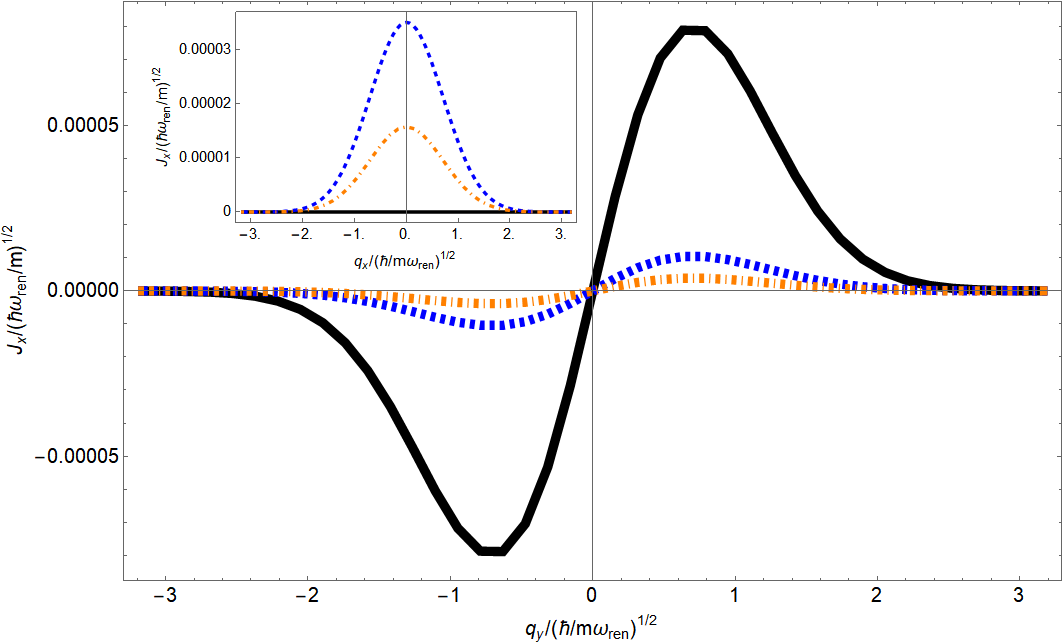}
\includegraphics[scale=0.32]{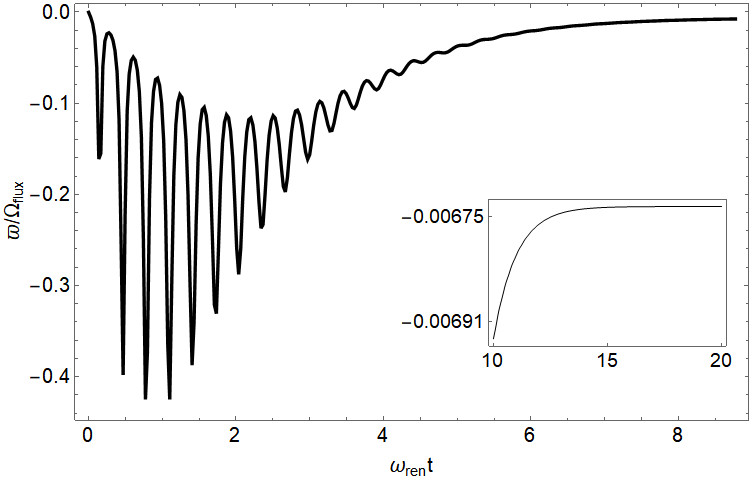}
\caption{(color online). (Left) The x-component of the flow density as a function of the y-coordinate at three fixed values of the x-coordinate $q_{x}$: the solid black, dashed blue and dot-dashed orange lines correspond to $q_{x}=0$, $q_{x}=0.13(\hbar/m\omega_{\text{ren}})^{1/2}$, and $q_{x}=0.17(\hbar/m\omega_{\text{ren}})^{1/2}$ respectively. Similarly, the inset depicts the x-component of the flow density as a function of the X-coordinate $q_x$ at three fixed values of $q_{y}$: the solid black, dashed blue and dot-dashed orange lines correspond to $q_{y}=0$, $q_{y}=0.13(\hbar/m\omega_{\text{ren}})^{1/2}$, and $q_{y}=0.17(\hbar/m\omega_{\text{ren}})^{1/2}$ respectively. (Right) Vorticity in the low-temperature regime as a function of time. The inset illustrates the asymptotic evolution of the vorticity. We have fixed $\omega_{\text{ren}}=11$, $\gamma=0.1\omega_{\text{ren}}$,  $\Omega_{\text{flux}}=0.01\omega_{\text{ren}}$ and $\beta=100(\hbar\omega_{\text{ren}})^{-1}$. Additionally, we have employed the asymotitc values of the thermal covariance matrix given by equations from (\ref{TCVCLTS1}) to (\ref{TCVCLTS3})  with $\omega_c=10\omega_{\text{ren}}$. \label{Fig2HS}} 
\end{figure*}

We now compute the flow density of the single particle after replacing the result (\ref{CVMHSLT}) in Eq. (\ref{MDFDSP}). This yields 
\begin{equation}
    \vect J(\vect q,t)=\frac{n(\vect q,t)}{m V_{\text{Brow}}^{(qq)}(t)
    }\Big(V_{\text{Brow}}^{(qp)}(t)\ \vect q+V_{\text{flux}}(t)\ \vect z\times \vect q \Big),
    \label{HSFDF}
\end{equation}
where we can recognize the flux-carrying contribution with a vortex flow. Interestingly, the latter completely determines the asymptotic value of the flow density, i.e.
\begin{equation}
\vect J(\vect q,\infty)=\frac{n(\vect q,\infty)\sigma_{\text{flux}}(\infty)}{m \sigma_{\text{Brow}}^{(qq)}(\infty)}\ \vect z\times \vect q,
\nonumber
\end{equation}
which manifests the formation of a non-vanishing vortex flow at thermal equilibrium. This result is also illustrated by figure (\ref{Fig1HS}), which depicts the flow density at three different times in the low-temperature regime. By paying attention, one may observe that, while the conventional Brownian diffusion dominates at the first stage of the quantum kinetics (see the left panel), the vortex flow prevails in the asymptotic time (see the right panel). Moreover, the left panel of figure (\ref{Fig2HS}) illustrates the x-component of the flow density, denoted by $J_{x}(\vect q,t)$, in the asymptotic time. One may clearly appreciate that this component exponentially decreases when we move away from the center, and more importantly, it is an odd function in $q_{y}$, which reflects the vorticity of the flow. Notably, its behavior resemblances the Lamb-Oseen profile found in dry chiral active fluids \cite{banerjee20171}, as well as its 2D pattern recalls the nonequilibrium stationary Lorentz flow for the Brownian motion in presence of external magnetic fields \cite{abdoli20201,abdoli20202}. Let us emphasize that the vortex flux-carrying Brownian flow is consequences of purely quantum effects, and, though it is not show here, $\vect J_{\text{flux}}(\vect q,\infty)$ vanishes in the high temperature limit in line with previous discussions (recall that $\sigma_{\text{flux}}(\infty)$ cancels in the classical regime). Furthermore, we must stress out that the heat current (\ref{HCD1}) is null in the asymptotic time (i.e. $\vect J_{K}(\vect q,\infty)=\vect 0$) as expected.

From the definition (\ref{MDFDSP}) of the flux density, it is immediate to obtain the stream velocity field after replacing the result (\ref{HSFDF}), i.e.
\begin{equation}
\vect u(\vect q,t)=\frac{1}{m V_{\text{Brow}}^{(qq)}(t)
    }\Big(V_{\text{Brow}}^{(qp)}(t)\ \vect q+V_{\text{flux}}(t)\ \vect z\times \vect q \Big).
\label{HSVFU}
\end{equation}
Additionally, by substituting Eq.(\ref{HSVFU}) in the definition (\ref{DFVORTSP}), we directly arrive to the expression for the fluid vorticity, that is
\begin{equation}
\varpi(\vect q, t)=\frac{2V_{\text{flux}}(t)}{m V^{(qq)}_{\text{Brow}}(t)}.
\label{HSVCF}
\end{equation}
Equation (\ref{HSVCF}) reveals that the flux-carrying effects give rise to an uniform vorticity flux in the plane. In other words, the motion of the fluid in the bulk becomes rotational. This could be expected by paying attention to the vorticity balance equation: since the flux diffusion coefficient goes quadratic with $\Omega_{\text{flux}}$ (see Eq.(\ref{EMDF1ARLT})), the first term in Eq.(\ref{BEV2SP}) representing the environmental torque will dominate in the weak coupling (i.e. $\Omega_{\text{flux}}/\gamma_{\text{Brow}}<1$) and low temperature regime. On the other side, it turns out that $\vect P_{K}\propto n \vect I_{2}$, see Eq. (\ref{EDKC2}) below, and thus, the forth term in the right hand side of (\ref{BEV2SP}) cancels (or equivalently, the second term in the right hand side of Eq.(\ref{EDC2DR}) vanishes). As anticipated in the previous section, the vortex flow thus arises out of the environmental torque combined with the transverse transport process characterized by $D_{\text{flux}}(\infty)$. Figure (\ref{Fig2HS}) showcases its time evolution for a fixed flux-carrying strength. One may see a highly oscillatory behavior with a varying amplitude upper bounded by $\Omega_{\text{flux}}$ at the beginning, whereas the fluid vorticity asymptotically decays to a constant non-zero value (see the inset) as a consequence of the underlying dissipative effects. Collectively, Figs. (\ref{Fig1HS}) and (\ref{Fig2HS}) prove the generation of a disipationless vortex flow in the single-particle scenario. This stands in contrast to the conventional Brownian motion in presence of an external magnetic field, in which there is no fluxes at thermal equilibrium \cite{abdoli20201,abdoli20202,vuijk20191}. We notice that the small strength of both the steady flow density and the fluid vorticity is in agreement with the subsidiary condition (\ref{SDC1}), which indirectly establishes that the flux-carrying effects must remain perturbative in comparison with the dissipative effects (otherwise the quantum kinetics (\ref{QBE1SP}) would deviate from the low-lying description provided by (\ref{EAE}) \cite{valido20191,valido20201}).

Finally, we draw attention to the kinetic pressure and the kinetic energy density. In the single-particle scenario, these quantities takes the form, respectively,
\begin{align}
    \vect P_{K}(\vect q,t)&=\frac{V_{\text{Brow}}^{(qq)}(t)V_{\text{Brow}}^{(pp)}(t)-\big(V_{\text{flux}}(t)\big)^2-\big(V_{\text{Brow}}^{(qp)}(t)\big)^2 }{m V_{\text{Brow}}^{(qq)}(t)} \ n(\vect q,t) \ \vect I_{2}, \label{STKC2} \\
    e_{K}(\vect q, t)&=\frac{1}{m V_{\text{Brow}}^{(qq)}(t)}\Big(V_{\text{Brow}}^{(qq)}(t)V_{\text{Brow}}^{(pp)}(t)-\big(V_{\text{flux}}(t)\big)^2-\big(V_{\text{Brow}}^{(qp)}(t)\big)^2\Big). \label{EDKC2}
\end{align}
To obtain the expression (\ref{STKC2}) and (\ref{EDKC2}) we follow the same procedure as for Eq. (\ref{HSFDF}): we carried out the Gaussian integral over momentum space, once (\ref{HSVFU}) is substituted in Eqs. (\ref{STKCSP}) and (\ref{BECESP}). By comparing Eqs.  (\ref{STKC2}) and (\ref{EDKC2}), one may realize that the hydrostatic pressure is given by $p_{K}(\vect q,t)=e_{K}(\vect q, t)n(\vect q,t)$. Interestingly, this coincides with the Boyle's law characteristic of 2D ideal gases \cite{schekochihin20201}, where $e_{K}(\vect q,t)/k_{B}$ is the so-called kinetic temperature \cite{lagos20111}. In the asymptotic time limit, the energy density takes the following form at leading order in the flux-attachment effects,
\begin{equation}
    e_{K}(\vect q, \infty)=\frac{\omega_{\text{ren}}\hbar}{2}\coth\big(\hbar\omega_{\text{ren}}\beta/2\big)-\frac{2\omega_{\text{ren}} \sigma_{\text{flux}}(\infty)}{\coth\big(\hbar\omega_{\text{ren}}\beta/2\big)},\label{EDKC2A}
\end{equation}
where the second term in the right hand side is just due to the flux-carrying contribution. Clearly, the latter eventually cancels in the high temperature limit in agreement with Eq. (\ref{EMSF1ARHT}). By replacing this result in the aforementioned Boyle's law, one may see that the flux-carrying contribution manifests as a screen mechanism. This result can be intuitively understood by recalling that the CS action produce repulsive effects that challenges with the confining harmonic potential \cite{valido20191}: the flux-carrying particle is drifted away by the vortex flow, which may effectively reduce the hydrostatic pressure. Furthermore, this result is consistent with the observation that (\ref{HSFDF}) must represent a dissipationless flow. Besides, the standard equipartition theorem for 2D Brownian particles (just having translational degrees of freedom) is recovered in the high temperature limit since the kinetic temperature approaches to the environmental temperature.


\section{General formalism: the non-equilibrium generating functional}\label{Sec_QKT}
In this section we illustrate the derivation of the quantum kinetic equation of the flux-carrying Brownian motion in general dissipative scenarios. Our strategy basically consists of obtaining the real-time effective action $S$ and the associated nonequilibrium generating functional $\mathcal{Z}_{\text{Brow-flux}}$, and then, we perform the Wigner-Weyl transform in the context of quantum path integrals to turn the problem to the aforementioned phase-space framework. Recall that we consider the tensor-product state $\hat \rho_{0}\otimes \hat \rho_{\beta}$ at initial time $t_0$ between the particle system and the MCS environment \cite{hu19921,weiss20121}, where $\hat\rho_{\beta}$ is $e^{-\beta \hat H_{MCS}}$ up to normalization and $\hat H_{MCS}$ denotes the free Hamiltonian of the MCS environment. Owing to the separated property of the initial joint state, the so-called Wick rotation \cite{weiss20121} leads us from $Z_{\text{Brow-flux}}$ to $\mathcal{Z}_{\text{Brow-flux}}$ (i.e. we may pass from $S^{(E)}$ to $-iS$ by doing $\tau \rightarrow it$). From this point we can follow the path integral formalism introduced in Refs.\cite{calzetta20031,boyanovsky20051,anisimov20091} to recast the nonequilibrium generating functional in the following convenient form (which we derive in App. \ref{app1})
\begin{align}
\mathcal{Z}_{\text{Brow-flux}}=\int_{\mathbb{R}^{4N}} d^{4N}\vect x\int_{\mathbb{R}^{4N}} d^{4N}\vect x_{i}\  W_{0}(\vect x_{i},t_{0})&\int \mathcal{D} \vect \Xi  \ \mathcal{P}[\vect \Xi]\ \delta\Big[ \dot{\vect x}(t)\label{NEGF1} \\
&+\vect \Phi\cdot\vect x(t)   +\int_{t_{0}}^{t}d s \ \vect \Psi(t-s)\cdot\vect x(s)- \vect \Xi(t) \Big],  \nonumber
\end{align}
where $W_{0}$ denotes the Wigner function of the $N$-particle flux-carrying Brownian system associated to $\hat \rho_{0}$. Here, $\delta[\bullet]$ stands for the functional Dirac delta function in the phase space, and $\vect \Xi$ represents a quantum  Gaussian noise fully characterized by the functional probability distribution \cite{calzetta20001,anisimov20091},
\begin{equation}
  \mathcal{P}[\vect \Xi]=\text{exp}\bigg(-\frac{1}{2} \int_{t_{0}}^{t}ds\int_{t_{0}}^{t}ds'\ \vect \Xi^{T}(s)\cdot\vect N^{-1} (s,s')\cdot\vect \Xi(s')\bigg), \label{FPDF}
\end{equation}
where $\vect N$ corresponds to the $4N\times 4N$ noise matrix, i.e. $\vect N(t,t')= \left\langle\left\lbrace   \vect \Xi(t), \vect \Xi(t')\right\rbrace \right\rangle_{\hat \rho_{\beta}}$ with $\left\langle\left\lbrace \bullet \right\rbrace \right\rangle_{\hat \rho_{\beta}}$ denoting the average over the environmental canonical equilibrium state $\hat \rho_{\beta}$ and $\left\lbrace \vect A, \vect B\right\rbrace =\vect A^{T}\cdot\vect B+\vect B^{T}\cdot\vect A$. Owing to the environmental equilibrium conditions, it is found that both $\left\langle\left\lbrace   \vect \Xi(t)\right\rbrace \right\rangle_{ \hat \rho_{\beta}}$ cancels and the two-point autocorrelation function of the fluctuating force satisfies a fluctuation-dissipation relation \cite{roura19991,calzetta20031}. This can be compactly expressed in terms of a $2N$ real vector $\vect \xi(t)$ in the phase space as follows 
\begin{eqnarray}
\left\langle\left\lbrace   \vect \Xi(t), \vect \Xi(t')\right\rbrace \right\rangle_{ \hat \rho_{\beta}}= \left( \begin{array}{cc}
        \vect 0_{2N}& \vect 0_{2N} \\
          \vect 0_{2N}&\left\langle\left\lbrace   \vect \xi(t), \vect \xi(t')\right\rbrace \right\rangle_{\hat \rho_{\beta}}
   \end{array} \right), \ \ \ \ t'\le t \label{FDRM}
\end{eqnarray}
where its matrix elements are given by \cite{valido20191}
\begin{align}
   \big(\left\langle \left\lbrace \vect \xi (t),\vect \xi (t')\right\rbrace  \right\rangle_{\hat \rho_{\beta}}\big)_{il}&=\frac{2\hbar }{\pi}\sum_{\nu=1,2}\int_{0}^{\infty}d\omega\ \coth\left(\frac{\hbar \omega\beta}{2}\right) \bigg[\delta_{\alpha\nu}\bigg(1-\Big(\frac{\kappa}{\omega}\Big)^2\bigg)\cos(\omega (t-t'))\nonumber \\
   &+\epsilon_{\alpha\nu} \Big(\frac{\kappa}{\omega}\Big)\sin(\omega (t-t'))\bigg]h_{\nu\lambda}(\omega,\Delta\bar{\vect q}_{ij}) ,\ \ \ \lambda\leq\alpha,
   \label{FDRM1}
\end{align}
with $l=\frac{1}{2}(i-(\alpha+\lambda-1))+1$ (with $i,l=1,\cdots, N$ and $\alpha,\lambda =1,2$), and $\vect 0_{2N}$ denoting the $2N\times 2N$ null matrix (i.e. every element is equal to the zero). Notice that $\vect \xi $ is the phase-space counterpart of the quantum operator $\hat{\vect \xi}_{\text{Brow}}+\hat{\vect \xi}_{\text{flux}}$ appearing in the generalized Langevin equation (\ref{QLEMASP}) in the single particle scenario: indeed, we shall see that Eq. (\ref{FDRM1}) retrieves the extended fluctuation-dissipation relation (\ref{FDRF}) in the weak coupling regime. Additionally, we have introduced the $2\times 2$ matrix which plays the role of an extended spectral density \cite{valido20131},
\begin{equation}
   \vect  h(\omega,\Delta\bar{\vect q}_{ij})=\frac{\hbar e^2}{8\pi L^2}\sum_{\vect k\in \mathbb{R}^2}\omega_{\vect k}f^2(|\vect k|)\cos(\vect k \cdot \Delta\bar{\vect q}_{ij})\delta(\omega-\omega_{\vect k})\left(\begin{array}{cc}
      \mathcal{P}_{11}(\vect k)   &  1\\
       1  & \mathcal{P}_{22}(\vect k) 
    \end{array}\right) ,\label{ESDD}
\end{equation}
where $\mathcal{P}_{\alpha\lambda}$ is the transverse projective operator in momentum space \cite{boyanovsky20051}, i.e. $\mathcal{P}_{\alpha\lambda}(\vect k)=(\delta_{\alpha\lambda}|\vect k|^2-k_\alpha k_\lambda)/|\vect k|^2$. Here, $0<e$ determines the coupling strength to the MCS environment, and $f(|\vect k|)\in\mathbb{R}$ represents the usual spherically symmetric smooth form factor from non-relativistic quantum electrodynamics \cite{buenzli20071}, which prevents from the ultraviolet catastrophe. It is important to note that the spectral density (\ref{ESDD}) explicitly depends on the distance $\Delta\bar{\vect q}_{ij}$ between system particles. In other words, our microscopic description (\ref{EAE}) takes into account the non-local self-interactions carried out by the common MCS environment. Concretely, it is known that an effective environmental-mediated coupling is established when the average distance between the system particles is sufficiently small in comparison with the time scale of the environmental memory effects \cite{valido20131}. The latter statement can be expressed as $|\Delta\bar{\vect q}_{ij}|\omega_{c}/c \ll 1$, with $\omega_{c}$ being the largest environmental frequency that significantly contributes to the dissipative dynamics (e.g., $\omega_{c}$ correspond to the high-frequency cutoff in Drude-model of the spectral density \cite{grabert19841}). In the opposite scenario (i.e. $|\Delta\bar{\vect q}_{ij}|\omega_{c}/c \gg 1$), the system particles can be considered in contact with independent environments.

Importantly, Eq.(\ref{NEGF1}) encodes the open quantum system dynamics in a generating functional form \cite{calzetta20001}: the time evolution is computed from a functional integral over all possible stochastic phase-space trajectories of the $N$-particle system, where the functional Dirac delta function ensures that a non-vanishing weight is only attributed to those trajectories obeying the quantum stochastic equations of motion. In other words, the nonequilibrium generating functional (\ref{NEGF1}) tells us that the open system dynamics of the flux-carrying Brownian particles is encoded by $2N$ generalized Langevin equations that can be compactly expressed in a matricial form as follows: while $\vect \Xi(t)$ represents the fluctuating force vector, the free evolution and the memory kernel are respectively given by the $4N\times 4N$ matrices
\begin{equation}
   \vect \Phi=\left( \begin{array}{cc}
        \vect 0_{2N}& -\frac{\vect I_{2N} }{m} \\
       \vect U_{\text{ren}}  &   \vect 0_{2N}
   \end{array} \right), \label{HPM}
\end{equation}
and 
\begin{equation}
   \vect \Psi(t-t')=\left( \begin{array}{cc}
        \vect 0_{2N}& \vect 0_{2N} \\
       \vect \Sigma(t-t')&   \vect 0_{2N}
   \end{array} \right), \label{MKM}
\end{equation}
Recall the matrix element of $\vect U_{\text{ren}}$ are compactly given by (\ref{RPME}), and we have further introduced the $2N\times 2N$ self-energy matrix,
\begin{equation}
  \big(\vect \Sigma(t)\big)_{il}=\Sigma_{\alpha\lambda}(t,\Delta\bar{\vect q}_{ij} ), \ \ \ \lambda\leq\alpha,
\end{equation}
where $l=\frac{1}{2}(i-(\alpha+\lambda-1))+1$ (with $i,l=1,\cdots, N$ and $\alpha,\lambda =1,2$), and its coefficients are obtained from,
\begin{align}
   \Sigma_{\alpha\lambda}(t,\Delta\bar{\vect q}_{ij})=&-\frac{2}{m\pi}\Theta(t-|\Delta\bar{\vect q}_{ij}|) \sum_{\nu=1,2}\int_{0}^{\infty}d\omega\  \bigg[\delta_{\alpha\nu}\bigg(1-\Big(\frac{\kappa}{\omega}\Big)^2\bigg)\sin(\omega t)\nonumber \\
   &-\epsilon_{\alpha\nu} \Big(\frac{\kappa}{\omega}\Big)\cos(\omega t)\bigg]h_{\nu\lambda}(\omega,\Delta\bar{\vect q}_{ij}), \label{FDRM2}
\end{align}
with $\Theta(t)$ denoting the Heaviside step function. For future treatment, it is important to bear in mind that the diagonal components in the position coordinates are obtained from a frequency-dependent sine Fourier transform, whereas its off-diagonal elements are given by a frequency-dependent cosine Fourier transform. As expected, Eq. (\ref{FDRM2}) exactly returns the retarded friction kernel from standard Brownian motion \cite{weiss20121} when flux-carrying effects are switched off (i.e. $\kappa\rightarrow 0$).

The solution of the generalized Langevin equation encapsulated in the nonequilibrium generating functional (\ref{NEGF1}) reads \cite{fleming20111}
\begin{equation}
    \vect x(t)=\vect x_{h}(t)+\int_{t_{0}}^{t}ds \ \vect K_{R}(t-s)\cdot\vect \Xi(s),
    \nonumber
\end{equation}
where $\vect x_{h}$ corresponds to the homogeneous solution and $\vect K_R$ is the $4N\times4N$ retarded kinetic propagator matrix, which can be compactly expressed as \cite{fleming20111}
\begin{equation}
  \vect K_{R}(t)= \left( \begin{array}{cc}
       m\dot{\vect G}_{R}(t)  & \vect G_{R}(t) \\
       m^2\ddot{\vect G}_{R}(t)  & m\dot{\vect G}_{R}(t)
    \end{array}\right),
    \label{RKPE}
\end{equation}
in terms of the $2N\times2 N$ retarded Green's function matrix $\vect G_{R}(t)$. As the action functional governing the open qunatum system dynamics takes a quadratic form in the system particle coordinates (see Eqs. (\ref{EASBR}) and (\ref{EASF})), the retarded Green's function as well. The latter means that $\vect G_{R}(t)$ can be computed by appealing to real-time Fourier transform methods \cite{valido20191}, so it is convenient to express this in terms of its frequency-dependent Green's function $\tilde{\vect G}_{R}(\omega)$. The latter is obtained via analytic continuation from the imaginary-time Fourier transforms of the dynamical susceptibilities (we referee the interesting reader to App. \ref{app1}). Concretely, we find 
\begin{equation}
\tilde{\vect G}_{R}^{-1}(\omega)=\vect I_{N}(-\omega^2-i0^{+})+\vect U_{\text{ren}}+\tilde{\vect \Sigma}(\omega+i0^{+}),
\label{FTRGF}
\end{equation}
where we have introduced the real-time Fourier transform of the self-energy (notice that the infinitesimal imaginary part $0^{+}$ enforces causality),
\begin{equation}
\tilde{\Sigma}_{\alpha\lambda}(\omega,\Delta\bar{\vect q}_{ij}) =\tilde{\Gamma}_{\alpha\lambda}(\omega,\Delta\bar{\vect q}_{ij})+\tilde{\Lambda}^{||}_{\alpha\lambda}(\omega,\Delta\bar{\vect q}_{ij})+\tilde{\Lambda}^{||}_{\alpha\lambda}(\omega,\Delta\bar{\vect q}_{ij}),
   \label{DKNGF}
\end{equation}
for $i,j=1,\cdots, N$ and $\alpha,\lambda=1,2$, where
\begin{align}
\tilde{\Gamma}_{\alpha\lambda}(\omega,\Delta\bar{\vect q}_{ij})&=-\frac{\delta_{\alpha\lambda}}{m}\bigg[i\ \text{sgn}(\omega)h_{\lambda\lambda}(|\omega|,\Delta\bar{\vect q}_{ij})+\omega^2\mathcal{H}\bigg(\text{sgn}(\omega')\frac{h_{\lambda\lambda}(|\omega'|,\Delta\bar{\vect q}_{ij}))}{|\omega'|^2}(\omega)\bigg)\bigg],\label{FTSECI} \\
\tilde{\Lambda}^{||}_{\alpha\lambda}(\omega,\Delta\bar{\vect q}_{ij})&=\frac{\delta_{\alpha\lambda}}{m}\Big(\frac{\kappa}{\omega}\Big)^2\bigg[i\ \text{sgn}(\omega)h_{\lambda\lambda}(|\omega|,\Delta\bar{\vect q}_{ij})+\omega^2\mathcal{H}\bigg(\text{sgn}(\omega')\frac{h_{\lambda\lambda}(|\omega'|,\Delta\bar{\vect q}_{ij}))}{|\omega'|^2}(\omega)\bigg)\bigg]\label{FTSECII}, \\
\tilde{\Lambda}^{\perp}_{\alpha\lambda}(\omega,\Delta\bar{\vect q}_{ij})&=-i\epsilon_{\alpha\lambda}\Big(\frac{\kappa}{m\omega}\Big)\bigg[i\ \text{sgn}(\omega)h_{\lambda\lambda}(|\omega|,\Delta\bar{\vect q}_{ij})+\omega^2\mathcal{H}\bigg(\text{sgn}(\omega')\frac{h_{\lambda\lambda}(|\omega'|,\Delta\bar{\vect q}_{ij}))}{|\omega'|^2}(\omega)\bigg)\bigg].
\label{FTSECIII}
\end{align}
Expressions (\ref{FTSECII}) and (\ref{FTSECIII}) are, respectively, the real-time Fourier transforms of the longitudinal and transverse dynamical susceptibilities introduced in Sec. \ref{Sub_MICm}, whereas Eq.(\ref{FTSECI}) corresponds to the usual dissipation kernel of the conventional Brownian motion \cite{weiss20121,grabert19881}. As similarly occurs in the Caldeira-Leggett model, we would like to emphasize that the precise form of all the dynamical susceptibilities  (\ref{FTSECI}), (\ref{FTSECII}) and (\ref{FTSECIII}) is fixed by the choice of the spectral density (\ref{ESDD}).

Now, by starting from the generalized Langevin equation characterized by the free evolution matrix (\ref{HPM}), the memory kernel matrix (\ref{MKM}) and the fluctuating force vector $\vect \Xi(t)$ satisfying the fluctuation-dissipation relation (\ref{FDRM}), one may follow the procedure presented in \cite{fleming20111} to obtain the desired expression for the quantum master equation in the terms of the Wigner distribution function. Up to doing this, we find the quantum kinetic equation
\begin{equation}
    \frac{\partial W(\vect x,t)}{\partial t}= \Bigg(\Big(\frac{\partial}{\partial \vect x}\Big)^{T} \cdot \vect \Phi\cdot  \vect x+\Big(\frac{\partial}{\partial \vect x}\Big)^{T}\cdot \vect \gamma(t)\cdot  \vect x +\Big(\frac{\partial}{\partial \vect x}\Big)^{T}\cdot \vect D(t) \cdot  \frac{\partial}{\partial \vect x}\Bigg)W(\vect x,t), \label{QMFPE}
\end{equation}
where $\vect \gamma(t)$ and $\vect D(t)$ are the so-called pseudo-Hamiltonian and diffusion matrices, respectively. These are determined from the retarded kinetic propagator (\ref{RKPE}) as follows \cite{fleming20111},
\begin{eqnarray}
\vect \gamma(t)&=&-(\dot{\vect K}_{R}(t)\cdot  \vect K_{R}^{-1}(t)+\vect \Phi) ,\label{QMEGP} \\
\vect D(t)&=&\frac{1}{2}\big(\{\vect \gamma(t)+\vect \Phi, \vect \sigma(t)\}+\dot{\vect \sigma}(t)\big), \label{QMEDP}
\end{eqnarray}
where $\vect \sigma(t)$ stands for the thermal covariance matrix (as previously introduced), i.e.
\begin{equation}
\vect \sigma(t)=   \int_{t_{0}}^{t}ds\int_{t_{0}}^{t}ds'\ \vect K_{R}(t-s)\cdot \left\langle\left\lbrace   \vect \Xi(s), \vect \Xi(s')\right\rbrace \right\rangle_{\hat \rho_{\beta}} \cdot  \vect K_{R}^{T}(t-s').
\label{SCVM}
\end{equation}
Despite the parity and time-reversal symmetry breaking, we note that $\vect \sigma(t)$ and $\vect D(t)$ are symmetric real matrices by construction. For our later purposes, it is convenient to decompose the diffusion matrix in terms of the decoherence and system-to-bath diffusion submatrices, that is
\begin{equation}
 \vect D(t)=\left( \begin{array}{cc}
     \vect O_{N} & \vect D^{(qp)}(t) \\
     \vect D^{(qp)}(t) & \vect D^{(pp)}(t)
 \end{array}    \right) ,
 \nonumber
\end{equation}
where $\vect D^{(qp)}(t)$ and $\vect D^{(pp)}(t)$ are $2N\times 2N$ symmetric real matrices, and are referred to as the anomalous and decoherence diffusion tensors, respectively. This block decomposition was employed employed in the analysis of the quantum master equation (\ref{QBE1SP}) in the single particle scenario.

Equation (\ref{QMFPE}) represents the so-called Kramers equation which takes account direct coupling between system particles or environmental-mediated interactions, as well as non-Markovian effects. This will describe a great diversity of many-body phenomena related to the (linear) Brownian motion consistent with the MCS electrodynamics in the low-energy regime \cite{valido20191} (e.g. thermalization, diffusive as well as nondiffusive processes). To circumvent the immense complication of solving this general many-particle problem, in the following section we shall focus the attention in the scenario in which the system and MCS environment are weakly coupled, such that we can employ a suitable Breit-Wigner approximation of the retarded Green's function (\ref{FTRGF}). This will lead us to a Fokker-Planck type equation for the flux-carrying Brownian particles that goes beyond the Markovian treatment widely used in quantum optics and atomic physics \cite{agarwal19711,devega20171}.
\subsection{Weak system-environment coupling regime: the Breit-Wigner approximation}\label{Subsec_FPPC}

The retarded Green’s function (\ref{FTRGF}) may display an intricate mixture of ”particle” poles and brunch cut singularities in the complex frequency plane \cite{rammer20071}. It is well known that $\tilde{\vect G}_{R}^{-1}(\omega)$ manifests sharply peakeds characterized by quasiparticle poles in the weak system-environment coupling regime \cite{boyanovsky20051,anisimov20091,alamoudi19991}. In the present work, we focus the attention on the open system dynamics which is mainly dominated by such quasiparticle poles. This is amount to approximate $\tilde{\vect G}_{R}(\omega)$ by a Breit-Wigner resonance shape \cite{valido20191,boyanovsky20051,kuzemsky20171}, i.e. $\tilde{\vect G}_{R}(\omega)\approx\tilde{\vect G}_{BW}(\omega+i0^{+})$ with
\begin{equation}
\tilde{\vect G}_{BW}^{-1}(\omega)= m\big(-\omega^2 \vect I_{2N}-i\omega\vect \Gamma+\vect \Omega^{\circ 2}\big)\circ\vect Z^{-1},
\label{RGFBWF}
\end{equation}
where we have introduced the friction tensor $\vect \Gamma$ and the quasi-particle resonance matrix $\vect \Omega$, i.e. 
\begin{align}
\vect \Gamma&= -\vect Z \circ\text{Im}\ \tilde{\vect \Sigma}(\vect \Omega)\circ\vect \Omega^{\circ-1},
\label{BWFMK}\\
 \vect \Omega^{\circ 2}&=\frac{1}{m}\vect U_{\text{ren}}+\text{Re}\ \tilde{\vect \Sigma}(\vect \Omega),
   \label{BWPRM}
\end{align}
with $\vect Z$ denoting a renormalization matrix, i.e.
\begin{equation}
  \vect Z^{\circ-1}=\vect{\mathcal{I}}-\frac{\partial \text{Re}\ \tilde{\vect \Sigma}(\omega)}{\partial \omega^2}\Bigg|_{\vect \Omega}.
\label{BWRNT}
\end{equation}
Here the symbol $\circ$ denotes the Haddamard product (e.g. $\vect a\circ\vect b=a_{ij}b_{ij}$), and $\vect{\mathcal{I}}$ is the $2N\times 2N$ all-ones matrix (i.e. every element is equal to the unit).  In order to obtain the expressions (\ref{BWFMK}) and (\ref{BWPRM}) we approximate the real and imaginary parts of the self-energy as usually in the context of the Breit-Wigner approximation, i.e.
\begin{align}
\text{Im}\ \tilde{\vect \Sigma} (\omega)&=\text{Im}\ \tilde{\vect \Sigma}(\vect \Omega)-\vect Z^{\circ-1} \circ\vect \Gamma\circ(\omega\vect{\mathcal{I}}-\vect\Omega),   \label{IMSE} \\
\text{Re}\ \tilde{\vect \Sigma}(\omega)&=\text{Re}\ \tilde{\vect \Sigma}(\vect \Omega)+\big(\vect{\mathcal{I}}-\vect Z^{\circ-1}\big)\circ(\omega^2\vect I_{2N}-\vect\Omega^{\circ 2}), \label{RESE} 
\end{align}
and thus, we substitute (\ref{IMSE}) and (\ref{RESE}) in Eq.(\ref{FTRGF}) to recast the retarded Green's function in the form of Eq.(\ref{RGFBWF}), as desired. The specific expressions of $\vect \Gamma$ and $\vect \Omega$ are obtained after substituting the real and imaginary parts of the frequency-dependent retarded self-energy in terms of the spectral density (\ref{ESDD}) by means of the Fourier transform of the dynamical susceptibilities (c.f. Eqs. (\ref{FTSECI})-(\ref{FTSECIII})). Upon doing this we get,
\begin{align}
 \Omega_{il}^{2}&= \frac{1}{m}(\vect U_{\text{ren}})_{il}-\frac{\delta_{\alpha\lambda}\Omega_{il}^2}{m}\bigg(1-\Big(\frac{\kappa}{\Omega_{il}}\Big)^2\bigg)\mathcal{H}\bigg(\frac{\text{sgn}(\omega)}{|\omega|^2}h_{\alpha\lambda}(|\omega|,\Delta\bar{\vect q}_{i\frac{1}{2}(i-(\alpha+\lambda-1))+1})\bigg)(\Omega_{il})
 \nonumber \\
 &+\epsilon_{\alpha\lambda} \Big(\frac{\kappa}{m\Omega_{il}}\Big)\ \text{sgn}(\Omega_{il})h_{\alpha\lambda}(|\Omega_{il}|,\Delta\bar{\vect q}_{i\frac{1}{2}(i-(\alpha+\lambda-1))+1}), \ \ \  \lambda\leq\alpha \label{PTSP1} 
 \end{align}
 and
 \begin{align}
\Gamma_{il}&=\frac{Z_{il}}{m\Omega_{il}}\bigg[\delta_{\alpha\lambda}\bigg(1-\Big(\frac{\kappa}{\Omega_{il}}\Big)^2\bigg)\text{sgn}(\Omega_{il})h_{\alpha\lambda}(|\Omega_{il}|,\Delta\bar{\vect q}_{i\frac{1}{2}(i-(\alpha+\lambda-1))+1})\nonumber \\
&+\epsilon_{\alpha\lambda}\kappa  \Omega_{il} \ \mathcal{H}\bigg(\frac{\text{sgn}(\omega)}{|\omega|^2}h_{\alpha\lambda}(|\omega|,\Delta\bar{\vect q}_{i\frac{1}{2}(i-(\alpha+\lambda-1))+1})\bigg)(\Omega_{il})\bigg], \ \ \  \lambda\leq\alpha
\label{PTSP2}
\end{align}
where $i,l=1,\cdots, N$; $\alpha,\lambda=1,2$; and $\mathcal{H}$ stands for the usual Hilbert transform (see Eq.(\ref{DFHT}) in App. \ref{app1}). We would like to emphasize that the Breit-Wigner ansatz (\ref{RGFBWF}) relies on general physical grounds, since it is equivalent to consider the most general time-local memory kernel (\ref{MKM}) of the generalized Langevin equation which holds causality (i.e. $ \vect \Psi(t-t')\propto \delta(t-t')\vect \Psi$). Indeed, the form of the friction tensor (\ref{BWFMK}) and the quasi-particle resonance matrix (\ref{BWPRM}) are very general as we have not made any specific choice of the spectral density.

Before going into the quantum master equation, it is useful to discuss in detail the algebraic properties of the matrices $\vect \Omega$ and $\vect \Gamma$ that are consistent with expressions (\ref{PTSP1}) and (\ref{PTSP2}). On one side, the first and second terms in (\ref{PTSP1}) are akin to the usual situation appearing in the conventional Brownian motion (i.e. they represent both symmetric and diagonal elements in the spatial coordinates), so they can be grouped together to retrieve a $2N\times 2N$ real matrix, say $\vect \Omega_{\text{ren}}$, which contains the usual quasi-particle frequencies \cite{boyanovsky20051,anisimov20091}. On the other side, the third term stems from the transverse dynamical susceptibilities (\ref{FTSECIII}) (notice that it is linear in the CS coupling constant), which also suggests the introduction of a $2N \times 2N$ real matrix, namely $\vect \Omega_{\text{flux}}^{\circ 2}$. Hence, we can put the quasi-particle resonance matrix into the convenient form
\begin{equation}
    \vect \Omega^{\circ 2}=\vect \Omega_{\text{ren}}^{\circ 2}-\vect \Omega_{\text{flux}}^{\circ 2}. \label{MOPOT}
\end{equation}
The Levi-Civita symbol appearing in the second term of the right hand side of Eq. (\ref{PTSP1}) dictates that $\vect \Omega_{\text{flux}}^{\circ 2}$ is skew-symmetric under spatial coordinates transposition. Hence, the most general expression of $\vect \Omega_{\text{flux}}^{\circ 2}$ that we can envisage is
\begin{equation}
    \vect \Omega_{\text{flux}}^{\circ 2}=\Omega_{\text{flux}}^{2}\left( \begin{array}{ccccc}
       \vect A_{1} &\vect A_{2} &\vect A_{3}&\cdots&\vect A_{N}\\
        \vect A_{2} &\vect A_{1} &\vect A_{N}&\cdots&\vect A_{2N-2}\\
        \vdots & && \vdots&\vdots \\
        \vect A_{N} &\vect A_{2N-2} &\vect A_{3N-5}&\cdots&\vect A_{1}
    \end{array}\right), \label{MOFlux}
\end{equation}
where $(A_i)_{\alpha\beta}=\epsilon_{\alpha\beta}a_{i}\in \mathbb{R}$, and $\Omega_{\text{flux}}$ retains the CS constant and, thereby, it denotes the strength of the flux-carrying effects introduced in Sec. \ref{Sub_COMP2DV}. It is important to realize in (\ref{MOFlux}) that the off-diagonal submatrices accounts for the environmental-mediated interactions among the flux-carrying Brownian particles. 

In the same way, one may elucidate the general form of the friction matrix $\vect \Gamma$. Similarly, the first term in Eq.(\ref{PTSP2}) corresponds to a diagonal contribution in the spatial coordinates that closely resemblances the standard Brownian motion, so this could be identified with the usual dissipative matrix, denoted by  $\vect \Gamma_{\text{Brow}}$. Accordingly, this is a $2N\times 2N$ real matrix that can be expressed as,
\begin{equation}
    \vect \Gamma_{\text{Brow}}=\left( \begin{array}{ccccc}
       \vect\Gamma_{1} &\vect \Gamma_{2} &\vect \Gamma_{3}&\cdots&\vect \Gamma_{N}\\
        \vect \Gamma_{2} &\vect \Gamma_{1} &\vect \Gamma_{N}&\cdots&\vect \Gamma_{2N-2}\\
        \vdots & && \vdots&\vdots \\
        \vect \Gamma_{N} &\vect \Gamma_{2N-2} &\vect \Gamma_{3N-5}&\cdots&\vect \Gamma_{1}
    \end{array}\right), \label{MODis}
\end{equation}
with 
\begin{equation}
    \vect \Gamma_{i}=\left(\begin{array}{cc}
        \gamma_{i}^{(1)}  &  0\\
        0 & \gamma_{i}^{(2)}
    \end{array}\right), \nonumber
\end{equation}
where $\gamma_{i}^{(x)}$ and $\gamma_{i}^{(y)}$ are interpreted as the common friction coefficients in each spatial coordinate. We note that the friction matrix (\ref{MODis}) contemplates the most general dissipative scenario within the Breit-Wigner approximation: that is, it encompasses the anisotropic case as well as the common environment situation where an environmental-mediated interaction arises among the flux-carrying Brownian particles due to interparticle friction. On the other side, for visualizing the role played by the second term in (\ref{PTSP2}), we resort to the expression (\ref{FDRM2}) for the memory kernel. By paying attention one may realize that such term is involved in the Fourier cosine transform in the frequency plane. It turns out that the transformation rules imply that it must cancels in order the memory kernel (\ref{MKM}) becomes time local. Concretely, the Fourier cosine transform just retrieves the Dirac delta function given a frequency-independent function, which has vanishing Hilbert transform \cite{gradshteyn20141}. To be self-consistent the Breit-Wigner approximation (\ref{RGFBWF}), we are thus left with $\vect \Gamma=\vect \Gamma_{\text{Brow}}$. In words, this means that $\vect \Gamma$ is identical to the friction matrix obtained from the conventional Brownian motion up to an appropriate rescaling of the friction coefficients (notice from (\ref{PTSP2}) that they are effectively diminished by the longitudinal dynamical susceptibility). This is in agreement with the fact that the flux-carrying influence coming from the transverse dynamical susceptibility represents a dissipationless environmental mechanism. In particular, recalling that we treat the isotropic damped harmonic particle in Sec. \ref{Sec_QBE}, the renormalized potential, friction and flux-carrying matrices take the form
\begin{eqnarray} 
 \vect \Omega_{\text{ren}}=\omega_{\text{ren}}\left(\begin{array}{cc}
         1 & 0 \\
        0 & 1
    \end{array}\right),
  \ \
\vect \Gamma_{\text{Brow}}=\gamma_{\text{Brow}}\left(\begin{array}{cc}
         1 & 0 \\
        0 &  1
    \end{array}\right), \text{and}
    \ \
 \vect \Omega_{\text{flux}}^{\circ 2}=\Omega_{\text{flux}}^{2}\left(\begin{array}{cc}
        0  & 1 \\
        -1 & 0
    \end{array}\right).
    \label{CH2}
 \end{eqnarray}

By doing the frequency-dependent integration once inserted (\ref{MOPOT}) and (\ref{MODis}) in (\ref{FDRM2}) according to expressions (\ref{PTSP1}) and (\ref{PTSP2}), one finds that the memory kernel within the Breit-Wigner approximation becomes $\vect \Psi(t-t')=\delta(t-t')(\vect  \Psi_{ \text{Brow}}+\vect \Psi_{\text{flux}})$, for $t\ge t'$, with 
\begin{equation}
 \vect \Psi_{ \text{Brow}}=\left( \begin{array}{cc}
        \vect 0_{2N}& \vect 0_{2N} \\
       \vect 0_{2N}  &   \vect \Gamma_{\text{Brow}}
   \end{array} \right), \ \  \vect \Psi_{\text{flux}}=\left( \begin{array}{cc}
        \vect 0_{2N}& \vect 0_{2N}  \\
      - \vect m\circ\vect \Omega_{\text{flux}}^{\circ 2}  &    \vect 0_{2N}
   \end{array} \right), 
   \label{SELFEE}
\end{equation}
where we have absorbed the renormalization term (\ref{BWRNT}) into the renormalized mass matrix $\vect m$ (i.e. $\vect m=m\vect Z^{-1}$). Alternatively, this result can be derived by appealing to the fact that the second and third terms of the retarded Green's function (\ref{RGFBWF}) identify with the Fourier transform of the memory kernel. Hence, one may verify that the result of the back Fourier transform to the real-time domain replaced in (\ref{MKM}) yields  (\ref{SELFEE}). Similarly, by substituting the coefficients of the friction matrix $\vect \Gamma_{\text{Brow}}$ and the flux-carrying contribution $\vect \Omega_{\text{flux}}$ deduced from (\ref{MOFlux}) and  (\ref{MODis}) in the two-time correlation function (\ref{FDRM1}), after doing the Fourier transform back to the real-time domain we arrive to the fluctuation-dissipation relation (see App. \ref{app1} for further details)
\begin{equation}
\left\langle\left\lbrace   \vect \Xi_{BW}(t), \vect \Xi_{BW}(t')\right\rbrace \right\rangle_{W_{\beta}}=\frac{2m}{\beta}\bigg(\frac{d }{dt}\bigg(\coth\bigg(\frac{\pi (t-t')}{\hbar \beta}\bigg)\bigg) \vect \Upsilon_{\text{Brow}} +\coth\bigg(\frac{\pi (t-t')}{\hbar \beta}\bigg) \vect \Upsilon_{\text{flux}}\bigg),
\label{FDTHBW}
\end{equation}
where we have introduced the noise matrices due to the Brownian and flux-carrying contributions,
\begin{equation}
    \vect \Upsilon_{\text{Brow}}= \left(\begin{array}{cc}
        \vect 0_{2N}& \vect 0_{2N} \\
       \vect 0_{2N}& \vect \Gamma_{\text{Brow}}
       \end{array} \right), \ \ \ \ \   \vect \Upsilon_{\text{flux}}= \left(\begin{array}{cc}
        \vect 0_{2N}& \vect 0_{2N} \\
       \vect 0_{2N}&-\vect\Omega_{\text{flux}}^{\circ 2}
       \end{array} \right). \label{FNMatr}
\end{equation}
In the single particle scenario, one may verify that the second term of the right hand side of Eq. (\ref{FDTHBW}) retrieves the extended fluctuation-dissipation relation (\ref{FDRF}) after replacing (\ref{CH2}) in (\ref{FNMatr}). Equation (\ref{FDTHBW}) manifests that the open quantum system dynamics remains non-Markovian despite the Breit-Wigner approximation. This is one of the important aspects that distinguishes the quantum kinetic approach presented here from the usual Boltzmann equation \cite{boyanovsky20051}. As previously mentioned, the strict Markovian limit occurs for time scales $1\ll  (t-t')/\hbar \beta$ (such that $\coth(\pi (t-t')/\hbar \beta)\rightarrow \text{sgn}(t-t')$). This directly yields,
\begin{eqnarray}
\left\langle\left\lbrace   \vect \Xi_{BW}(t), \vect \Xi_{BW}(t')\right\rbrace \right\rangle_{W_{\beta}}=\frac{2m}{\beta} \Big(2\delta(t-t') \vect \Upsilon_{\text{Brow}}+\text{sgn}(t-t') \vect\Upsilon_{\text{flux}}\Big),
\label{FDTHBWSML}
\end{eqnarray}
which returns Eq.(\ref{FDRFML}) in the single-particle scenario, as expected. By paying attention to (\ref{FDTHBWSML}), one may recognize the first term in the right hand side with the usual Gaussian white noise characteristic of the conventional Brownian motion, whilst the second term corresponds to the flux noise discussed in Sec. \ref{Sub_COMP2DV}. 

Finally, we conclude the analysis of the Breit-Wigner approximation (\ref{RGFBWF}) discussing its validity. Essentially, this demands that the system-environment coupling is weak in comparison with the particle bare frequencies \cite{alamoudi19991}. Furthermore, as outlined in the Sec. \ref{Sec_FCBM}, we must also impose that the dissipative effects dominate the quantum kinetics against the flux-carrying effects to be our treatment physically consistent with the microscopic description (\ref{EAE}) \cite{valido20191}. These conditions can be compactly rephrased into the following inequality,
\begin{equation}
   \vect \Omega_{\text{flux}}\circ\vect \Omega_{\text{ren}}^{\circ-1}< \vect \Gamma_{\text{Brow}}\circ\vect \Omega^{\circ-1}_{\text{ren}}\ll 1 .\label{SDC1}
\end{equation}
The subsidiary condition (\ref{SDC1SPS}) for the single particle scenario is directly computed from (\ref{SDC1}) after substituting the renormalized potential, friction, and flux-carrying matrices given by (\ref{CH2}).

Returning the attention to the quantum master equation (\ref{QMFPE}), we now focus on the generalized Langevin equation associated to (\ref{NEGF1}) within the discussed Breit-Wigner approximation. This is directly obtained from the approximated Green's function (\ref{RGFBWF}) after substituting the expressions (\ref{MOPOT}), (\ref{MOFlux}) and (\ref{MODis}). By doing the Fourier transform back to the real time domain, this returns
\begin{equation}
\dot{\vect x}+\big(\vect \Phi_{BW}+\vect \Psi_{ \text{Brow}}+\vect \Psi_{\text{flux}}\big)\cdot\vect x =\vect \Xi_{BW},
\label{QLEMA}
\end{equation}
with the free evolution dynamics described by the $4N\times 4N$ real matrix,
\begin{equation}
   \vect \Phi_{BW}=\left( \begin{array}{cc}
        \vect 0_{2N}& -\frac{\vect I_{2N}}{m} \\
       \vect m\circ\vect \Omega_{\text{ren}}^{\circ 2}  &  \vect 0_{2N}
   \end{array} \right). \label{FPTALE}
\end{equation}
In particular, one may verify that we get the approximated Langevin equation (\ref{QLEMASP}) discussed in Sec. \ref{Sub_COMP2DV} from the expression (\ref{QLEMA}) after substituting the renormalized potential, friction and flux-carrying matrices (\ref{CH2}) in (\ref{SELFEE}) and (\ref{FPTALE}). Regarding the retarded kinetic propagator (\ref{RKPE}) within the Breit-Wigner approximation, this is replaced by $\vect K_{R}(t)\approx \vect K_{BW}(t)$, with
\begin{equation}
    \vect K_{BW}(t) =\Theta(t) e^{-(\vect \Phi_{BW}+\vect \Psi_{\text{Brow}}+\vect \Psi_{\text{flux}})t}. \label{GFBWA1}
\end{equation}
Now, inserting (\ref{GFBWA1}) in Eq.(\ref{QMEGP}) directly gives the pseudo-Hamiltonian matrix in the weak coupling regime, that is
\begin{equation}
    \vect \gamma_{BW}=\left(\begin{array}{cc}
       \vect  0_{2N} & \vect 0_{2N} \\
       -\vect m\circ\vect \Omega_{\text{flux}}^{\circ 2}  & \vect \Gamma_{\text{Brow}}
    \end{array}\right). \label{QMEGPBW}
\end{equation}
To derive the diffusion matrix (\ref{QMEDP}) in the weak system-environment coupling regime, it is better to split up the kinetic operator into the conventional Brownian part and the flux contribution as suggested by  the effective action in (\ref{ZIFBP}). Let us emphasize that this can be done by virtue of the subsidiary condition (\ref{SDC1}), which guarantees that $\Omega_{\text{flux}}$ plays the role of a perturbative parameter. More precisely, we may employ the Baker-Hausdorff formula to expand (\ref{GFBWA1}) in Taylor series such that the retarded kinetic propagator can be cast in the following convenient form,
\begin{equation}
    \vect K_{BW}(t)=\vect K_{\text{Brow}}(t)-\vect K_{\text{Brow}}(t)\cdot \vect L_{\text{flux}}(t)+\mathcal{O}\big(\Omega_{\text{flux}}^4\big),\label{GFBWA2}
\end{equation}
where we identify the first term with the usual Brownian kinetic operator \cite{fox19781}, i.e. $\vect K_{\text{Brow}}(t)=\Theta(t) e^{-(\vect \Phi_{BW}+\vect \Psi_{\text{Brow}})t}$, while the second term contains the flux contribution, i.e.
\begin{align}
\vect L_{\text{flux}}(t)=& \sum_{n=0}^{\infty}\frac{t^{n+1}}{(n+1)!}\big[(\vect \Phi_{BW}+\vect \Psi_{\text{Brow}})^{n},\vect \Psi_{\text{flux}}\big],
\label{LFE}
\end{align}
where we have introduced the notation
\begin{equation}
   \big[(\vect \Phi_{BW}+\vect \Psi_{\text{Brow}})^{n},\vect \Psi_{\text{flux}}\big]=\big[\underbrace{\vect \Phi_{BW}+\vect \Psi_{\text{Brow}},\cdots\big[\vect \Phi_{BW}+\vect \Psi_{\text{Brow}},\big[\vect \Phi_{BW}+\vect \Psi_{\text{Brow}}}_{n\ \text{times}},\vect \Psi_{\text{flux}}\big]\big]\cdots\big].
   \nonumber
\end{equation}
Using the block decomposition of $\vect \Phi_{BW}$, $\vect \Psi_{\text{Brow}}$ and $\vect \Psi_{\text{flux}}$ (recall they are composed of $2\times 2$ submatrices which are either diagonal or anti-diagonal), one may appreciate that $\vect L_{\text{flux}}(t)$ is a block matrix as well. More precisely, it can be partitioned in terms of $2\times 2$ submatrices which are skew-symmetric and anti-diagonal similarly to $\vect \Omega_{\text{flux}}^{\circ 2}$. The latter is the hallmark of the time-reversal and parity symmetries breaking. This is immediate to see from the first term in the right-hand side of (\ref{LFE}), whereas we must appeal to the fact that the product between an anti-diagonal matrix and a diagonal matrix is an anti-diagonal matrix. On the contrary, the conventional Brownian influence of the retarded kinetic propagator $\vect K_{\text{Brow}}(t)$ is a block matrix composed of $2\times 2$ symmetric, diagonal submatrices. By substituting Eq. (\ref{GFBWA2}) and the approximated fluctuation-dissipation relation in the expression (\ref{SCVM}) for the thermal covariance matrix, one finds that the latter can be rewritten as follows
\begin{equation}
\vect \sigma(t)=\vect \sigma_{\text{Brow}}(t)+\vect \sigma_{\text{flux}}(t)+\mathcal{O}(\Omega_{\text{flux}}^4),   
\label{SigmaMDD}
\end{equation}
where we identify the standard Brownian part,
\begin{align}
\vect \sigma_{\text{Brow}}(t)=\frac{2m\hbar}{\pi}\int_{0}^{\infty}d\omega\int_{t_0}^{t}ds\int_{t_0}^{t}ds'\ & \omega\cos(\omega (s-s'))\coth\left(\frac{\hbar \omega\beta}{2}\right) \label{SigmaMD1} \\
&\times \big(\vect K_{\text{Brow}}(t-s) \cdot \vect \Upsilon_{\text{Brow}}\cdot \vect K_{\text{Brow}}^{T}(t-s')\big), \nonumber
\end{align}
and the flux-carrying contribution, 
\begin{align}
    \vect \sigma_{\text{flux}}(t)=\frac{2m\hbar}{\pi}\int_{0}^{\infty}&d\omega \int_{t_0}^{t}ds\int_{t_0}^{t}ds'\ \sin(\omega (s-s'))\coth\left(\frac{\hbar \omega\beta}{2}\right)  \label{SigmaMD2}\\
   & \times\bigg(\vect K_{\text{Brow}}(t-s) \cdot \vect \Upsilon_{\text{flux}}\cdot  \vect K_{\text{Brow}}^{T}(t-s')\nonumber \\
&-\Big[\vect K_{\text{Brow}}(t-s)\cdot \vect \Upsilon_{\text{Brow}} \cdot \vect  L_{\text{flux}}^{T}(t-s') \cdot  \vect K_{\text{Brow}}^{T}(t-s') \nonumber \\
&+\vect K_{\text{Brow}}(t-s)\cdot \vect L_{\text{flux}}(t-s) \cdot \vect \Upsilon_{\text{Brow}}\cdot \vect K_{\text{Brow}}^{T}(t-s') \Big]\bigg). \nonumber
\end{align} 
To obtain the above expressions we have employed the alternative fluctuation-dissipation relation (\ref{AppFDRM1}) deduced from the Breit-Wigner approximation. Finally we insert the result (\ref{SigmaMDD}) in Eq.(\ref{QMEDP}) to obtain the desired expression for the diffusion matrix, i.e.
 \begin{eqnarray}
 \vect D_{BW}(t)&=& \vect D_{\text{Brow}}(t)+\vect D_{\text{flux}}(t)  +\mathcal{O}\big(\Omega_{\text{flux}}^4\big),
 \label{DCM}
 \end{eqnarray}
 where
 \begin{align}
 \vect D_{\text{Brow}}(t)&= \frac{1}{2}\big(\left\lbrace \vect \Phi_{BW}+\vect \Psi_{\text{Brow}} , \vect \sigma_{\text{Brow}}(t)\right\rbrace+\dot{\vect \sigma}_{\text{Brow}}(t)\big),  \label{DCM1} \\
 \vect D_{\text{flux}}(t)&= \frac{1}{2}\big(\left\lbrace\vect \Phi_{BW}+ \vect \Psi_{\text{Brow}} , \vect \sigma_{\text{flux}}(t)\right\rbrace+ \left\lbrace \vect \Psi_{\text{flux}} , \vect \sigma_{\text{Brow}}(t)\right\rbrace+\dot{\vect \sigma}_{\text{flux}}(t)\big).
 \label{DCM2}
 \end{align}
Clearly, Eqs. (\ref{SigmaMDD}) and (\ref{DCM}) manifest that the flux-carrying effects represent a second-order correction to the thermal covariance and diffusion matrix from the conventional Brownian contribution. At this point, it is also worthy to realize that $\vect \sigma_{\text{flux}}(t)$ and $\vect D_{\text{flux}}(t)$ can be partitioned into $2\times 2$ block matrices which are anti-diagonal.  This result can be seen by following an identical algebraic argument as to deduce the specific form of $\vect L_{\text{flux}}(t)$. In stark contrast, $\vect \sigma_{\text{Brow}}(t)$ and $\vect D_{\text{Brow}}(t)$ consist of $2\times 2$ block diagonal submatrices. These algebraic properties of the thermal and diffusion matrices will be exploited in Sec. \ref{subsecHD} to obtain the explicit expressions of the hydrodynamic equations.
 
Now, by replacing the expressions (\ref{QMEGPBW}) and (\ref{DCM}) in Eq.(\ref{QMFPE}), after some manipulation we end up with the general expression of the quantum kinetic equation of the $N$ flux-carrying Brownian particles for weak system-environment coupling, 
\begin{equation}
\bigg(\frac{\partial }{\partial t}+\Big(\frac{\partial}{\partial \vect q}\Big)^{T}\cdot  \big(\vect m^{-1}\circ \vect I_{2N}\big)\cdot  \vect p-\Big(\frac{\partial}{\partial \vect p}\Big)^{T}\cdot \big(\vect m\circ\vect \Omega^{\circ 2}_{\text{ren}}\big)\cdot  \vect q\bigg)W(\vect x,t)=C_{FP}\big[W\big], \label{QBE1}
\end{equation}
where $C_{FP}[W]= C_{\text{Brow}}\big[W\big]+C_{\text{flux}}\big[W\big]$ is a Fokker-Planck-type collision operator where we can readily distinguish the Brownian and flux-carrying contributions (up to higher-order terms in $\Omega_{\text{flux}}$), i.e.
 \begin{align}
C_{\text{Brow}}\big[W\big]&=\Big(\frac{\partial}{\partial \vect p}\Big)^{T}\cdot \vect \Gamma_{\text{Brow}}\cdot  \big(\vect p W(\vect x,t)\big)+ \bigg(2\Big(\frac{\partial}{\partial \vect q}\Big)^{T}\cdot  \vect D^{(qp)}_{\text{Brow}}(t)   \nonumber \\
&+\Big(\frac{\partial}{\partial \vect p}\Big)^{T}\cdot  \vect D^{(pp)}_{\text{Brow}}(t)\bigg)\cdot  \frac{\partial}{\partial \vect p}W(\vect x,t),
\label{COBROW}
\end{align}
and,
 \begin{align}
C_{\text{flux}}\big[W\big]&=-\Big(\frac{\partial}{\partial \vect p}\Big)^{T}\cdot  \big(\vect m\circ\vect \Omega^{\circ 2}_{\text{flux}}\big)\cdot  \big(\vect q W(\vect x,t)\big)+\bigg(2\Big(\frac{\partial}{\partial \vect q}\Big)^{T}\cdot  \vect D^{(qp)}_{\text{flux}}(t)\nonumber \\
&+\Big(\frac{\partial}{\partial \vect p}\Big)^{T}\cdot  \vect D^{(pp)}_{\text{flux}}(t)\bigg)\cdot  \frac{\partial}{\partial \vect p}W(\vect x,t).
\label{COFLUX}
\end{align}
Equation (\ref{COBROW}) is the familiar collision operator which models the (two-dimensional) conventional Brownian motion \cite{chavanis20102,fleming20111,agarwal19711,vacchini20091,caldeira19832}, while Eq.(\ref{COFLUX}) represents a new collision operator that encodes all the flux-carrying corrections (up to second order in $\Omega_{\text{flux}}$). As anticipated in Sec. \ref{Sec_FCBM}, the first-term in the right-hand side of (\ref{COFLUX}) is characteristic of an array of environmental rotational forces acting on the system particles, whereas the second term is responsible for a cross-diffusion process between transversal spatial coordinates. In this line, expression (\ref{QBE1}) can be thought of as the quantum Kramers equation (\ref{QBE1}) of a chiral fluid composed of an intricate ensemble of frictional, rotational harmonic oscillators that effectively interacts through the environment via a rotational force and a pseudomagnetic flux noise. Recall that this collision term has no counterpart on the Kramers equation of the either conventional Brownian motion in presence of an external/synthetic magnetic field \cite{lagos20111,czopnik200111,jimenez20061,cobanera20161,ghorashi20131,santos20171,halder20171} or active Brownian particles endowed with an intrinsic torque \cite{han20201,hargus20201,hargus20201,Klymko20171,epstein20191}.

A final remark in this section is about the subsidiary condition (\ref{SDC1}) guarantees that the real part of the self-energy in Eq.(\ref{QLEMA}) is positive definite, or equivalently,
\begin{equation}
    \vect \Phi_{BW}+\vect\Psi_{\text{Brow}}+\vect\Psi_{\text{flux}}>0.
    \label{SDC1PD}
\end{equation}
This amounts to the retarded kinetic propagator will eventually decay and wipe out any dependence of the asymptotic state with the initial conditions \cite{valido20191}. One may realize the latter from paying attention to Eq.(\ref{GFBWA1}) after substituting (\ref{SDC1PD}). This immediately implies that the system is asymptotically stable, or in other words, the open quantum system governed by Eq.(\ref{QBE1}) behaves ergodic \cite{weiss20121}. Combined with the fluctuation dissipation relation (\ref{FDTHBW}), the system composed of flux-carrying Brownian particles are thus enforced to reach certain thermal equilibrium state characterized by the inverse temperature $\beta$ of the MCS environment. Furthermore, the subsidiary condition ensures that (\ref{QBE1}) preserves the complete positivity of the quantum state along the time evolution. As it is well-known from open quantum system theory, we must impose the so-called positivity condition \cite{vacchini20091}. In our case, this condition requires that the diffusion matrix is positive definite, i.e. the system-to-bath diffusion remains perturbative in comparison with the decoherence coefficients. From a dimensional analysis of Eq.(\ref{DCM2}) (once substituted the expressions (\ref{MOFlux}) and (\ref{MODis})), one may realize that such condition is guaranteed as long as the conventional Brownian part preserves it, since both $|\vect D^{(pp)}_{\text{flux}}(t)|$ and $|\vect D^{(qp)}_{\text{flux}}(t)|$ are second-order corrections to the standard Brownian motion. In what follows we focus the attention to the parameter domain where the subsidiary condition (\ref{SDC1}) is always satisfied for the choices of the bare confining potential $\vect U$ and the spectral density (\ref{ESDD}).

\subsection{Hydrodynamic description}\label{subsecHD}
In this section we complete the kinetic theory of the flux-carrying Brownian particles by addressing the time evolution of the number density, stream velocity, and energy flow in the usual low-density scenario in which environment-mediated particle collisions are rare \cite{chavanis20081,chavanis20082}. We shall concentrate on a homogeneous and isotropic system, which is composed of structurally identical particles. This permits us to follow the standard procedure to obtain a quantum Boltzmann-type equation for flux-carrying Brownian particles by appealing to the molecular chaos assumption \cite{chavanis20062,chavanis20061,schieve20091}. We start by defining the reduced probability distributions \cite{mayorga20021},
\begin{equation}
    W^{(n)}(\underbrace{\vect q_{1},\vect p_{1}}_{\vect x_{1}},\cdots,\vect q_{n},\vect p_{n},t)=V^{n}\int_{\mathbb{R}^{4(N-n)}}  W(\vect x, t) \ \underbrace{d^{2}\vect q_{n+1}d^{2} \vect p_{n+1}}_{d^4\vect x_{n+1}}\cdots d^{2}\vect q_{N}d^{2}\vect p_{N},
    \label{RWGF}
\end{equation}
where we have renormalized as $W^{(n)}(\vect x_{1},\cdots,\vect x_{n},t)=V^{n}W(\vect x)$. As we are dealing with structurally simple identical particles, the Wigner function $W(\vect x,t)$ is symmetric upon switching any pair of particles \cite{chavanis20062,schieve20091,rudyak19951,sonnenburg19911}. As a consequence, $W^{(n)}(\vect x_{1},\cdots,\vect x_{n},t)$ is a symmetric function as well, since the functional form of (\ref{RWGF}) is independent of any particular choice of $n$ particles \cite{ferzinger19721}. Starting from the $N$-particle Kramers equation (\ref{QBE1}), one can derive a coupled hierarchy of equations for the set of $n$-particle Wigner functions, which yields the BBGKY  hierarchy. This is done by integrating over the coordinates of $(N-n)$ remaining particles and dropping the integral terms evaluated at infinity (this terms vanishes rapidly at the boundary of integration since $W(\vect x,t)$ is normalized by definition). In particular, one arrives at the single-particle Wigner function
\begin{align}
\bigg(\frac{\partial }{\partial t}&+\frac{1}{m}\vect p_1\cdot\Big(\frac{\partial}{\partial \vect q_1}\Big)^{T}  -\Big(\frac{\partial}{\partial \vect p_1}\Big)^{T}\cdot  \big(\vect m \circ\vect \Omega^{\circ 2}_{\text{ren}}\big)_{11}\cdot  \vect q_1\bigg)W^{(1)}(\vect x_{1},t) \nonumber \\
&-\frac{N-1}{V}\Big(\frac{\partial}{\partial \vect p_1}\Big)^{T}\cdot  \big(\vect m \circ\vect \Omega^{\circ 2}_{\text{ren}}\big)_{12} \cdot  \int_{\mathbb{R}^{4}} d^4\vect x_{2}  \ \vect q_{2} W^{(2)}(\vect x_{1},\vect x_2)=
C_{FP}^{(1)}\big[W^{(1)}\big]\nonumber \\
&+C_{\text{flux}}^{\text{hyd}}\big[W^{(2)}\big] +C_{\text{Brow}}^{\text{hyd}}\big[W^{(2)}\big],\label{SPKE}
\end{align}
where $\big(\vect \Omega^{\circ 2}_{\text{ren}}\big)_{12}$ denotes the $2\times 2$ real symmetric submatrix of $\vect \Omega^{\circ 2}_{\text{ren}}$ that represents the renormalized potential interaction between the first and second particles, and $C_{FP}^{(1)}\big[\bullet\big]$ is the projection of the collision operator (\ref{QBE1}) into the phase space supported by the first particle. The rest of collision operators in the right-hand side of (\ref{SPKE}) describes collective effects or hydrodynamics collision events arising from environmental-mediated interactions among flux-carrying Brownian particles, namely
\begin{align}
C_{\text{Brow}}^{\text{hyd}}\big[W^{(2)}\big]&=\frac{N-1}{V}\int_{\mathbb{R}^{4}}  d^4\vect x_{2} \bigg[\Big(\frac{\partial}{\partial \vect p_1}\Big)^{T} \cdot \big(\vect \Gamma_{\text{Brow}}\big)_{12} \cdot \vect p_2  \nonumber \\
&+ \bigg(2\Big(\frac{\partial}{\partial \vect q_1}\Big)^{T}\cdot  \big(\vect D^{(qp)}_{\text{Brow}}(t)\big)_{12} +\Big(\frac{\partial}{\partial \vect p_1}\Big)^{T}\cdot  \big(\vect D^{(pp)}_{\text{Brow}}(t)\big)_{12}\bigg)\cdot  \frac{\partial}{\partial \vect p_2}\bigg] W^{(2)}(\vect x_{1},\vect x_2) , \label{SPKE1}
\end{align}
and 
\begin{align}
C_{\text{flux}}^{\text{hyd}}\big[W^{(2)}\big]&=-\frac{N-1}{V}\int_{\mathbb{R}^{4}}  d^4\vect x_{2} \bigg[ \Big(\frac{\partial}{\partial \vect p_1}\Big)^{T}\cdot  \big(\vect m\circ\vect \Omega^{\circ 2}_{\text{flux}}\big)_{12}\cdot   \vect q_2 \nonumber \\
&-\bigg(2\Big(\frac{\partial}{\partial \vect q_1}\Big)^{T}\cdot  \big(\vect D^{(qp)}_{\text{flux}}(t)\big)_{12}+\Big(\frac{\partial}{\partial \vect p_1}\Big)^{T}\cdot  \big(\vect D^{(pp)}_{\text{flux}}(t)\big)_{12}\bigg)\cdot   \frac{\partial}{\partial \vect p_2}\bigg] W^{(2)}(\vect x_{1},\vect x_2), \label{SPKE2}
\end{align}
with $\big(\vect D^{(pp)}_{\text{Brow}/\text{flux}}(t)\big)_{12}$ and $\big(\vect D^{(qp)}_{\text{Brow}/\text{flux}}(t)\big)_{12}$ being the $4\times 4$ diffusion submatrices supported by the phase spaces of the first and second particle. Here we are interested in the hydrodynamics for a dilute flux-carrying Brownian gas, so that we can follow the standard approach in kinetic gas theory to truncate the BBGKY hierarchy at the lowest level. This consists of taking the thermodynamic limit in the sense of Bogoliubov \cite{chavanis20111,rudyak19951,mayorga20021}: $N\rightarrow \infty$ for $V\rightarrow \infty$, while $N/V=n_0$ remain constant. At leading order in the low density $n_{0}$, we may implement the mean field approximation which consists of factorizing the two-particle Wigner function as follows \cite{mayorga20021,chavanis20081,chavanis20082},
\begin{equation}
    W^{(2)}(\vect x_1,\vect x_2,t)\approx W^{(1)}(\vect x_1,t)W^{(1)}(\vect x_2,t). \label{FATPD}
\end{equation}
Substituting the mean field ansatz (\ref{FATPD}) in the expressions (\ref{SPKE1}) and (\ref{SPKE2}) yields
\begin{align}
\tilde{C}_{\text{Brow}}^{\text{hyd}}\big[W^{(1)}\big]&\approx n_0 \bigg[\Big(\frac{\partial}{\partial \vect p_1}\Big)^{T} \cdot \big(\vect \Gamma_{\text{Brow}}\big)_{12} \cdot  \vect \rho_{p_2}(\vect x_{1})   \nonumber \\
&+ \bigg(2\Big(\frac{\partial}{\partial \vect q_1}\Big)^{T}\cdot  \big(\vect D^{(qp)}_{\text{Brow}}(t)\big)_{12} +\Big(\frac{\partial}{\partial \vect p_1}\Big)^{T}\cdot  \big(\vect D^{(pp)}_{\text{Brow}}(t)\big)_{12}\bigg)\cdot  \vect \rho_{\partial_{p_2}}(\vect x_{1}) \bigg] W^{(1)}(\vect x_{1}),  \label{FATPD1}
\end{align}
and 
\begin{align}
\tilde{C}_{\text{flux}}^{\text{hyd}}\big[W^{(1)}\big]&\approx-n_{0}\bigg[ \Big(\frac{\partial}{\partial \vect p_1}\Big)^{T}\cdot  \big(\vect m\circ\vect \Omega^{\circ 2}_{\text{flux}}\big)_{12}\cdot    \vect \rho_{\vect q_2}(\vect x_{1}) \nonumber \\
&-\bigg(2\Big(\frac{\partial}{\partial \vect q_1}\Big)^{T}\cdot  \big(\vect D^{(qp)}_{\text{flux}}(t)\big)_{12}+\Big(\frac{\partial}{\partial \vect p_1}\Big)^{T}\cdot  \big(\vect D^{(pp)}_{\text{flux}}(t)\big)_{12}\bigg)\cdot   \vect \rho_{\partial \vect p_2}(\vect x_{1})\bigg] W^{(1)}(\vect x_{1}),  \label{FATPD2}
\end{align}
where all the hydrodynamics contribution is retained by the auxiliary functions
\begin{align}
    \vect \rho_{\vect s}=\int_{\mathbb{R}^{4}}   \vect s\ W^{(1)}(\vect x_2,t) \ d^4\vect x_{2},
\end{align}
where $s$ is shorthand for $\vect q_{2}$, $\vect p_2$, and $\frac{\partial}{\partial \vect p_2}$. We note that the low-density assumption must be taken with the understanding that, though the hydrodynamical effects may be eventually neglected, the interparticle potential interaction still relevant. The latter is encoded by the mean-field harmonic potential, i.e.
\begin{align}
    \Phi_{\text{ren}}(\vect q_1)\approx\frac{1}{2}\  \vect q_1\cdot\big(\vect m\circ\vect \Omega^{\circ 2}_{\text{ren}}\big)_{11}\cdot  \vect q_1 +  \int_{\mathbb{R}^{4}}  \vect q_1\cdot\big(\vect m\circ\vect \Omega^{\circ 2}_{\text{ren}}\big)_{11}\cdot  \vect q_2 \ W^{(1)}(\vect x_2,t) \ d^4\vect x_2. \label{HPIREN}
\end{align}
By replacing (\ref{FATPD1}), (\ref{FATPD2}) and (\ref{HPIREN}) in (\ref{SPKE}), we are left with the following mean-field kinetic equation within the low density and weak system-environment coupling limit,
\begin{align}
\bigg(\frac{\partial }{\partial t}+\frac{1}{m}\vect p_1\cdot\Big(\frac{\partial}{\partial \vect q_1}\Big)^{T}  &-\Big(\frac{\partial \Phi_{\text{ren}}}{\partial \vect q_1}\Big)^{T}\cdot\frac{\partial}{\partial \vect p_1}\bigg)W^{(1)}(\vect x_{1},t) =
C_{FP}^{(1)}\big[W^{(1)}\big]\nonumber \\
&+\tilde{C}_{\text{flux}}^{\text{hyd}}\big[W^{(1)}\big]+\tilde{C}_{\text{Brow}}^{\text{hyd}}\big[W^{(1)}\big]+\mathcal{O}(n_0^2)\label{SPKE3},
\end{align}
where the first line looks similar to the extended Kramers equation (\ref{QBE1}) associated to the single particle. In order to avoid a misleading interpretation, it is important to point out that the Brownian and flux-carrying contributions to the diffusion matrix involved in the collision operators in (\ref{SPKE}) must be obtained from (\ref{DCM1}) and (\ref{DCM2}) (rather than from the expressions (\ref{DBrowM}) and (\ref{DfluxM}) for the single particle scenario). Just when the effective environmental-mediated interaction is significantly small (e.g. the system particles are sufficiently separated in comparison with the time scales of the environmental memory effects), the hydrodynamics collision terms (\ref{SPKE1}) and  (\ref{SPKE2}) are negligible and  $C_{FP}^{(1)}$ is determined from (\ref{DBrowM}) and (\ref{DfluxM}). 

Having settled an approximated evolution equation for the single-particle Wigner function, we shall go through the steps of deriving the hydrodynamic conservation laws characteristic of the flux-carrying Brownian motion. More precisely, we have applied the standard procedures from the hydrodynamics description \cite{mayorga20021,schieve20091,kreuzer19811} in the next section to derive extended balance equations up to terms of order $\mathcal{O}(n_{0}\Omega_{\text{flux}^2})$.

\subsection{Quantum balance equation in the dilute scenario and weak coupling regime}\label{Sub_QBE}
In the $N$-particle scenario, we define the (single-particle) number density as usual \cite{sonnenburg19911,Klymko20171}
\begin{equation}
 n (\vect q_{1},t)= \int_{\mathbb{R}^2} W^{(1)}(\vect q_{1},\vect p_{1},t)\ d^2\vect p_{1},
\label{NDDF}
\end{equation}
which satisfies the continuity equation (\ref{BEN1SP}) introduced in Sec. \ref{Sub_QHSP}. Similarly, the hydrodynamic velocity or (single-particle) flow density for the flux-carrying Brownian motion \cite{lagos20111, mayorga20021,chavanis20101},
\begin{equation}
    \vect J(\vect q_{1},t)=n(\vect q_{1},t) \vect u(\vect q_{1},t)=\frac{1}{m_{1}}\int_{\mathbb{R}^2}  \vect p_{1} W^{(1)}(\vect q_{1},\vect p_{1},t)\ d^2\vect p_{1}, \label{MDFD}
\end{equation}
with $\vect u$ being the stream velocity and $m_{1}=(\vect m)_{11}$. By introducing the material or hydrodynamic derivative \cite{kreuzer19811}, i.e.
\begin{equation}
    \frac{d}{dt}=\frac{\partial}{\partial t}+\vect u \cdot \frac{\partial}{\partial\vect  q_{1}},
    \nonumber
    \end{equation}
the continuity equation (\ref{BEN1SP}) can be alternatively expressed as 
\begin{equation}
    \frac{dn}{dt}+n\frac{\partial}{\partial \vect q_{1}}\cdot\vect u=0.
    \label{BEN2}
\end{equation}
Notice that the number and flow densities (\ref{NDDF}) and (\ref{MDFD}) identically coincides with the previous (\ref{NDDFSP}) and (\ref{MDFDSP}) definitions after replacing $W \rightarrow W^{(1)}$.

We now take the partial derivative of the definition (\ref{MDFD}) by replacing the mean-field Kramers equation (\ref{SPKE3}). By carrying out integration by parts over momentum space, we arrive to the following expression
\begin{align}
\frac{\partial (n\vect u)}{\partial t}&+\frac{1}{m_{1}}\Big(\frac{\partial}{\partial \vect q_{1}}\Big)^{T}\cdot  \Big(\vect P_{K}+m_{1}n\vect u \vect u \Big)= -\frac{1}{m_{1}}\Big(\frac{\partial}{\partial \vect q_{1}}\Big)^{T}\cdot \vect P_{\Phi_{\text{ren}}}-\frac{1}{m_{1}}\big(\vect m\circ\vect \Gamma_{\text{Brow}}\big)_{11}\cdot  (n\vect u)    \nonumber \\ 
&-\frac{n_0}{m_{1}}\big(\vect m\circ\vect \Gamma_{\text{Brow}}\big)_{12}\cdot   \delta\vect{ J}+\frac{1}{m_{1}} \big(\vect m\circ\vect \Omega^{\circ 2}_{\text{flux}}\big)_{11} \cdot  (n \vect q_{1}) +\frac{n_{0}}{m_{1}} \big(\vect m\circ\vect \Omega^{\circ 2}_{\text{flux}}\big)_{12} \cdot  \delta\vect{ n}_{\vect q_2}\nonumber\\
&-\frac{2}{m_{1}}\Big(\big(\vect D^{(qp)}_{\text{Brow}}(t)\big)_{11}^{T}+\big(\vect D^{(qp)}_{\text{flux}}(t)\big)_{11}^{T} \Big)\cdot  \frac{\partial n}{\partial \vect q_1}    \nonumber \\
&-\frac{2n_{0}}{m_{1}}\Big(\big(\vect D_{\text{Brow}}^{(qp)}(t)\big)_{12}^{T}+\big(\vect D^{(pp)}_{\text{Brow}}(t)\big)_{12}^{T} +\big(\vect D_{\text{flux}}^{(qp)}(t)\big)_{12}^{T}+\big(\vect D^{(pp)}_{\text{flux}}(t)\big)_{12}^{T}\Big) \cdot  \delta\vect{ n}_{\partial\vect p_2} \nonumber \\
&+\mathcal{O}(n_{0}^2), \label{BEM}
\end{align}
where we have recognized the familiar kinetic pressure \cite{mayorga20021,sonnenburg19911,lagos20111,Klymko20171},
\begin{equation}
\vect P_{K}(\vect q_{1},t)=\frac{1}{m_{1}}\int_{\mathbb{R}^{2}} (\vect p_{1}-m_{1}\vect u)\circ(\vect p_{1}-m_{1}\vect u)W^{(1)}(\vect q_1,\vect p_1,t)  \ d^2\vect p_{1},
\label{STKC}
\end{equation}
whereas the potential contribution is now given by \cite{mayorga20021},
\begin{align}
\vect P_{\Phi_{\text{ren}}}(\vect q_{1},t)=\vect I_{2}\int_{\mathbb{R}^{2}} \Phi_{\text{ren}}(\vect x_{1})  W^{(1)}(\vect q_1,\vect p_1,t) \ d^2\vect p_1 .\nonumber\nonumber
\end{align}
Furthermore, we identify the hydrodynamic contribution, coming from the hydrodynamic collision operators (\ref{SPKE1}) and (\ref{SPKE2}), which produces a distortion of the hydrodynamic momentum, i.e.
\begin{equation}
\delta \vect J=\int_{\mathbb{R}^{2}} \ \vect \rho_{\vect p_2}(\vect x_1) W^{(1)}(\vect q_1,\vect p_1,t) \ d^2\vect p_{1}, \nonumber 
\end{equation}
as well as,
\begin{eqnarray}
\delta\vect{ n}_{\vect q_2}&=&\int_{\mathbb{R}^{2}} \ \vect \rho_{\vect q_{2}}(\vect x_{1}) \ W^{(1)}(\vect q_1,\vect p_1,t)\ d^2\vect p_1,  \nonumber \\
\delta\vect{ n}_{\partial\vect p_2} &=&\int_{\mathbb{R}^{2}}  \  \vect \rho_{\partial\vect p_2}(\vect x_1) \ W^{(1)}(\vect q_1,\vect p_1,t) \ d^2\vect p_1.      \nonumber 
\end{eqnarray}
Equation (\ref{BEM}) can be further simplified by recalling that the Brownian and flux-carrying contributions constitute, respectively, the symmetric and antisymetric parts of the anomalous diffusive tensor (as remarked in Sec. \ref{Subsec_FPPC}), i.e.
\begin{equation}
\big(\vect D_{\text{Brow}}^{(qp)}(t)\big)_{11}=\left(\begin{array}{cc}
    D_{\text{Brow}}^{(q_xp_x)}(t) & 0 \\
   0  & D_{\text{Brow}}^{(q_yp_y)}(t)
\end{array} \right),
\ \ \ \
\big(\vect D_{\text{flux}}^{(qp)}(t)\big)_{11}=\left(\begin{array}{cc}
   0  &  D_{\text{flux}}(t)\\
   -D_{\text{flux}}(t)  & 0
\end{array} \right) . 
\label{MDCHD}
\end{equation}
We would like to remark that the diffusion coefficients (\ref{MDCHD}) reduce to (\ref{DBrowM}) and (\ref{DfluxM}) in the single-particle scenario equipped with an isotropic harmonic potential. By inserting (\ref{MDCHD}) and combining the hydrodynamic derivative once replaced the continuity equation (\ref{BEN1SP}), we obtain the final expression for the stream velocity balance equation at leading order in the low density limit (i.e. omitting the hydrodynamic collision operators), 
\begin{equation}
n\frac{d \vect u}{dt}+\frac{1}{m}\Big(\frac{\partial}{\partial \vect q_{1}}\Big)^{T} \cdot\vect T=- \gamma_{\text{Brow}} (n\vect u)+\Omega_{\text{flux}}^2 \ \vect z\times(n \vect q_{1}) +\mathcal{O}(n_{0}),
\label{ENVST}
\end{equation}
where the stress tensor $\vect T$ coincides with (\ref{ENVST1}) after replacing the difussive coefficients according to (\ref{MDCHD}). For the single particle scenario, we can rewrite $\vect q_{1}\rightarrow \vect q$ and $(\partial/\partial\vect  q_{1})^{T}\rightarrow\nabla$, as well as $D^{(q_yp_y)}_{\text{Brow}}(t)=D^{(q_xp_x)}_{\text{Brow}}(t)=D^{(qp)}_{\text{Brow}}(t)$ and $D_{\text{flux}}(t)$ are given by Eqs. (\ref{DBrowM}) and (\ref{EMDF1}). Combined with the fact that the hydrodynamics collision effects vanishes, Eq. (\ref{ENVST}) directly retrieves (\ref{ENVSTSP}).

We can go further and obtain the balance equation for the kinetic energy density \cite{mayorga20021,chavanis20061,lagos20111} by starting from the definition (\ref{ENVST1}) once we have replaced $W\rightarrow W^{(1)}$. We substitute the time derivative of the kinetic energy density by expression (\ref{SPKE3}) and integrate over the momentum variables, as well as we appeal to the remark related to the diagonal and anti-diagonal properties of the decoherence diffusion matrices $\big(\vect D_{\text{Brow}}^{(pp)}(t)\big)_{11}$ and $\big(\vect D_{\text{flux}}^{(pp)}(t)\big)_{11}$. After some transformation and integration by parts, we find at leading order in the low density limit
\begin{align}
\frac{\partial (ne_{K})}{\partial t}&=-\vect P_{K}\circ\Big(\Big(\frac{\partial}{\partial \vect q_{1}}\Big)^{T} \cdot \vect u\Big) -\Big(\frac{\partial}{\partial \vect q_{1}}\Big)^{T}\cdot\big(\vect J_{K}-(ne_{K})\vect u \big) \label{BECE} \\
&-2\gamma_{\text{Brow}}(ne_{K}) + \frac{n}{m}\Big( D_{\text{Brow}}^{(p_{x}p_{x})}(t)+ D_{\text{Brow}}^{(p_{y}p_{y})}(t)\Big) +\mathcal{O}(n_{0}),\nonumber
\end{align}
where $\vect J_{K}(\vect q,t)$ is the energy flow vector density or heat current provided by (\ref{HCD1}). As expected the above expression coincides with (\ref{BECESP}) upon disregarding the hydrodynamic collision events. 

We finally examine the hydrodynamics conservation laws for the fluid vorticity and the circulation flux \cite{thorne20171,jackiw20041}, previously introduced in Sec. \ref{Sub_QHSP} (see Eqs. (\ref{DFVORTSP}) and (\ref{EDC1SP})). By taking the time derivative of the vortitcity into the mean-field Kramers equation (\ref{SPKE3}), after some manipulation we get  
\begin{align}
     \frac{\partial\varpi}{\partial t}&=-\Big(\vect u\cdot\Big(\frac{\partial}{\partial \vect q_{1}}\Big)^{T}\Big)\varpi-\varpi \Big(\Big(\frac{\partial}{\partial \vect q_{1}}\Big)^{T}\cdot \vect u\Big) -\frac{1}{m n^2}\Big[\Big(\frac{\partial}{\partial \vect q_{1}}\Big)^{T}\cdot (\vect P_{K}+\vect P_{\Phi_{\text{ren}}})\times\Big(\frac{\partial}{\partial \vect q_{1}}\Big)^{T} n\Big]_{z}-\gamma_{\text{Brow}}\varpi +2\Omega_{\text{flux}}^2  \nonumber \\
      &+\frac{1}{m n}\big(D^{(q_xp_x)}_{\text{Brow}}(t) -D^{(q_yp_y)}_{\text{Brow}}(t)  \big) \bigg(\frac{\partial^2 n}{\partial q_{1,x}\partial q_{1,y}}-\frac{1}{n}\frac{\partial n}{\partial q_{1,x}}\frac{\partial n}{\partial q_{1,y}}\bigg) \nonumber \\
      &+\frac{2D_{\text{flux}}(t)}{m n}\bigg(\frac{\partial^2n}{\partial \vect q_{1}^{2}} -\frac{1}{n}\Big(\Big(\frac{\partial}{\partial \vect q_{1}}\Big)^{T}n\Big)^2\bigg)+\mathcal{O}(n_{0}).
    \label{BEV1}
\end{align}
where we have employed the identities (\ref{MVDI}). Next, introducing the hydrodinamic deviate by means of the continuity equation (\ref{BEN2}), yields the vorticity balance equation,
\begin{align}
    \frac{d\varpi}{d t}&= 2\Omega_{\text{flux}}^2-\gamma_{\text{Brow}}\varpi+\frac{\varpi}{n}\frac{dn}{d t}-\frac{1}{m n^2}\big[\nabla\cdot (\vect P_{K}+\vect P_{\Phi_{\text{ren}}})\times\nabla n\big]_{z}  \nonumber \\
    &+\frac{1}{m n}\big(D^{(q_xp_x)}_{\text{Brow}}(t) -D^{(q_yp_y)}_{\text{Brow}}(t)  \big) \bigg(\frac{\partial^2 n}{\partial q_{1,x}\partial q_{1,y}}-\frac{1}{n}\frac{\partial n}{\partial q_{1,x}}\frac{\partial n}{\partial q_{1,y}}\bigg) \nonumber \\
    &+\frac{2D_{\text{flux}}(t)}{m n}\bigg(\frac{\partial^2n}{\partial \vect q_{1}^{2}} -\frac{1}{n}\Big(\Big(\frac{\partial}{\partial \vect q_{1}}\Big)^{T}n\Big)^2\bigg) +\mathcal{O}(n_{0}).
    \label{BEV2}
\end{align}
Beside the hydrodynamic collision effects, one may notice that the second line in the right hand side of (\ref{BEV2}) is absent in its single-particle counterpart (\ref{BEV2SP}). This is because we considered an isotropic damped harmonic particle (see Eq.(\ref{CH2})), and thus, the Brownian diffusive coefficient is identical in both degrees of freedom. Similarly, we obtain the balance equation for the vorticity flux defined by Eq. (\ref{EDC1SP}). After some manipulation, we arrive at the following expression
\begin{align}
    \frac{d\psi}{d t}&= 2\pi\Omega_{\text{flux}}^2-\gamma_{\text{Brow}}\psi -\oint_{\mathcal{C}}\frac{dP}{n} -\frac{2}{m}\oint_{\mathcal{C}}\frac{1}{n}\bigg(D^{(q_{x}p_{x})}_{\text{Brow}}(t)\frac{\partial n}{\partial q_{1,x}}dq_{1,x}+D^{(q_{y}p_{y})}_{\text{Brow}}(t)\frac{\partial n}{\partial q_{1,y}}dq_{1,y}\bigg)\nonumber \\
    &-\frac{2D_{\text{flux}}(t)}{m}\oint_{\mathcal{C}}\frac{1}{n} \Big(\frac{\partial}{\partial \vect q_{1}}\Big)^{T} n\times d\vect q_{1} +\mathcal{O}(n_{0}).
    \label{EDC2}
\end{align}
One may readily see that Equation (\ref{EDC2DR}) is directly obtained from (\ref{EDC2}) after replacing $D^{(q_yp_y)}_{\text{Brow}}(t)=D^{(q_xp_x)}_{\text{Brow}}(t)=D^{(qp)}_{\text{Brow}}(t)$ and ignoring the hydrodynamics collision effects.

\section{Summary and concluding remarks}\label{OCR}

Motivated by recent progresses in the study of chiral fluids, we developed the quantum kinetic theory of flux-carrying Brownian particles by following a generic perturbative approach for weak system-environment coupling. This permitted us to obtain the quantum kinetic equation and hydrodynamic equations up to second-order in the flux-carrying effects. Compared with the well-known Kramers equation from the conventional Brownian motion, the kinetic equation here displays an additional collision term encoding both a flux-like noise responsible for non-vanishing antisymmetric components in the diffusive matrix (as well as stress tensor), and a rotational drift which arises out of an environmental torque acting upon the flux-carrying Brownian particles. We argue that this equation encompasses a wide class of previously treated chiral-fluid effects: for instance, the rotational drift resemblances the active torque in chiral active fluids \cite{han20201,hargus20201,hargus20201,Klymko20171,epstein20191}, whereas the antrysimetric diffusive coefficients are responsible for a nondiffusive flow which is common to conventional Brownian particles subject to a Lorentz force \cite{abdoli20202,abdoli20201}. However, we further showed that the flux-carrying effects give rise to an unconventional fluid dynamics at equilibrium conditions. Concretely, we illustrate in the single-particle scenario that a dissipationless vortex flow is eventually reached in the thermal equilibrium without the need of an external magnetic field or an intrinsic angular momentum. This is in contrast to most instances of parity violating or time-reversal breaking fluids \cite{lucas20141,kaminski20141,vuijk20191,banerjee20171}. Essentially, while the (nonequilibrium) vortex dynamics of the vast majority of known chiral fluids is due to either an external/synthetic magnetic field or an intrinsic torque (which represents a continuous injection of energy), the equilibrium vortex dynamics of the flux-carrying Brownian particles relies on the fact that there exists a balance between both the environmental torque and the aforementioned flux-like noise which is established by an extended fluctuation-dissipation relation. In addition, the relevant hydrodynamic quantities, including the fluid vorticity and vorticity flux, and the diffusive matrix are explicitly computed in the simple particle scenario near thermal equilibrium. Conversely, we find out that the flux-carrying effects do not influence either the kinetic energy or particle density, so that the usual Boyle's law is retained up to a reformulation of the kinetic temperature.

In spite of recent progresses in the study of chiral fluids, the implications of chirality on the kinetics as well as the hydrodynamics remains largely unexplored under dominant dissipative and noise effects \cite{burmistrov20191,kaminski20141}, which are ubiquitous in real life experiments and applications. In this context, the flux-carrying Brownian motion constitutes an ideal testing ground to get useful insight of the hydrodynamic transport properties and electromagnetic response of open quantum systems whose microscopic components violates both time-reversal and parity invariance. More directly, there are several appealing directions in which the present work could be extended. For instance, one may expect that the flux-carrying Brownian fluid could support odd viscosity in the quantum domain: this would show that the latter is not exclusive to electronic states with non-trivial adiabatic Berry curvatures. It is therefore of particular interest to compute the viscosity tensor when short-range interparticle interactions become important.

\acknowledgments
The author warmly thanks D. Alonso for enlightening discussions. The author acknowledges support from Spanish project PGC2018-094792-B-100 (MCIU/AEI/FEDER,EU).

\appendix


\section{Nonequilibrium generating functional}\label{app1}

In this appendix we briefly illustrate the derivation of the nonequilibrium generating functional (\ref{NEGF1}) starting from the microscopic Lagrangian (\ref{EAE}) of the $N$ flux-carrying Brownian particles. The latter is expressed in terms of the auxiliary coupling coefficients,
\begin{align}
 g_{\alpha}(\vect k, \bar{\vect q}_{i})&=-\frac{e \sqrt{m_{\vect k}\hbar  } }{2\pi L|\vect k|\omega_{\vect k}}f(\vect k)\bigg(\sum_{\lambda=1,2}\omega_{\vect k}\sin(\vect k \cdot \bar{\vect q}_{i})\epsilon_{\alpha\lambda} k_{\lambda} +\kappa k_{\alpha}\cos(\vect k \cdot \bar{\vect q}_{i})\bigg), \label{gCC} \\
 l_{\alpha}(\vect k, \bar{\vect q}_{i})&=-\frac{e \sqrt{m_{\vect k}\hbar  } }{2\pi L|\vect k|\omega_{\vect k}}f(\vect k)\bigg(\sum_{\lambda=1,2}\omega_{\vect k}\cos(\vect k \cdot \bar{\vect q}_{i})\epsilon_{\alpha\lambda} k_{\lambda} -\kappa k_{\alpha}\sin(\vect k \cdot \bar{\vect q}_{i})\bigg),\label{lCC} 
\end{align}
and the renormalized potential,
\begin{equation}
\phi_{\alpha\lambda}(\Delta\bar{\vect q}_{ij})= \frac{e^2\kappa^2 \hbar}{m(2\pi L)^2}\sum_{\vect k\in \mathbb{R}^2}\frac{f^2(|\vect k|)k_{\alpha} k_{\lambda}}{|\vect k|^2\omega^2_{\vect k}}\cos(\vect k \cdot \Delta\bar{\vect q}_{ij}),  \label{QRP1}
\end{equation}
recall that $0<e$ determines the coupling strength to the MCS environment, $L$ is a characteristic length of the MCS environment, and $f(|\vect k|)\in\mathbb{R}$ represents the usual spherically symmetric smooth form factor from non-relativistic quantum electrodynamics \cite{buenzli20071}. Provided that the coupled system-environment complex, composed of the 2D harmonic particles and the environmental MCS field, is in a canonical equilibrium state at inverse temperature $\beta=1/k_{B}T$, the partition function governing the statistical mechanics is given by Eq.(\ref{ZIFBP}) \cite{valido20201} (after integrating out the environmental degrees of freedom), where $S_{\text{Brow}}^{(E)}$ is the familiar effective action describing the conventional Brownian motion. This can be expressed as usual \cite{weiss20121,grabert19881},
\begin{equation}
S_{\text{Brow}}^{(E)}[\{\vect q\}]= S_{\text{free}}^{(E)}[\{\vect q\}]+\sum_{i,j=1}^{N}\sum_{\alpha,\lambda=1,2}\int_0^{\hbar\beta}d\tau \ \int_{0}^\tau d\tau' \ \text{Re}\ \check{\Gamma}_{\alpha\lambda}(\tau-\tau',\Delta\bar{\vect q}_{ij})q_i^{\alpha}(\tau)q_j^{\lambda}(\tau'), \label{EASBR}
\end{equation}
where $\text{Re}\ \bullet$ stands for the real part of $\bullet$ ($\text{Im}\ \bullet$ denotes the imaginary part) and
\begin{equation}
    S_{\text{free}}^{(E)}[\{\vect q\}]=\sum_{i,j=1}^{N}\sum_{\alpha,\lambda=1,2}\int_0^{\hbar\beta}d\tau \ \int_{0}^\tau d\tau' \  \bigg(\delta_{ij}\delta_{\alpha\lambda}\frac{\partial^2}{\partial\tau^2}+U_{\alpha\lambda}^{ij}\bigg)\delta(\tau-\tau')q_i^{\alpha}(\tau)q_j^{\lambda}(\tau'), \label{EASFREE}
\end{equation}
with $U_{\alpha\lambda}^{ij}$ takes account any harmonic interparticle interaction as well as the confining harmonic potential. Here, $\check{\Gamma}_{\alpha\lambda}$ coincides with the imaginary-time Fourier transform of the well-known dissipation kernel or dynamical susceptibility characteristic of the standard Brownian motion \cite{weiss20121,grabert19881}, i.e.
\begin{equation}
    \check{\Gamma}_{\alpha\lambda}(\tau,\Delta\bar{\vect q}_{ij})=-\frac{\delta_{\alpha\lambda}}{\pi}\int_{0}^{\infty}d\omega \ \frac{h_{\lambda\lambda}(\omega,\Delta\bar{\vect q}_{ij})}{\omega^2} \frac{\partial^2B_\omega(\tau)}{\partial\tau^2},\label{SDCI}
\end{equation}
where $B(\tau)$ is the imaginary-time boson propagator at thermal equilibrium \cite{weiss20121},
\begin{equation}
   B_{\omega}(\tau)=(1+n(\omega,\beta^{-1}))e^{-\omega\tau}+n(\omega,\beta^{-1})e^{\omega\tau},
    \label{ITBP}
\end{equation}
whereas $n(\omega,\beta^{-1})$ is the single-particle Bose distribution and $\vect h(\omega,\Delta\bar{\vect q}_{ij})$ is the extended spectral density given by (\ref{ESDD}).

The Euclidean action emerging from the flux attachment is given by Eq.(\ref{EASF}) \cite{valido20201}, where we have introduced the imaginary-time Fourier transforms of the longitudinal dynamical susceptibility
\begin{equation}
\check{\Lambda}^{||}_{\alpha\lambda}(\tau,\Delta\bar{\vect q}_{ij})=\delta_{\alpha\lambda}\frac{\kappa^2}{\pi}\int_{0}^{\infty}d\omega \ \frac{h_{\lambda\lambda}(\omega,\Delta\bar{\vect q}_{ij})}{\omega^2} B_\omega(\tau), \label{SDCII} 
\end{equation}
as well as the transverse dynamical susceptibility
\begin{equation}
\check{\Lambda}^{\perp}_{\alpha\lambda}(\tau,\Delta\bar{\vect q}_{ij})=-\epsilon_{\alpha\lambda}\frac{i\kappa}{\pi}\int_{0}^{\infty}d\omega \ \frac{h_{\lambda\lambda}(\omega,\Delta\bar{\vect q}_{ij})}{\omega^2} \frac{\partial B_\omega(\tau)}{\partial\tau}, \label{SDCIII}
\end{equation}
with $i$ being the imaginary unit.

Now, we assume that the initial state of the coupled system-environment complex is the usual product state \cite{weiss20121,caldeira19832,caldeira20141,caldeira19831}, that is $\hat \rho_{0}\otimes \hat \rho_{\beta}$ with $\hat\rho_{\beta}$ being a canonical thermal equilibrium state of the MCS environment with inverse temperature $\beta$. It can be shown that the nonequilibrium generating function in the path integral formalism can be cast as follows \cite{calzetta19881,weiss20121}
\begin{align}
\mathcal{Z}_{\text{Brow-flux}}[\vect J^{+},\vect J^{-}]&=\int d^{2N}\vect q_{f}\int d^{2N}\vect q_{i}^{+}d^{2N}\vect q_{i}^{-}\rho(\vect q_{i}^{+},\vect q_{i}^{-},t_{0})\int_{\vect q^{+}(t_{0})=\vect q^{+}_{i}}^{\vect q^{+}(t)=\vect q^{+}_{f}} \mathcal{D} \underline{\vect q}^{+}\int_{\vect q^{-}(t_{0})=\vect q^{-}_{i}}^{\vect q^{-}(t)=\vect q^{-}_{f}} \mathcal{D}\underline{\vect q}^{-} \nonumber \\
\times&\Bigg[\ \text{exp}\Big(\frac{i}{\hbar}\int_{t_0}^{t} ds\Big(\vect J^{+}(s)\cdot\vect q^{+}(s)-\vect J^{-}(s)\cdot\vect q^{-}(s)\Big) \Big)\nonumber \\
&\times\text{exp}\Big(\frac{i}{\hbar}\Big(\mathcal{S}_{\text{Sys}}[\{\vect q^{+}\}]-\mathcal{S}_{\text{Sys}}[\{\vect q^{-}\}]\Big)\Big)\mathcal{F}_{IF}[\{\vect q^{+}\},\{\vect q^{-}\}]\Bigg],
\label{App_NGF}
\end{align}
where $\mathcal{F}_{IF}$ is the so-called Feynman-Vernon influence functional of the MCS environment, and $\mathcal{S}_{\text{Sys}}$ is the real-time action given by the $N$-particle Lagrangian (\ref{ALSys}). As the microscopic model (\ref{EAE}) is quadratic in the coordinates of the system and environment particles, the latter can be written as \cite{roura19991}
\begin{align}
\mathcal{F}_{IF}[\{\vect q^{+}\},\{\vect q^{-}\}]&=\text{exp}\bigg(\frac{i}{\hbar}\bigg(\frac{1}{2}\int_{t_0}^{t}d s\int_{t_0}^{t} ds' \ \vect X^{-}(s) \cdot\vect H(s,s')\cdot\vect X^{+}(s') \nonumber \\
&+\frac{i}{2}\int_{t_0}^{t}ds\int_{t_0}^{t} ds'\ \vect X^{-}(s)\cdot \vect N(s,s')\cdot\vect X^{-}(s')\bigg)\bigg),   
\end{align}
with $\vect X^{\pm}=\vect q^{+}\pm \vect q^{-}$, whereas $\vect H$ and $\vect N$ are the so-called dissipative and noise kernels, respectively. In our case, the former coincides with the real-time retarded self-energy \cite{roura19991,calzetta20031} (or equivalently, with the generalized environmental susceptibility \cite{valido20191}), i.e. $ \vect H(t,t')=\vect \Sigma(t-t')$, which is given by
\begin{equation}
\Sigma_{\alpha\lambda}(t,\Delta\bar{\vect q}_{ij}) =\Gamma_{\alpha\lambda}(t,\Delta\bar{\vect q}_{ij})+\Lambda^{||}_{\alpha\lambda}(t,\Delta\bar{\vect q}_{ij})+\Lambda^{||}_{\alpha\lambda}(t,\Delta\bar{\vect q}_{ij}),
   \label{DKNGF}
\end{equation}
for $i,j=1,\cdots, N$ and $\alpha,\lambda=1,2$. Expression (\ref{DKNGF}) leads to Eq. (\ref{FDRM2}) after replacing the dynamical susceptibilities. To show the latter we begin with the frequency-dependent Fourier transforms of the dynamical susceptibilities, which can be computed from the back imaginary-time Fourier transforms (\ref{SDCI}), (\ref{SDCII}) and (\ref{SDCIII}) via the analytic continuation \cite{weiss20121}, e.g.
\begin{equation}
  \tilde{\Gamma}_{\alpha\lambda}(\omega,\Delta\bar{\vect q}_{ij})=\lim_{\epsilon\rightarrow 0^{+}}\bar{\Gamma}_{\alpha\lambda}(-i\omega+\epsilon,\Delta\bar{\vect q}_{ij}),
  \nonumber
\end{equation}
with 
\begin{equation}
  \bar{\Gamma}_{\alpha\lambda}(s,\Delta\bar{\vect q}_{ij})=  \frac{1}{m}\int_{0}^{\hbar\beta}d\tau \ \check{\Gamma}_{\alpha\lambda}(\tau,\Delta\bar{\vect q}_{ij})\ e^{-is\tau}.
  \nonumber
\end{equation}
Specifically, the back imaginary-time Fourier transforms express as follows \cite{valido20201}
\begin{align}
\bar{\Gamma}_{\alpha\lambda}(s,\Delta\bar{\vect q}_{ij})&=\delta_{\alpha\lambda}\frac{2s^2}{m\pi}\int_{0}^{\infty}d\omega'\ \frac{h_{\lambda\lambda}(\omega',\Delta\bar{\vect q}_{ij})}{\omega'}\frac{1}{s^2+\omega'^2},\label{SECI} \\
\bar{\Lambda}^{||}_{\alpha\lambda}(s,\Delta\bar{\vect q}_{ij})&=\delta_{\alpha\lambda}\frac{2\kappa^2}{m\pi}\int_{0}^{\infty}d\omega'\ \frac{h_{\lambda\lambda}(\omega',\Delta\bar{\vect q}_{ij})}{\omega'}\frac{1}{s^2+\omega'^2}\label{SECII}, \\
\bar{\Lambda}^{\perp}_{\alpha\lambda}(s,\Delta\bar{\vect q}_{ij})&=\epsilon_{\alpha\lambda}\frac{2\kappa s}{m\pi}\int_{0}^{\infty}d\omega'\ \frac{h_{\alpha\lambda}(\omega',\Delta\bar{\vect q}_{ij})}{\omega'}\frac{1}{s^2+\omega'^2},\label{SECIII}
\end{align}
where we have made used of the formal definition of the imaginary-time photon propagator \cite{weiss20121}. Starting from Eqs. (\ref{SECI})-(\ref{SECIII}), the analytic continuation can be readily performed by means of the useful identity,
\begin{align}
&\lim_{\epsilon \rightarrow
0^{+}}\frac{2}{\pi}\int_{0}^{\infty}d\omega'\frac{h_{\alpha\lambda}(\omega',\Delta\bar{\vect q}_{ij})}{\omega'}\frac{1}{\omega'^2-\omega^2-i\epsilon\ \text{sgn}(\omega)}= \nonumber \\
&=\frac{i}{\omega^2}\bigg(\Theta(\omega)h_{\alpha\lambda}(\omega,\Delta\bar{\vect q}_{ij})-\Theta(-\omega)h_{\alpha\lambda}(-\omega,\Delta\bar{\vect q}_{ij})\bigg)\nonumber \\
&+\mathcal{H}\bigg(\Theta(\omega')\frac{h_{\alpha\lambda}(\omega',\Delta\bar{\vect q}_{ij})}{\omega'^2}-\Theta(-\omega')\frac{h_{\alpha\lambda}(-\omega',\Delta\bar{\vect q}_{ij})}{(-\omega')^2}\bigg)(\omega),
\label{ASID}
\end{align}
where $\mathcal{H}(\bullet)(\omega)$ denotes the Hilbert transform, i.e.
\begin{equation}
    \mathcal{H}\big(f(\omega')\big)(\omega)=\frac{1}{\pi}\text{P}\int_{-\infty}^{\infty}\frac{f(\omega')}{\omega'-\omega}d\omega',
    \label{DFHT}
\end{equation}
with $\text{P}$ being the principal part. To obtain this identity one employed the Sokhotski-Plemelj formula, i.e. $1/(\omega+i\epsilon)=-i\pi\delta(\omega)+\text{P}(1/\omega)$ for $\omega\in \mathbb{R}$ \cite{schieve20091,fleming20111}. Notice that Eq.(\ref{ASID}) is consistent with the Kramers-Kronig relations, which agrees with the fact that the retarded self-energy is an analytic function in the upper-half complex plane. Upon replacing the above identity in (\ref{SECI})-(\ref{SECIII}), we arrive at the frequency-dependent Fourier transform of the dynamical susceptibilities (\ref{FTSECI}), (\ref{FTSECII}) and (\ref{FTSECIII}). Once we have carried out the Fourier transform back to the real time domain of  Eqs. from (\ref{FTSECI}) to (\ref{FTSECIII}), Eq. (\ref{FDRM2}) is immediately obtained after replacing these results in (\ref{DKNGF}). 

On the other side, since the initial state of the MCS environment is in a canonical equilibrium state, (\ref{DKNGF}) is related to the noise kernel through a fluctuation-dissipation relation \cite{boyanovsky20051,calzetta20031,roura19991}. This relation was obtained in the frequency domain in \cite{valido20191}, which reads 
\begin{align}
 \tilde{N}_{\alpha\lambda}^{ij}(\omega,\omega')=-\frac{\hbar}{4\pi}\delta(\omega-\omega')\left(1+2n(\omega, \beta^{-1}) \right)&\big( \text{Im}\ \tilde{\Gamma}_{\alpha\lambda}(\omega,\Delta\bar{\vect q}_{ij})+ \text{Im}\ \tilde{\Lambda}^{||}_{\alpha\lambda}(\omega,\Delta\bar{\vect q}_{ij})-i \text{Re}\ \tilde{\Lambda}^{\perp}_{\alpha\lambda}(\omega,\Delta\bar{\vect q}_{ij}) \big). 
\label{FDTI}    
\end{align}
As before, the real-time Fourier transform of Eq.(\ref{FDTI}) retrieves expression (\ref{FDRM1}) after some manipulation. Finally, we get (\ref{NEGF1}) after taking the Wigner-Weyl transform in (\ref{App_NGF}) for $\vect J^{+}=\vect J^{-}=\vect 0$ and reproducing the program introduced in \cite{calzetta20031,boyanovsky20051,anisimov20091}.

Now we briefly show the computation of the fluctuation-dissipation relation (\ref{FDTHBW}) after performing the Breit-Wigner approximation. We start substituting the expressions of the friction matrix $\vect \Gamma_{\text{Brow}}$ and the flux-carrying contribution $\vect \Omega_{\text{flux}}$ in the fluctuation-dissipation relation (\ref{FDRM1}), which retrieves
\begin{align}
 \left\langle \left\lbrace \vect \xi (t),\vect \xi (t')\right\rbrace  \right\rangle_{\hat \rho_{\beta}}=\frac{2m\hbar }{\pi}\int_{0}^{\infty}d\omega & \coth\left(\frac{\hbar \omega\beta}{2}\right) \big(\omega  \cos(\omega (t-t'))\vect \Gamma_{\text{Brow}}+\sin(\omega (t-t'))\vect \Omega_{\text{flux}}^{\circ 2}\big). \label{AppFDRM1}
\end{align}
To obtain the above expression we have replaced the spectral density by means of the relations (\ref{PTSP1}) and (\ref{PTSP2}). Clearly, the fluctuation-dissipation relation consists of two parts: the Brownian part is obtained from the cosine Fourier transform, whereas the flux-carrying contribution involves a sine Fourier transform. The latter transform yields
\begin{equation}
    \int_{0}^{\infty}d\omega \ \coth\left(\frac{\hbar \omega\beta}{2}\right) \sin(\omega t)=\frac{\pi}{\hbar\beta}\coth\Big(\frac{\pi t}{\hbar \beta}\Big),\label{App_ST}
\end{equation}
while the cosine transform,
\begin{equation}
    \int_{0}^{\infty}d\omega \ \omega\coth\left(\frac{\hbar \omega\beta}{2}\right) \cos(\omega t)=-\frac{\pi}{(\hbar\beta)^2 \sinh^2\big(\frac{\pi t}{\hbar \beta}\big)}+\frac{2\pi}{\hbar \beta}\delta(t).\label{App_CT}
\end{equation}
Equation (\ref{App_CT}) was extracted from \cite{oConnell20121}, while (\ref{App_ST}) is deduced by means of the identity
\begin{equation}
    \int_{0}^{\infty}d\omega \ \omega\coth\left(\frac{\hbar \omega\beta}{2}\right) \cos(\omega t)=\frac{d}{dt}\int_{0}^{\infty}d\omega \ \coth\left(\frac{\hbar \omega\beta}{2}\right) \sin(\omega t),
    \nonumber
\end{equation}
where we have employed $\frac{d}{dx}\text{sgn}(x)=2\delta(x)$. Replacing (\ref{App_ST}) and (\ref{App_CT}) in (\ref{AppFDRM1}) returns the expression (\ref{FDTHBW}).


\section{Diffusive coefficients, thermal covariance matrix and time evolution of the covariance matrix}\label{App_QBE}

In this appendix we illustrate the explicit expression of the auxiliary function involved in Eq.(\ref{EMSF1}) from Sec. \ref{Sec_QBE}, the computation of the asymptotic values (\ref{EMDF1ARHT})-(\ref{EMDF1ARLT}) from Sec. \ref{Subsec_QKLT}, and the derivation of the time-dependent covariance matrix (\ref{CVMHSLT}) from Sec. \ref{subsecQHALT}. 

Starting from Eq.(\ref{SigmaMD2}) once replaced (\ref{CH2}), the flux-carrying contribution to the covariance matrix in the single particle scenario can be determined by making use of the symbolic computation handled by MATHEMATICA. Specifically, we determine the auxiliary function appearing in (\ref{EMSF1}), i.e.
\begin{align}
    f(t,\omega)&= \frac{\omega e^{-t \gamma_{\text{Brow}} } }{m\left(\left(\omega ^2-\omega _{\text{ren}}^2\right)^2-\gamma_{\text{Brow}} ^2
   \omega ^2\right)}\Bigg(\frac{\gamma_{\text{Brow}} ^2 \omega ^2+\left(\omega ^2-\omega_{\text{ren}}^2\right)^2}{\gamma_{\text{Brow}} ^2-4 \omega _{\text{ren}}^2} \Bigg(\left(1+e^{t \gamma_{\text{Brow}} }\right) \omega  \left(\gamma_{\text{Brow}} ^2-4 \omega_{\text{ren}}^2\right) \nonumber \\
   &-2 e^{\frac{t \gamma_{\text{Brow}} }{2}} \omega  \cos (t \omega ) \cosh \left(\frac{1}{2} t
   \sqrt{\gamma_{\text{Brow}} ^2-4 \omega_{\text{ren}}^2}\right) \left(\gamma_{\text{Brow}} ^2-4 \omega _{\text{ren}}^2\right)\nonumber \\
   &-2e^{\frac{t \gamma_{\text{Brow}}}{2}} \left(\omega ^2+\omega _{\text{ren}}^2\right) \sin (t \omega ) \sinh
   \left(\frac{1}{2} t \sqrt{\gamma_{\text{Brow}} ^2-4 \omega_{\text{ren}}^2}\right) \sqrt{\gamma_{\text{Brow}} ^2-4
   \omega_{\text{ren}}^2}\Bigg)    \nonumber \\
   &-\frac{\gamma_{\text{Brow}}  \omega  }{\left(\gamma_{\text{Brow}}^2-4   \omega_{\text{ren}}^2\right)^{3/2}}\Bigg)\Bigg(\sinh \left(t \sqrt{\gamma_{\text{Brow}} ^2-4 \omega _{\text{ren}}^2}\right) \omega ^6-2 t \sqrt{\gamma_{\text{Brow}} ^2-4
   \omega _{\text{ren}}^2} \omega ^6 \nonumber \\
   &+3 \gamma_{\text{Brow}} ^2 \sinh \left(t \sqrt{\gamma_{\text{Brow}} ^2-4 \omega_{\text{ren}}^2}\right) \omega ^4-6 \omega _{\text{ren}}^2 \sinh \left(t \sqrt{\gamma_{\text{Brow}} ^2-4 \omega_{\text{ren}}^2}\right) \omega ^4-2 t \gamma_{\text{Brow}} ^2 \sqrt{\gamma_{\text{Brow}} ^2-4 \omega _{\text{ren}}^2} \omega ^4\nonumber \\
   &+2 t\omega _{\text{ren}}^2 \sqrt{\gamma_{\text{Brow}} ^2-4 \omega _{\text{ren}}^2} \omega ^4-\gamma_{\text{Brow}}  \sqrt{\gamma_{\text{Brow}} ^2-4
   \omega _{\text{ren}}^2} \omega ^4+\gamma_{\text{Brow}} ^4 \sinh \left(t \sqrt{\gamma_{\text{Brow}} ^2-4 \omega_{\text{ren}}^2}\right) \omega ^2\nonumber \\
   &+6 \omega _{\text{ren}}^4 \sinh \left(t \sqrt{\gamma_{\text{Brow}} ^2-4 \omega_{\text{ren}}^2}\right) \omega ^2-4 \gamma_{\text{Brow}} ^2 \omega _{\text{ren}}^2 \sinh \left(t \sqrt{\gamma_{\text{Brow}} ^2-4
   \omega _{\text{ren}}^2}\right) \omega ^2\nonumber \\
   &+2 t \omega _{\text{ren}}^4 \sqrt{\gamma_{\text{Brow}} ^2-4 \omega_{\text{ren}}^2} \omega ^2+2 e^{t \gamma_{\text{Brow}} } \gamma_{\text{Brow}} ^3 \sqrt{\gamma_{\text{Brow}} ^2-4 \omega _{\text{ren}}^2} \omega
   ^2+\gamma_{\text{Brow}} ^3 \sqrt{\gamma_{\text{Brow}} ^2-4 \omega _{\text{ren}}^2} \omega ^2 \nonumber \\
   &-2 t \gamma_{\text{Brow}} ^2 \omega _{\text{ren}}^2
   \sqrt{\gamma_{\text{Brow}} ^2-4 \omega _{\text{ren}}^2} \omega ^2-8 e^{t \gamma_{\text{Brow}} } \gamma_{\text{Brow}}  \omega _{\text{ren}}^2
   \sqrt{\gamma_{\text{Brow}} ^2-4 \omega _{\text{ren}}^2} \omega ^2-6 \gamma_{\text{Brow}}  \omega _{\text{ren}}^2 \sqrt{\gamma_{\text{Brow}} ^2-4
   \omega _{\text{ren}}^2} \omega ^2 \nonumber \\
   &+\gamma_{\text{Brow}}  \sqrt{\gamma_{\text{Brow}} ^2-4 \omega _{\text{ren}}^2} \left(\gamma_{\text{Brow}} ^2
   \omega ^2+\left(\omega ^2-\omega _{\text{ren}}^2\right)^2\right) \cosh ^2\left(\frac{1}{2} t
   \sqrt{\gamma_{\text{Brow}} ^2-4 \omega _{\text{ren}}^2}\right) \nonumber \\
   &+\gamma_{\text{Brow}}  \sqrt{\gamma_{\text{Brow}} ^2-4 \omega _{\text{ren}}^2}
   \left(\gamma_{\text{Brow}} ^2 \omega ^2+\left(\omega ^2-\omega _{\text{ren}}^2\right)^2\right) \sinh
   ^2\left(\frac{1}{2} t \sqrt{\gamma_{\text{Brow}} ^2-4 \omega _{\text{ren}}^2}\right) \nonumber \\
   &-2 e^{\frac{t \gamma_{\text{Brow}} }{2}}
   \sqrt{\gamma_{\text{Brow}} ^2-4 \omega _{\text{ren}}^2} \Bigg(2 \omega ^2 \gamma_{\text{Brow}} ^3 \nonumber \\
   &-t \omega ^2 \left(\omega
   ^2+\omega _{\text{ren}}^2\right) \gamma_{\text{Brow}} ^2-8 \omega ^2 \omega _{\text{ren}}^2 \gamma_{\text{Brow}} -t \left(\omega
   ^2-\omega _{\text{ren}}^2\right)^2 \left(\omega ^2+\omega _{\text{ren}}^2\right)\Bigg) \cos (t \omega
   ) \cosh \left(\frac{1}{2} t \sqrt{\gamma_{\text{Brow}} ^2-4 \omega _{\text{ren}}^2}\right) \nonumber \\
   &-2 e^{\frac{t \gamma_{\text{Brow}}
   }{2}} \Bigg(\left(\omega ^2 \gamma_{\text{Brow}} ^4+\left(3 \omega ^4-4 \omega _{\text{ren}}^2 \omega
   ^2+\omega _{\text{ren}}^4\right) \gamma_{\text{Brow}} ^2+2 \left(\omega ^2-\omega _{\text{ren}}^2\right)^3\right)
   \cos (t \omega ) \nonumber \\
   &+\omega  \left(\gamma_{\text{Brow}} ^2-4 \omega _{\text{ren}}^2\right) \left(t \gamma_{\text{Brow}} ^2 \omega
   ^2+t \left(\omega ^2-\omega _{\text{ren}}^2\right)^2+2 \gamma_{\text{Brow}}  \left(\omega ^2+\omega_{\text{ren}}^2\right)\right) \sin (t \omega )\Bigg) \sinh \left(\frac{1}{2} t \sqrt{\gamma_{\text{Brow}} ^2-4
   \omega _{\text{ren}}^2}\right) \nonumber \\
   &-2 \omega _{\text{ren}}^6 \sinh \left(t \sqrt{\gamma_{\text{Brow}} ^2-4 \omega_{\text{ren}}^2}\right)+\gamma_{\text{Brow}} ^2 \omega _{\text{ren}}^4 \sinh \left(t \sqrt{\gamma_{\text{Brow}} ^2-4 \omega_{\text{ren}}^2}\right)-2 t \omega _{\text{ren}}^6 \sqrt{\gamma_{\text{Brow}} ^2-4 \omega _{\text{ren}}^2} \nonumber \\
   &-\gamma_{\text{Brow}}    \omega _{\text{ren}}^4 \sqrt{\gamma_{\text{Brow}} ^2-4 \omega _{\text{ren}}^2}\Bigg).
     \label{AEqf1}
\end{align}
Similarly, one may obtain the explicit expression of the auxiliary function $F(t,\omega)$ involved in (\ref{EMDF1}) with the help of such symbolic computation, which we prefer to skip here since the specific form of the latter is not crucial for the discussion from the main text.

Now, we briefly illustrate the derivation of Eqs. from (\ref{EMDF1ARHT}) to (\ref{EMSF1ARLT}). These are computed by taking the asymptotic time limit $t\rightarrow \infty$ in (\ref{EMSF1}) as well as (\ref{EMDF1}). In such limit the aforementioned auxiliary functions return the unit, i.e.
\begin{equation}
    \lim_{t\rightarrow\infty}F(t,\omega)=-\lim_{t\rightarrow\infty}f(t,\omega)=-\frac{1}{m},
    \label{AppATL}
\end{equation}
which replaced in (\ref{EMDF1}) retrieves after some manipulation
\begin{equation}
    D_{\text{flux}}(\infty)=-\frac{\hbar\Omega_{\text{flux}} ^2}{\pi}\int_{-\infty}^{\infty}  \frac{(\gamma_{\text{Brow}}\omega)^3   }{\left(\gamma_{\text{Brow}} ^2 \omega ^2+\left(\omega ^2-\omega_{\text{ren}}^2\right)^2\right)^2}\coth\left(\frac{\hbar \omega\beta}{2}\right)\ d\omega,
    \label{AEMDF1}
\end{equation}
where we have made use of the fact that the argument of the integrad is an odd function. Since the argument of the latter decays algebraically as faster as $\omega^{3}$ and has no brunch cut, the above integral can be analytically computed by means of contour integration methods. First, we replace the rational expression of the hyperbolic cotangent \cite{weiss20121,fleming20111}, i.e.
\begin{equation}
\coth\left(\frac{\hbar \omega\beta}{2}\right)=  \frac{2}{\hbar\omega\beta} +\frac{2}{\hbar \beta}\sum_{i=1}^{\infty}\Bigg(\frac{1}{\omega+i\frac{2\pi n}{\hbar\beta}}+\frac{1}{\omega-i\frac{2\pi n}{\hbar\beta}}\Bigg),
\nonumber
\end{equation}
which permits to split (\ref{AEMDF1}) into the classical and quantum contributions, i.e.
\begin{align}
    D_{\text{flux}}(\infty)&= -\frac{2\gamma_{\text{Brow}}^3\Omega_{\text{flux}} ^2}{\pi\beta}\int_{-\infty}^{\infty}  \frac{\omega^2   }{\left(\gamma_{\text{Brow}} ^2 \omega ^2+\left(\omega ^2-\omega_{\text{ren}}^2\right)^2\right)^2} d\omega \nonumber \\
    &-\frac{2\gamma_{\text{Brow}}^3\Omega_{\text{flux}} ^2}{\pi\beta}\sum_{i=1}^{\infty}\int_{-\infty}^{\infty} \Bigg[ \frac{\omega^3   }{\left(\gamma_{\text{Brow}} ^2 \omega ^2+\left(\omega ^2-\omega_{\text{ren}}^2\right)^2\right)^2\big(\omega+i\frac{2\pi n}{\hbar\beta}\big)}  \nonumber \\
    &+\int_{-\infty}^{\infty}  \frac{\omega^3   }{\left(\gamma_{\text{Brow}} ^2 \omega ^2+\left(\omega ^2-\omega_{\text{ren}}^2\right)^2\right)^2\big(\omega-i\frac{2\pi n}{\hbar\beta}\big)}\Bigg] d\omega.
    \label{AEMDF2}
\end{align}
Notice that the first line in (\ref{AEMDF1}) provides the value of the flux-carrying diffusive coefficient in the high temperature limit, while the sum term completely encodes the low temperature effects. The former has a degenerate simple pole in the upper-half complex plain given by Eq.(\ref{AppSPR}) aside its complex conjugate $-\eta^{\dagger}_{\text{Brow}}$. The residue associated takes the form,
\begin{align}
    \text{Res}(\eta_{\text{Brow}})&=\frac{\beta\gamma_{\text{Brow}}  \Omega_{\text{flux}} ^2 \hbar   \ 
   \text{csch}^2\left(\hbar  \eta_{\text{Brow}}^{\dagger}\beta/2\right)}{16 \pi  \left(\gamma_{\text{Brow}} ^2-4
   \omega_{\text{ren}}^2\right) \sqrt{4 \gamma_{\text{Brow}} ^2 \omega_{\text{ren}}^2-\gamma_{\text{Brow}} ^4}}  \Bigg(2\hbar  \eta_{\text{Brow}}^{\dagger}\sqrt{4 \gamma_{\text{Brow}} ^2 \omega_{\text{ren}}^2- \gamma_{\text{Brow}} ^4} \nonumber \\
   &+\frac{4 i }{\beta}\left(\gamma_{\text{Brow}}
   ^2-2 \omega_{\text{ren}}^2\right) \sinh \left(\hbar \eta_{\text{Brow}}^{\dagger}\beta\right)\Bigg),
   \nonumber
\end{align}
and its complex conjugate
\begin{align}
    \text{Res}(-\eta^{\dagger}_{\text{Brow}})&=\frac{\beta\gamma_{\text{Brow}}  \Omega_{\text{flux}} ^2 \hbar   \ 
   \text{csch}^2\left(\hbar  \eta_{\text{Brow}}\beta/2 \right)}{16 \pi \left(\gamma_{\text{Brow}} ^2-4
   \omega_{\text{ren}}^2\right) \sqrt{4 \gamma_{\text{Brow}} ^2 \omega_{\text{ren}}^2-\gamma_{\text{Brow}} ^4}} \Bigg(2i\hbar\eta_{\text{Brow}}  \sqrt{4 \gamma_{\text{Brow}} ^2 \omega_{\text{ren}}^2-\gamma_{\text{Brow}} ^4}\nonumber \\
   &+\frac{4 i}{\beta}\left(\gamma_{\text{Brow}}
   ^2-2 \omega_{\text{ren}}^2\right) \sinh \left(\hbar  \eta_{\text{Brow}}\beta\right)\Bigg).
   \nonumber
\end{align}
According to the Cauchy's residue theorem, the first line in (\ref{AEMDF2}) is directly given by $2i\pi(\text{Res}(\eta_{\text{Brow}})+\text{Res}(-\eta_{\text{Brow}}^{\dagger}))$. On the other hand, the low temperature contribution contains additional simple poles that corresponds to the so-called Matsubara frequencies $i \nu_{n}$, i.e. $\nu_{n}=\frac{2\pi}{\hbar \beta} n$ with $n=1,\cdots,\infty$. In this case, the associated residue simplifies to,
\begin{equation}
    \text{Res}(i\nu_{n})=\frac{16 i \pi ^2 \beta ^4 \gamma_{\text{Brow}} ^3 n^3 \Omega_{\text{flux}} ^2 \hbar ^5}{\left(\beta ^4 \omega_{\text{ren}}^4 \hbar ^4+16
   \pi ^4 n^4-4 \pi ^2 \beta ^2 n^2 \hbar ^2 \left(\gamma_{\text{Brow}} ^2-2 \omega_{\text{ren}}^2\right)\right)^2}.
    \nonumber
\end{equation}
Combining these results together, we find after some manipulation the asymptotic value of the flux-carrying diffusive coefficient, that is
 \begin{align}
     D_{\text{flux}}(\infty)&=2i\pi(\text{Res}(\eta_{\text{Brow}})+\text{Res}(-\eta_{\text{Brow}}^{\dagger}))+2i\pi\sum_{n=1}^{\infty} \text{Res}(i\nu_{n})\nonumber \\
     &=-\frac{\hbar\beta\Omega_{\text{flux}}^2}{16\zeta_{\text{Brow}}^3}\bigg[\frac{4}{\beta}\Big(\omega_{\text{ren}}^2-\frac{\gamma_{\text{Brow}}^2}{2} \Big)\Big( \coth\Big(\frac{\hbar\eta_{\text{Brow}}\beta}{2}\Big)+\coth\Big(\frac{\hbar\eta_{\text{Brow}}^{\dagger}  \beta}{2}\Big)\Big)  \nonumber \\
     &+i\hbar\gamma_{\text{Brow}} \zeta_{\text{Brow}}\Big(\eta_{\text{Brow}}\Big(\sinh\Big(\frac{\hbar\eta_{\text{Brow}} \beta}{2}\Big)\Big)^{-2}- \eta_{\text{Brow}}^{\dagger}\Big(\sinh\Big(\frac{\hbar\eta_{\text{Brow}}^{\dagger} \beta}{2}\Big)\Big)^{-2} \Big)\bigg]                                             \nonumber \\
     &-\frac{4\gamma_{\text{Brow}}^3\Omega_{\text{flux}}^2}{\beta}\sum_{n=1}^{\infty}\frac{\nu_{n}^3}{|\eta_{\text{Brow}}-i\nu_{n}|^4|\eta_{\text{Brow}}+i\nu_{n}|^4}, \label{EMDF1AR}
\end{align}
which returns the expressions (\ref{EMDF1ARHT}) and (\ref{EMDF1ARLT}) after taking the high and low temperature limits, respectively. Although it is not show here, one can repeat this procedure to obtain the time asymptotic expression of the thermal covariance matrix by starting from (\ref{EMSF1}) after replacing the result (\ref{AppATL}). This yields
\begin{align}
    \sigma_{\text{flux}}(\infty) &=\frac{\hbar\beta\Omega_{\text{flux}}^2}{8\zeta_{\text{Brow}}^3}\bigg[\frac{\gamma_{\text{Brow}}}{\beta}\Big( \coth\Big(\frac{\hbar\eta_{\text{Brow}}\beta}{2}\Big)+\coth\Big(\frac{\hbar\eta_{\text{Brow}}^{\dagger}  \beta}{2}\Big)\Big)  \nonumber \\
     &-i\hbar\zeta_{\text{Brow}}\Big(\eta_{\text{Brow}}\Big(\sinh\Big(\frac{\hbar\eta_{\text{Brow}} \beta}{2}\Big)\Big)^{-2}- \eta_{\text{Brow}}^{\dagger}\Big(\sinh\Big(\frac{\hbar\eta_{\text{Brow}}^{\dagger} \beta}{2}\Big)\Big)^{-2} \Big)\bigg]                                             \nonumber \\
     &-\frac{4\Omega_{\text{flux}}^2}{\beta \hbar}\sum_{n=1}^{\infty}\frac{\nu_{n}(\nu_{n}^4+\nu_{n}^2(\gamma_{\text{Brow}}^2+2\omega_{\text{ren}}^2)+\omega_{\text{ren}}^4)}{|\eta_{\text{Brow}}-i\nu_{n}|^4|\eta_{\text{Brow}}+i\nu_{n}|^4}, \label{EMSF1AR}
\end{align}
from which we obtain Eqs. (\ref{EMSF1ARHT}) and (\ref{EMSF1ARLT}) in the high and low temperature limits. One may contrast these with the asymptotic values of the diffusive and thermal covariance coefficients of the standard Brownian motion. These are well-known for weak system-environment coupling, for instance, in the Drude-model of the spectral density with a high-frequency $\omega_{c}$ \cite{grabert19841}. In particular, in the low-temperature limit and large cutoff (i.e. $1\ll \omega_{c}/\omega_{\text{ren}}$), these are given by \cite{fleming20111,lombardo20051}
\begin{align}
     D_{\text{Brow}}^{(pp)}(\infty)&=m\hbar\omega_{\text{ren}}\gamma_{\text{Brow}}+\mathcal{O}(\beta^{-1}), \label{DCLTS1} \\
     D_{\text{Brow}}^{(qp)}(\infty)&=-2\hbar\gamma_{\text{Brow}}\log\Big(\frac{\omega_{c}}{\omega_{ren}}\Big)+\mathcal{O}(\beta^{-1}),  \nonumber \\
     D_{\text{Brow}}^{(qq)}(\infty)&=0, \label{DCLTS2}
\end{align}
as well as
\begin{align}
     \sigma_{\text{Brow}}^{(qq)}(\infty)&=\frac{1}{m\omega_{\text{ren}}^2}\bigg(\frac{D_{\text{Brow}}^{(pp)}(\infty)}{2m\gamma_{\text{Brow}}}-D_{\text{Brow}}^{(qp)}(\infty)\Bigg), \label{TCVCLTS1} \\
     \sigma_{\text{Brow}}^{(pp)}(\infty)&=\frac{D_{\text{Brow}}^{(pp)}(\infty)}{2\gamma_{\text{Brow}}},\label{TCVCLTS2} \\
     \sigma_{\text{Brow}}^{(qp)}(\infty)&=0. \label{TCVCLTS3}
\end{align}

Finally, we analyse the time evolution of the single-particle scenario at late times. One can make use of standard Green's function methods to determine the system state $W(\vect x,t)$ at any time near the asymptotic equilibrium. Concretely, given an arbitrary initial state $W(\vect x,0)$, the system state is obtained from \cite{agarwal19711,ford20011},  
\begin{equation}
    W(\vect x, t)=\int_{\mathbb{R}^{4}} W(\vect x',0)\ \mathcal{K}_{R}(\vect x,\vect x',t) \ d^{4}\vect x',
    \label{ETWF}
\end{equation}
where $\mathcal{K}_R$ denotes the retarded kinetic propagator defined in the phase space, i.e.
\begin{equation}
    \mathcal{K}_{R}(\vect x,\vect x',t)=\frac{1}{(2\pi)^2\sqrt{\det(\vect \sigma_{BW}(t))}}\text{Exp}\bigg(-\frac{1}{2}\big(\vect x-\vect  K_{BW}(t)\cdot \vect x'\big)^{T}\vect \vect \Delta_{BW}^{-1}(t)\big(\vect x- \vect K_{BW}(t)\cdot \vect x'\big)\bigg), \nonumber
\end{equation}
with 
\begin{eqnarray}
\vect \Delta_{BW}(t)&=&\vect \sigma(\infty)-\vect K_{BW}(t)\cdot \vect \sigma(\infty)\cdot \vect K_{BW}^{T}(t),\nonumber
\end{eqnarray}
where $\vect K_{BW}(t)$ is given by Eq. (\ref{GFBWA1}), and $\vect \sigma(\infty)$ is obtained from expressions (\ref{CTMB}) and (\ref{CTMF}) after substituting the asymptotic values (\ref{TCVCLTS1}) to (\ref{TCVCLTS3}), and (\ref{EMSF1AR}). Considering an initial Gaussian state characterized by an arbitrary covariance matrix $\vect V(0)$, expression (\ref{ETWF}) hence returns a Gaussian state for all times with zero mean values and covariance matrix determined by \cite{agarwal19711,fleming20111}
\begin{equation}
    \vect V(t)=\vect \sigma(\infty)+\vect K_{BW}(t)\cdot( \vect V(0)-\vect \sigma(\infty))\cdot \vect K_{BW}^{T}(t).
    \label{CVEQ}
\end{equation}
Since the retarded kinetic propagator $\vect K_{BW}(t)$ is an exponential decaying function in time provided the subsidiary condition is satisfied (\ref{SDC1PD}), from Eq.(\ref{CVEQ}) is clear that $\vect V(t)\rightarrow \vect \sigma(\infty)$ for $t\rightarrow \infty$. This result is employed in Sec. \ref{subsecQHALT} to study the hydrodynamic properties of the single flux-carrying Brownian particle. In the low-temperature limit, the covariance matrix $\vect V(t)$ can be obtained from (\ref{CVEQ}) by appealing to Eqs. (\ref{TCVCLTS1})-(\ref{TCVCLTS3}) and (\ref{EMSF1ARLT}).

\bibliographystyle{apsrev4-1}
\bibliography{references}

\end{document}